\documentclass[preprint]{aastex}
\usepackage{amsmath}
\usepackage{amssymb}
\usepackage{bm}
\usepackage{ulem}
\usepackage{xspace}
\usepackage{graphicx}
\usepackage{subfigure}
\usepackage{longtable}
\shorttitle{X-ray Spectral Model of Reprocessing by AGN Torri with 
{\tt MONACO}}
\shortauthors{Furui et al.}

\begin{document}


\title{X-ray Spectral Model of Reprocessing by Smooth and Clumpy
Molecular Tori in Active Galactic Nuclei with the {\tt MONACO} framework}


\author{Shun'ya Furui\altaffilmark{1}, Yasushi
Fukazawa\altaffilmark{1,2,3}, Hirokazu Odaka\altaffilmark{4,5},
Toshihiro Kawaguchi\altaffilmark{6}, Masanori Ohno\altaffilmark{1,3},
Kazuma Hayashi\altaffilmark{1}}

\email{\texttt{fukazawa@hep01.hepl.hiroshima-u.ac.jp}}


\altaffiltext{1}{Department of Physical Science, Hiroshima University,
1-3-1 Kagamiyama, Higashi-Hiroshima, Hiroshima 739-8526, Japan}
\altaffiltext{2}{Hiroshima Astrophysical Science Center, Hiroshima University, 1-3-1 Kagamiyama, Higashi-Hiroshima, Hiroshima 739-8526, Japan}
\altaffiltext{3}{Core Research for Energetic Universe (Core-U), Hiroshima University, 1-3-1 Kagamiyama, Higashi-Hiroshima, Hiroshima 739-8526, Japan}
\altaffiltext{4}{Kavli Institute for Particle Astrophysics and Cosmology, Stanford University 2575 Sand Hill Rd, Menlo Park, CA 94025}
\altaffiltext{5}{Institute of Space and Astronautical Science (ISAS), Japan Aerospace Exploration Agency (JAXA), 3-1-1 Yoshinodai, Chuo, Sagamihara, Kanagawa, 252-5210, Japan}
\altaffiltext{6}{Department of Liberal Arts and Sciences, Sapporo Medical University, S1W17, Chuo-ku, Sapporo 060-8556, Japan}

%\altaffiltext{1}{Hiroshima Astrophysical Science Center, Hiroshima University, 1-3-1 Kagamiyama, Higashi-Hiroshima, Hiroshima 739-8526, Japan}
%\altaffiltext{3}{Department of Physics, University of Tokyo, 7-3-1 Hongo, Bunkyo, Tokyo, 113-0033, Japan}
%\altaffiltext{3}{Dublin Institute for Advanced Studies (DIAS), 31 Fitzwilliam Place, Dublin 2, Ireland}
%\altaffiltext{4}{Max-Planck-Institut f\"ur Kernphysik, Saupfercheckweg 1, Heidelberg 69117, Germany}

\newcommand{\degree}{$^\circ$~}


\begin{abstract}
  % To extract the physical interpretation of the
  % variability, a radiation model of the thermal and bulk
  % Comptonization of the accreted plasma has been constructed in the
  % framework of Monte Carlo simulations. The modeling of the
  % observational data indicated that thermal Comptonization plays an
  % important role in the formation of the spectrum featured by a hard
  % power law with a quasi-exponential cutoff at 20-30 keV.  We also
  % find a natural, positive correlation between the mass accretion rate
  % and the optical thickness of the accreted plasma. A synergy of the
  % Monte-Carlo code with the analytical solution for the accretion
  % column derived by \citet{Becker:1998} allowed to obtain
  % self-consistent sets of accretion column parameters (accretion rate,
  % column radius and height of the sonic point), which lead to the
  % observed range of luminosities. 

We construct an X-ray spectral model of reprocessing by a torus in an active 
galactic nucleus (AGN) with a Monte Carlo simulation framework {\tt MONACO}.
Two torus geometries of smooth and clumpy cases are considered and compared.
In order to reproduce a Compton shoulder accurately, {\tt MONACO} 
includes not only free electron scattering but also bound electron scattering.
Raman and Reyleigh scattering are also treated, and scattering cross
 sections dependent on chemical states of hydrogen and helium are
 included.
Doppler broadening by turbulence velocity can be implemented.
Our model gives consistent results with other available models,
 such as {\tt MYTorus}, except for differences due to different 
physical parameters and assumptions.
We studied the dependence on torus parameters for Compton shoulder, and
found that a intensity ratio of Compton shoulder to line core mainly
 depends on the column density, inclination angle, and metal abundance.
For instance, an increase of metal abundance makes
the Compton shoulder relatively weak.
Also, shape of Compton shoulder depends on the column density.
Furthermore, 
these dependences become different between smooth and clumpy cases.
Then, we discuss the possibility of {\it ASTRO-H} SXS spectroscopy of Compton
 shoulder in AGN reflection spectra.

\end{abstract}




%\bibliographystyle{apj}

%% square-bracket argument, for instan{ce, \object[\[WEG2004\] 14h-090]{90}).

\section{Introduction}

Active galactic nuclei (AGNs) have supermassive black holes (SMBHs) at
the center of each galaxy, and accretion discs surrounding the SMBHs
emit an enormous energy of $10^{42}$--$10^{47}$ $\mathrm{erg~s^{-1}}$  from region whose size is as small as the solar system.
Based on various observational results, a huge gas-dust structure, so-called ``torus'', presumably exists around the accretion disc (Antonucci \& Miller 1985), and hides the central engine with a substantial fraction from distant observers.
The torus is likely composed of inflowing materials from a parent galaxy
to an accretion disc and thus related to the evolution of SMBH. 
For X-ray radiation of AGN, we observe not only a direct (e.g.,
transmitted) component from accretion disc/corona but also a reprocessed
component by the torus.
The reprocessed component contains information on such as X-ray fluorescence, absorption and reflection (e.g. Awaki et al. 1991; Fukazawa et al. 2011).
Therefore, it is important to clarify what can be learned about the
state of matter (e.g., temperature, degree of ionization, and velocity
dispersion) via the observed X-ray spectra.

In order to estimate the state of materials in the tori, various X-ray spectral models have been constructed and made comparison with the observed data.
Among them, a commonly used model is {\tt pexrav} (Magdziarz \& Zdziarski 1995) in XSPEC.
This is the reflection model from a flat infinite disc, but fluorescence lines are not included in this model.
There is an improved model ({\tt pexmon}) (Nandra et al. 2007) by adding fluorescence lines to the {\tt pexrav} model.
Furthermore, more complicated models are available (Ikeda et al. 2009;
Murphy \& Yaqoob 2009; Brightman \& Nandra 2011), 
and Liu \& Li (2014, 2015) has begun to study
X-ray spectra from the AGN torus with Geant4,
a widely used Monte Carlo simulation library in fields ranging from 
high-energy particle physics to space science (Ivanchenko et al. 2003).
As seen in these models, more realistic shapes of the torus and physical
processes have come to be thought in late years, and they enable us to
estimate the state of materials in detail.
Brightman et al. (2015) reported some comparisons of models of
reprocessing by torus among 
{\tt pexrav}, {\tt MYTorus} (Murphy \& Yaqoob 2009), and the model of
Brightman \& Nandra (2011).
However, most of these models fall under any of the following: the distribution of material is smooth, only simple geometry is discussed, and the effect of velocity dispersion and/or scattering with bound electron are not considered.

To sustain the geometrical thickness of the torus, the velocity dispersion must be as large as a typical rotation velocity of the torus,  $\sim100~\mathrm{km~s^{-1}}$.
However, thermal velocity of these dusty gas cannot become fast up to this speed, because dust grains reach the sublimation temperature $T\sim1500~\mathrm{K}$ (Barvainis 1987; Laor \& Draine 1993). 
Therefore, the torus is most likely a group of many dusty clumps with a large velocity dispersion (clumpy torus), rather than a smooth mixture of gas and dust (smooth torus) (Krolik \& Begelman 1988).
Although simple geometries (e.g., smooth gas fulfilled in doughnuts)
have been applied in most torus models for X-ray studies, an actual
torus is likely (1) clumpy, and (2) has a concave shape at the innermost
region because of anisotropic emissions from the accretion disc (Kawaguchi \& Mori 2010; 2011).
Moreover, (3) the velocity dispersion of the innermost part is as high as thousands $\mathrm{km~s^{-1}}$ (Kawaguchi 2013), which inevitably affects the interpretations and predictions 
for future high-resolution spectroscopy, which will be realized by the
{\it ASTRO-H} X-ray observatory, scheduled for launch in 2016 
(Takahashi et al. 2014).

All these realistic geometrical configurations and kinematics should be
considered for the analysis of high-resolution spectra. In order to build
such a realistic spectral model, we must solve a problem of radiative
transfer which requires treatment of a discrete process of a photon,
competing processes, and multiple interactions in a complicated
geometry. This is obviously difficult by analytical methods, and we
therefore adopt a Monte Carlo approach as the {\tt MYTorus} project 
(Murphy \& Yaqoob 2009). 
As the first step of this project, this paper focuses on
the effects of the clumpiness and the velocity dispersion, though we
still use a simple geometry of the entire torus which is adopted in
{\tt MYTorus}. Our simulation model is constructed with a Monte Carlo
simulation framework, {\tt MONACO} (MONte Carlo simulation for 
Astrophysics and COsmology), which was originally developed and
verified in the context of X-ray reflection nebulae in the Galactic
Center region (Odaka et al. 2011).

In the Compton-thick reflection case, a conspicuous feature, so-called
Compton shoulder, is formed in adjacent to emission lines, after scattering of line photons.
Since the shape of Compton shoulder changes with various parameters such
as the state of scattering electron (free or bound, etc.) and the
kinematics of materials, the Compton shoulder is a strong probe to
investigate the state of matter around the black hole (Murphy \& Yaqoob 2009).
Matt (2002) presented the dependence of the Compton shoulder against the
column density and inclination angle, but only for reflector geometry of
sphere or plane.
Yaqoob and Murphy (2011) studied the Compton shoulder for the reflector
geometry of smooth torus, and mainly showed the dependence of the
Compton shoulder shape on the column density, inclination, incident
spectral slope, and velocity dispersion.
Our simulator based on {\tt MOCACO} includes accurate physical processes
considering Compton scattering by not only free electrons but also bound
electrons in atoms or molecules for hydrogen and helium.
Since these processes are not included properly in other works, 
we are able to investigate the shape of Compton shoulder more accurately
and in detail.
In addition, we for the first time study the Compton shoulder 
for the clumpy torus.

In this work, we present results of simulations on a smooth and a clumpy torus and discuss the effects of clumpiness.
We explain the torus geometry, the Monte Carlo code, and the simulation process in Section 2.
We present results of the reflection continuum and Compton shoulder in Section 3, followed by discussions in Section 4.

\section{Models and Calculation Methods}

\subsection{Torus Geometry and Material Properties}
\label{sec:model}

\begin{figure*}[htbp]
\centerline{
		\includegraphics[clip, width=15cm]{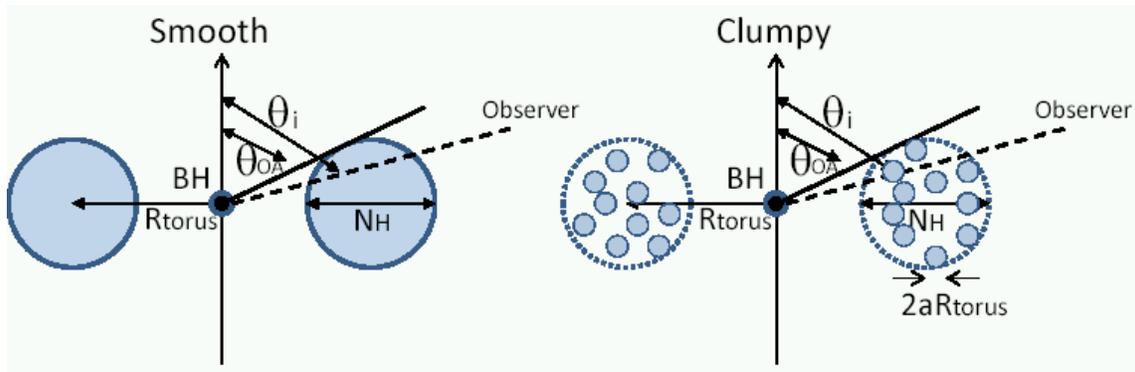}
}
	\caption{Torus geometry implemented in our model for smooth
	(left) and clumpy (right) cases. Inclination angle $\theta_i$,
	opening angle $\theta_{\rm OA}$, torus radius $R_{\rm torus}$,
	column density $N_{\rm H}$ are also shown.}
	\label{fig:torus_geom}
\end{figure*}

We construct both smooth and clumpy torus models for the Monte Carlo simulations.
Figure \ref{fig:torus_geom} shows the cross-section view of these tori: the smooth one is shown in the left panel and the clumpy one in the right.
In the both models, matter is distributed in a common torus geometry
that is determined by the opening angle $\theta_\mathrm{OA}$ and the
major radius $R_\text{torus}$ of the torus surrounding the central black hole.
The smooth torus has a uniform hydrogen number density $n_\mathrm{H}$ and therefore the column density measured along a line on the equatorial plane from the central black hole is given by $N_\mathrm{H} = 2n_\mathrm{H}R_\text{torus}\cos\theta_\mathrm{OA}$.
We assume chemical composition of solar values obtained by Anders \& Grevesse (1989), and elemental abundances of metals (lithium and heavier elements) can be scaled by a common factor $A_\text{metal}$ (metal abundance relative to the solar values).
In addition, the simulation is able to treat Doppler effects due to random motion of the torus material.
The gas motion is treated as micro-turbulence for simplisity, so the gas velocity has a Gaussian distribution with a standard deviation of $v_\text{turb}$, which is called a turbulent velocity.

The clumpy torus consists of a number of small spherical clumps all of which have common radius, density, and chemical composition.
These clumps are randomly distributed in the torus geometry identical to one adopted in the smooth torus model.
The positions of the clumps are randomly determined so as to keep 
uniformity of the torus.
Then, if there are geometrical overlaps between the clumps, the conflicting clumps are rearranged in order to define clumps as geometrically separate objects.
We introduce a clump scaling factor $a$ which determines the clump radius as $R_\text{clump}=aR_\text{torus}$.
The degree of clumpiness is characterized by a volume filling factor $f$, which is defined as a fraction of volume that is filled with the clumps in the whole torus volume.
The number of clumps is calculated to be $fV_\text{torus}/V_\text{clump}=(3/2)f\pi a^{-3} \cos^2\theta_\mathrm{OA}$, where we use that the torus volume is $V_\text{torus}=2\pi^2 R_\text{torus}{}^3\cos^2\theta_\mathrm{OA}$ and the clump volume is $V_\mathrm{clump}=(4/3)\pi R_\mathrm{clump}{}^3$.
The hydrogen number density $n_\mathrm{H}$ in a single clump should be enhanced by a factor of $f^{-1}$ from the number density averaged over the whole torus, namely we have a relation: $N_\mathrm{H} = 2fn_\mathrm{H}R_\text{torus}\cos\theta_\mathrm{OA}$. 

The average number of clumps along a radial equatorial
direction is calculated by the product of the clump density
in a unit volume [$f / (4 \pi R_{\rm clump}^3 / 3)$],
the cross section of one clump ($\pi R_{\rm clump}^2$)
and the length of the region where clumps are
located ($2 R_{\rm torus} \cos \theta_{\rm OA} $)
(e.g., Kawaguchi \& Mori 2011, equation A6).
With our fidicial parameters, it is
$7.5 (f/0.05) (a/0.005)^{-1} (\cos \theta_{\rm OA} / 0.5)$,
within the probable range of this quantity (5--15)
estimated by Nenkova et al. (2008).

Hydrogen can exist in a state of atom or molecule.
Although both forms can be treated as the same way for most of the photon interactions, only Rayleigh scattering for molecular hydrogen should be enhanced by a factor of two per electron because of coherent effects (Sunyaev \& Churazov, 1996).
Our simulation code is able to treat both atomic hydrogen and molecular hydrogen.
Most hydrogens within each clump likely form
molecules (e.g., Pier \& Voit 1995).
For instance, the temperature inside a clump
even at the innermost edge of the torus is around
500\,K or less (H{\"o}nig et al. 2006, their Figure 4 and
Table 1).
Although hard X-ray photons can reach deep into
each clump,
a large fraction ($33-95$\%) of hydrogen within
a clump form molecules (Krolik \& Lepp 1989, their
Table 2).
In this paper, we assume that all hydrogens
exist as H$_2$ molecules.

%Since xxx, most hydrogens probably form in molecules.
%\textit{Based on the results by Honig et al. (2006), their Figure 4 and Table 1, the temperature of the torus is likely $\sim$\,500\,K.}
%In this paper, we assume that all hydrogens exist as $\mathrm{H}_2$ molecules.

In short, our model is specified by the geometry of the torus and the material properties (i.e. the chemical composition and the turbulent velocity).
Essentially, the column density controls the total amount of the matter; thus, we do not need to care the absolute size of the system.
In Table~\ref{table:list-parameters}, we summarize independent model parameters of the smooth torus model as well as additional parameters of the clumpy torus model.

\begin{table}[htbp]
\caption{Independent Model Parameters}
\begin{center}
\begin{tabular}{cc}
\hline\hline
Common parameters & Range \\
\hline
Torus opening angle $\theta_\mathrm{OA}$ & 60$^\circ$ \\
Column density $N_\mathrm{H}$ & $1\times 10^{21}$--$1\times 10^{26}$ $\mathrm{cm}^{-2}$ \\
Metal abundance $A_\text{metal}$ & 0.1--10 solar \\
Turbulent velocity $v_\text{turb}$ & 0--3000 km s$^{-1}$ \\
\hline
Additional parameters of clumpy torus & Range \\
\hline
Clump scaling factor $a$ & 0.002--0.005 \\
Volume filling factor $f$ & 0.01--0.05 \\
\hline
\end{tabular}
\end{center}
\label{table:list-parameters}
\end{table}%

\subsection{Monte Carlo Code and Photon Interactions}

A Monte Carlo approach is suitable for solving radiative transfer in the AGN torus since multiple interactions must be treated in a complicated optically thick geometry.
For our simulations, we use {\tt MONACO} (Odaka et al.\ 2011), a Monte-Carlo calculation framework of X-ray radiation for general astrophysical problems.
This code utilizes the Geant4 toolkit library (Agostinelli et al.\ 2003; Allison et al.\ 2006) for tracking photons in a complicated geometry.
Although Geant4 already has its own physical process libraries mainly for radiation measurements, we do not use them but introduce our original physics implementations required for astrophysical purposes.
Several photon processes playing important roles in astrophysics---e.g.\ photoelectric absorption,
scattering, photoionization, photoexcitation, and Comptonization---are
included in {\tt MONACO} and we select necessary processes among them for the purpose of the simulation.
{\tt MONACO} is able to treat a variety of geometries composed of neutral matter as well as ionized plasma.
The Doppler shift and broadening due to bulk and random (thermal and turbulent) motions are also calculated in the photon tracking calculation.

Physics implementation used in this work was originally developed and was verified in the context of X-ray reflection from molecular clouds by Odaka et al.\ (2011).
As described in Section \ref{sec:model}, we assume that matter in the torus is cold and hydrogens are all in molecular forms.
In this condition, we need to consider photoelectric absorption followed by a fluorescence emission and scattering by an electron that is bound to hydrogen and helium as the interactions of photons with matter.
{\tt MONACO} uses the Evaluated Photon Data Library 97 (EPDL97)\footnote{https://www-nds.iaea.org/epdl97/}, which is distributed together with the Geant4 library, as cross section data of photoelectric absorption.
In our physics implementation, a K-shell fluorescent photon is generated with a probability of a fluorescence yield just after the photoelectric absorption.
Atomic properties relevant to the fluorescence, namely \ K-shell line energies, fluorescence yields, and K$\beta$-to-K$\alpha$ ratios are taken from Thompson et al.\ (2001), Krause et al.\ (1979), and Ertu\u{g}ral et al.\ (2007), respectively.
K$\alpha_2$-to-K$\alpha_1$ intensity ratios are fixed to 0.5.
When an Auger electron is generated, tracking is stopped.
Table \ref{table:kenergy} summarizes the energies of K-edge, K-lines, and the
fluorescence yields used in our model.

Scattering by electrons also plays an important role in generating spectral features particularly at the hard X-ray band above 10 keV and the Compton shoulders of the iron K line at 6.4 keV.
Since most electrons are bound to atoms or molecules in the condition of interest, photons are mostly scattered by electrons bound to hydrogen and helium.
Binding an electron to an atom or molecule alters the scattering process, though many previous studies of the AGN reflection had considered only Compton scattering by free electrons at rest.
The scattering process by a bound electron can be classified under three channels by difference
in the final state of the target electron: (1) Rayleigh scattering to the ground state, (2) Raman scattering to excited states, and (3) Compton scattering to free states (Sunyaev \& Churazov, 1996).
The shape of the Compton shoulder is greatly modified by the electron binding since the target electron is not at rest but has finite momentum in the atomic or molecular system.
Thus, accurate treatment of the scattering process by a bound electron is of great importance in evaluating detailed spectral features measured by high-resolution spectroscopy.
Details about the physics implementation are described in Odaka et al.\ (2011) (see appendix in this reference).

\begin{table}[htbp]
\caption{Energies of K-edge, K-lines, and fluorescence yields}
\begin{center}
\begin{tabular}{cccccccc}
\hline\hline
$Z^a$ & Element & K-edge$^b$ & K$\alpha_1^b$ & K$\alpha_2^b$ & K$\beta^b$ & $Y_{{\rm K}\alpha}^c$ & $Y_{{\rm K}\beta}^c$ \\
\hline
1 & H  &    13.6 &     &     &     &  &  \\ 
2 & He &    23.4 &     &     &     &  &  \\ 
3 & Li &    59.9 &   54.30 &     &     &  &  \\ 
4 & Be &   118.4 &  108.50 &     &     &  &  \\ 
5 & B  &   195.6 &  183.30 &     &     & 0.0017 &  \\ 
6 & C  &   291.0 &  277.00 &     &     & 0.0028 &  \\ 
7 & N  &   404.9 &  392.40 &     &     & 0.0052 &  \\ 
8 & O  &   537.3 &  524.90 &     &     & 0.0083 &  \\ 
9 & F  &   688.4 &  676.80 &     &     & 0.0130 &  \\ 
10 & Ne &   858.2 &  848.60 &  848.60 &     & 0.0180 &  \\ 
11 & Na &  1064.0 & 1040.98 & 1040.98 & 1071.10 & 0.0230 &  \\ 
12 & Mg  &  1294.5 & 1253.60 & 1253.60 & 1302.20 & 0.0296 & 0.0004 \\ 
13 & Al  &  1549.9 & 1486.70 & 1486.27 & 1557.45 & 0.0382 & 0.0008 \\ 
14 & Si &  1828.5 & 1739.98 & 1739.38 & 1835.94 & 0.0487 & 0.0013 \\ 
15 & P  &  2130.4 & 2013.70 & 2012.70 & 2139.10 & 0.0604 & 0.0026 \\ 
16 & S &  2455.9 & 2307.84 & 2306.64 & 2464.04 & 0.0736 & 0.0044 \\ 
17 & Cl &  2804.9 & 2622.39 & 2620.78 & 2815.60 & 0.0906 & 0.0063 \\ 
18 & Ar &  3177.6 & 2957.70 & 2955.63 & 3190.50 & 0.1068 & 0.0112 \\ 
19 & K  &  3583.3 & 3313.80 & 3311.10 & 3589.60 & 0.1258 & 0.0142 \\ 
20 & Ca &  4015.0 & 3691.68 & 3688.09 & 4012.70 & 0.1452 & 0.0178 \\ 
21 & Sc &  4465.8 & 4090.60 & 4086.10 & 4460.50 & 0.1668 & 0.0212 \\ 
22 & Ti &  4940.6 & 4510.84 & 4504.86 & 4931.81 & 0.1897 & 0.0243 \\ 
23 & V  &  5439.6 & 4952.20 & 4944.64 & 5427.29 & 0.2151 & 0.0278 \\ 
24 & Cr &  5957.6 & 5414.72 & 5405.51 & 5946.71 & 0.2424 & 0.0325 \\ 
25 & Mn &  6510.9 & 5898.75 & 5887.65 & 6490.45 & 0.2692 & 0.0388 \\ 
26 & Fe &  7083.4 & 6403.84 & 6390.84 & 7057.98 & 0.3003 & 0.0398 \\ 
27 & Co &  7680.7 & 6930.32 & 6915.30 & 7649.43 & 0.3275 & 0.0455 \\ 
28 & Ni &  8302.8 & 7478.15 & 7460.89 & 8264.66 & 0.3583 & 0.0477 \\ 
29 & Cu &  8943.2 & 8047.78 & 8027.83 & 8905.29 & 0.3873 & 0.0526 \\ 
30 & Zn &  9622.4 & 8638.86 & 8615.78 & 9572.00 & 0.4166 & 0.0574 \\ 
\hline
\multicolumn{8}{l}{$a$: atomic number} \\
\multicolumn{8}{l}{$b$: energies in unit of eV} \\
\multicolumn{8}{l}{$c$: fluorescence yield of K lines} \\
\end{tabular}
\end{center}
\label{table:kenergy}
\end{table}%

\subsection{Comparison with the {\tt MYTorus} model}

To demonstrate the performance of our simulation model generated by {\tt MONACO}, we make a comparison with a widely used X-ray spectral model for AGN torus studies that is also based on Monte Carlo simulations.
We select {\tt MYTorus} (Murphy \& Yaqoob, 2009) as a benchmark model since it assumes a simple, well-defined torus geometry with an opening angle of 60$^\circ$ and a metal abundance of 1 solar value, where the solar values are based on Anders \& Grevesse (1989).
Then, we performed Monte Carlo simulations with {\tt MONACO} to generate a spectral model for parameters of $\theta_\mathrm{OA}=60^\circ$, $A_\text{metal}=1.0$, and $N_\mathrm{H}=1 \times 10^{24}\ \mathrm{cm^{-2}}$.
The initial spectrum is assumed to be a power law with a photon index of 1.9 in an energy range of 2--300 keV, and $6.4\times 10^{8}$ incident photons were simulated in total to make the spectrum.

The spectrum emerging from the AGN torus system can be divided into two components, namely direct component and reprocessed component.
The direct component is composed of photons that are initially emitted at the central source and then escape from the system without any interaction.
The reprocessed component is a result of scattering and fluorescence following photoelectric absorption.
Since {\tt MONACO} is able to distinguish these two components for each
observed photon and {\tt MYTorus} also provides the two components separately, we make the comparison between the two models for each component to demonstrate effects of detailed implementations of the physical processes.

\begin{figure*}[htbp]
\centerline{
	\includegraphics[clip, width=3.5in]{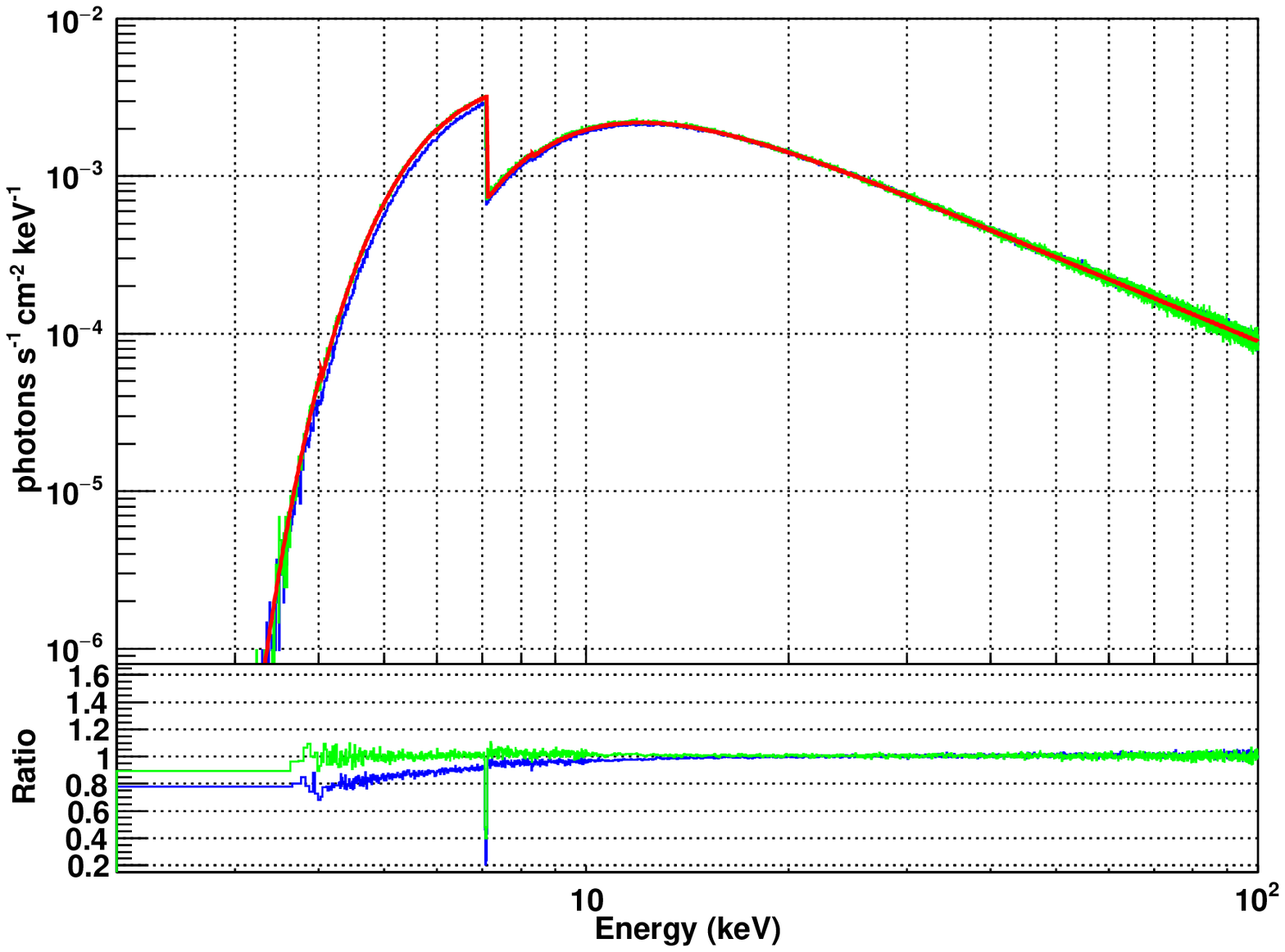}
	\hfil
	\includegraphics[clip, width=3.5in]{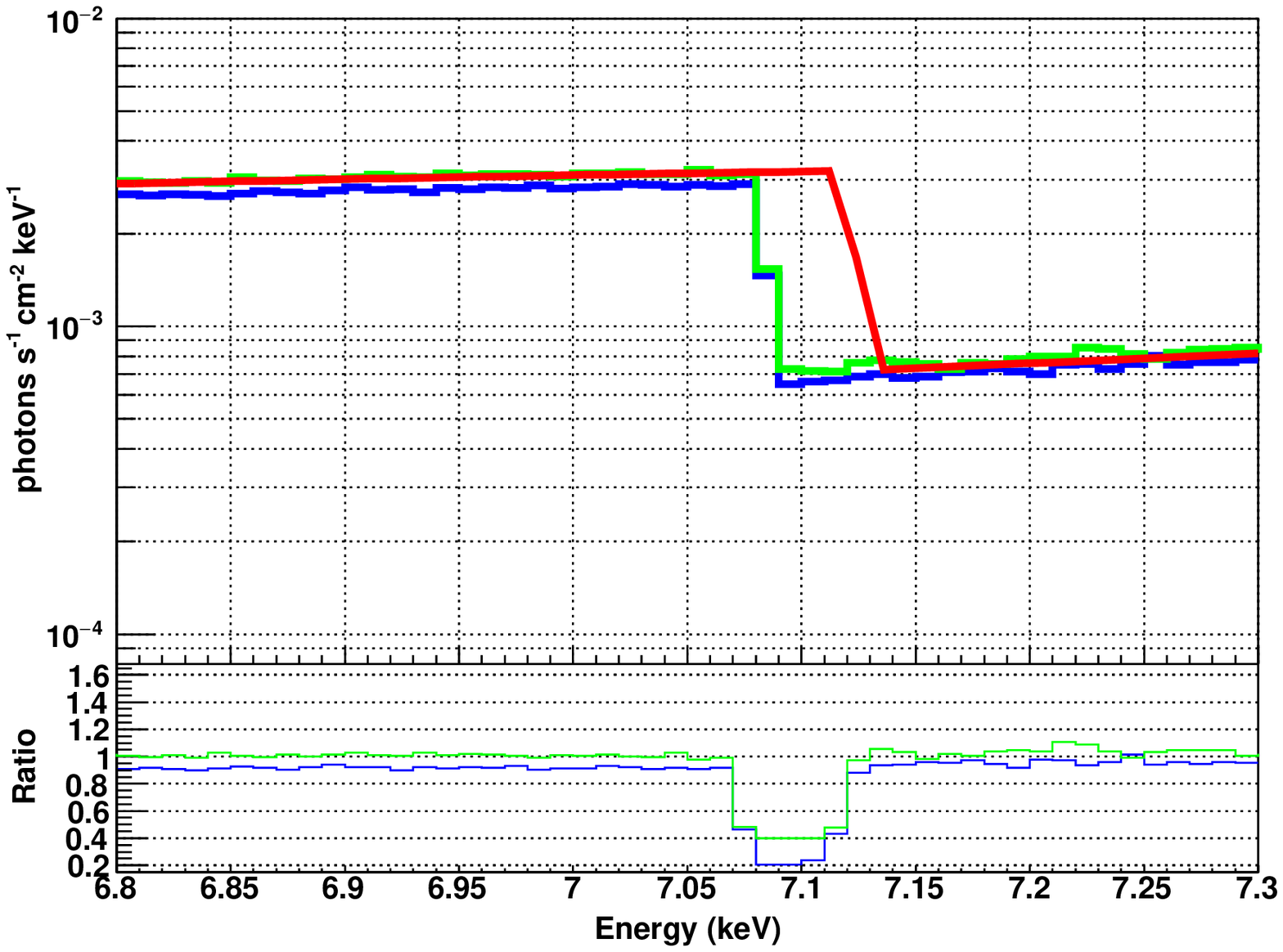}
}
	\caption{Spectra of the direct component for comparison between
 {\tt MONACO} and {\tt MYTorus}. The left panel is shown in 2--100 keV while
 the right panel is enlarged around the iron K-shell edge at 7
 keV. Three models are shown: the {\tt MYTorus} model (red), the free electron
 case generated with {\tt MONACO} (green), and the bound electron case
 generated with {\tt MONACO} (blue). (See text in detail).  The bottom
 panels in each figure are spectral ratios against {\tt MYTorus}. 
}
	\label{fig:comp_torus_dir}
\end{figure*}

In Figure~\ref{fig:comp_torus_dir}, we show spectra of the direct
component for comparison between {\tt MONACO} and {\tt MYTorus}.
These spectra are extracted by integrating escaping photons that have direction within $0.1 \le \cos\theta_i <0.2$ (or $81.4^\circ \pm 2.9$).
In this work, we assume that in the AGN torus all electrons responsible
for scattering are bound to atoms or molecules, while {\tt MYTorus} uses
Compton scattering by free electrons at rest.
In order to check consistency with {\tt MYTorus}, we compare three spectra:
{\tt MYTorus} model spectrum (red line in the figure), a {\tt MONACO} spectrum
assuming free electrons at rest (green), and a {\tt MONACO} spectrum assuming electrons bound to hydrogen molecules and helium atoms (blue).
The comparison shows excellent agreement among the three models above 10
keV within 5\%.
In the low energy band, our model of the bound electron case gives a
less direct component, and this can be understood as described below
for a similar comparison of the scattering component.
Also, another difference can be seen in the enlarged view around the iron K-shell edge at 7 keV.
The difference in energy of the iron edge comes from difference in the origin of the cross section data of photoelectric absorption used in the two models.
The {\tt MONACO} spectrum of the bound electron scattering shows slightly lower than that of the free electron scattering since a cross section of Rayleigh scattering is enhanced by a factor of two for a hydrogen molecule, resulting in insignificant reduction of the direct component.
Note that this direct component is easily calculated by the total cross section (absorption and scattering), which means that we do not need detailed Monte Carlo simulations in practice.
{\tt MYTorus} actually provides the direct component model without Monte Carlo simulations.
Here we have verified the Monte Carlo simulations in {\tt MONACO} by using the direct component which is easily calculated.

\begin{figure*}[htbp]
\centerline{
	\includegraphics[clip, width=3.5in]{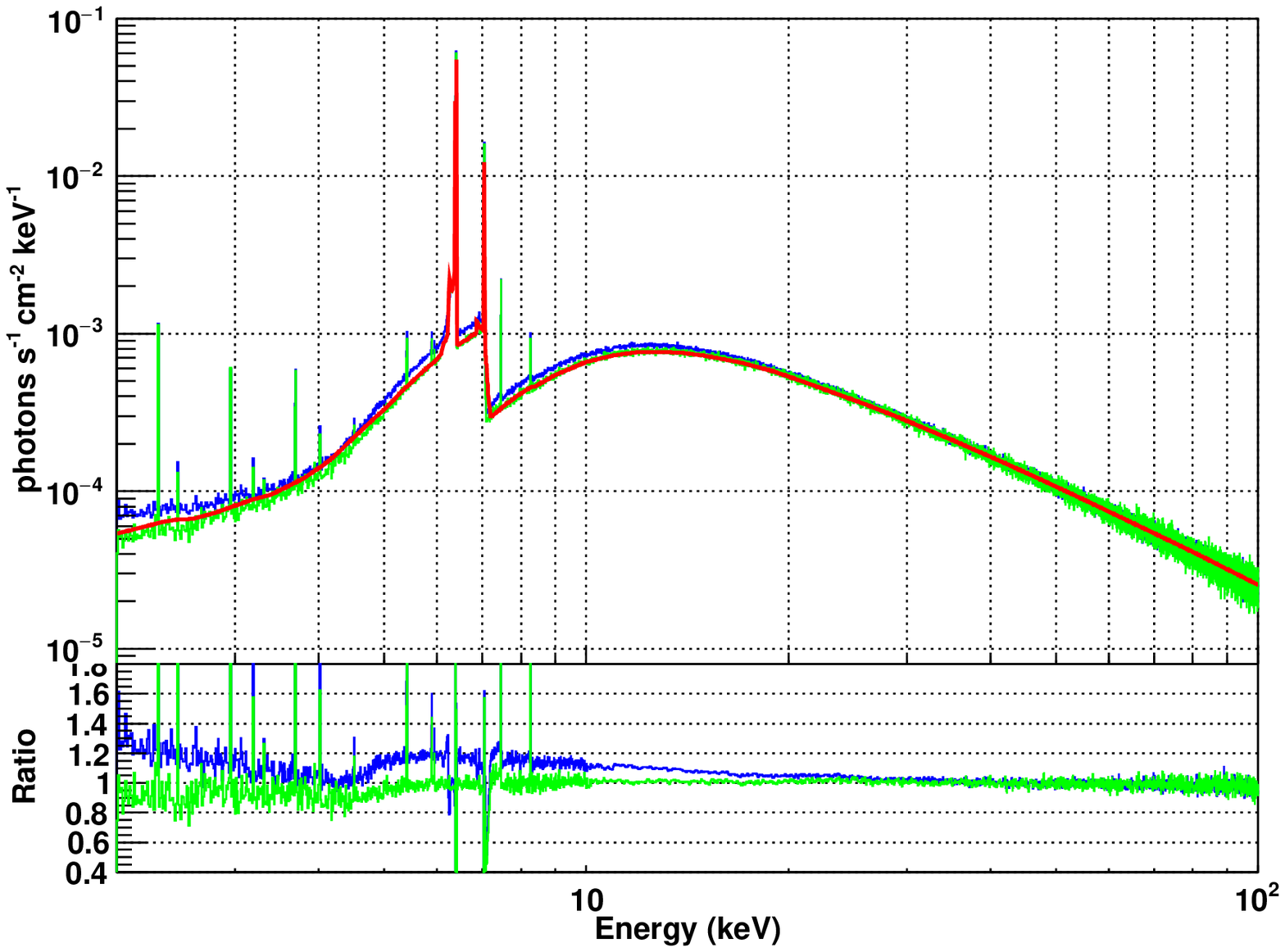}
	\hfil
	\includegraphics[clip, width=3.5in]{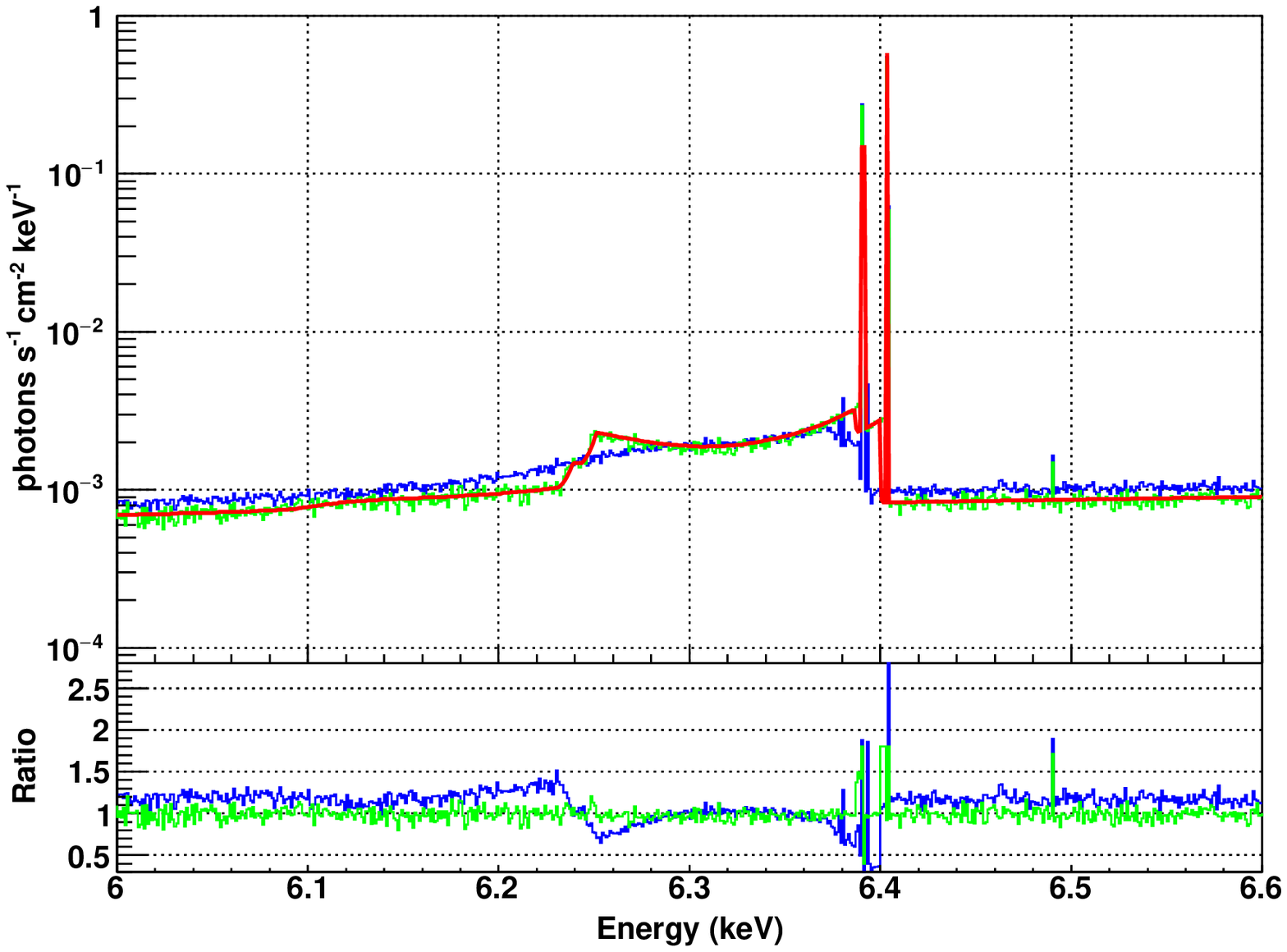}
}
\centerline{
	\includegraphics[clip, width=3.5in]{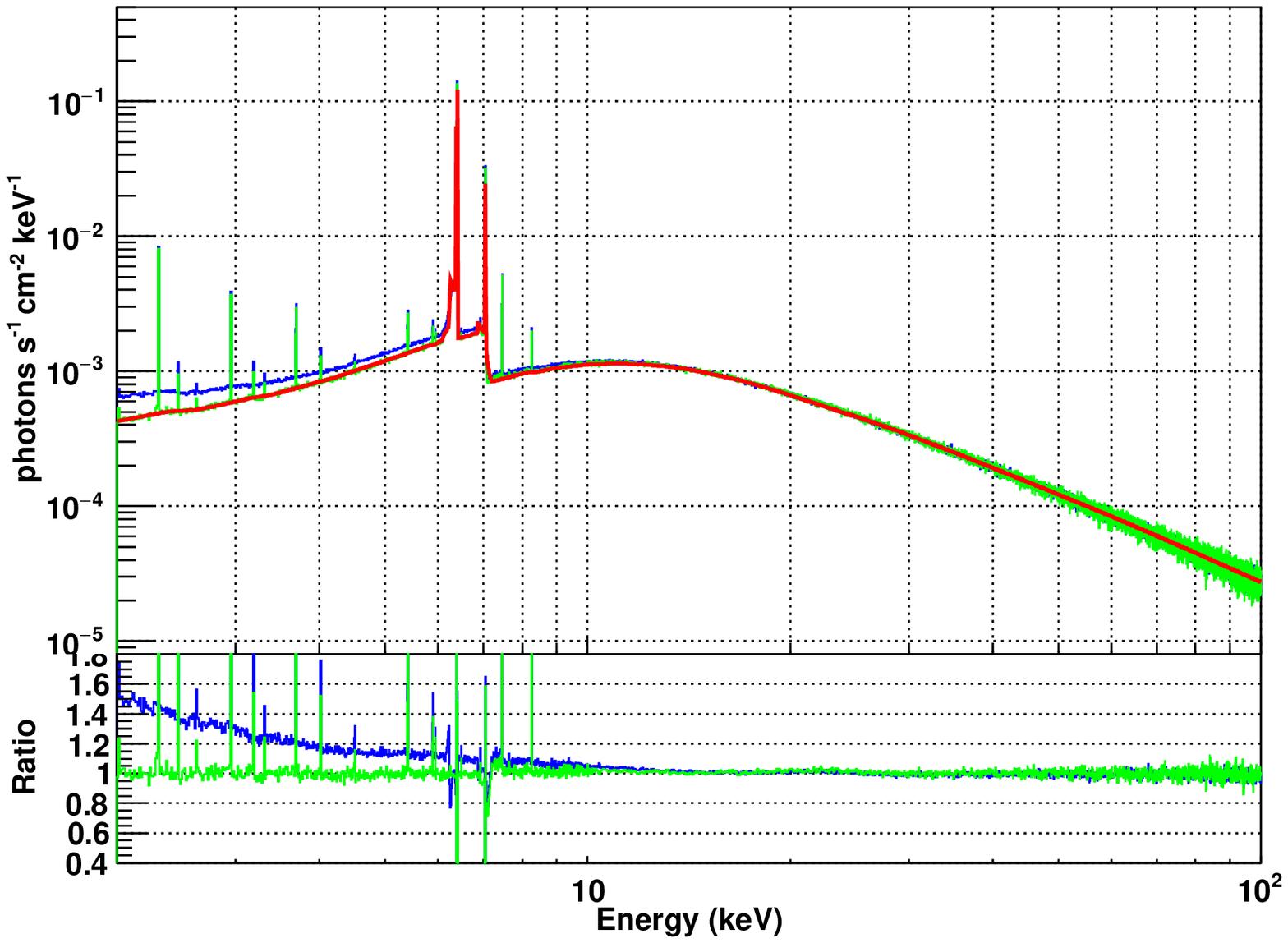}
	\hfil
	\includegraphics[clip, width=3.5in]{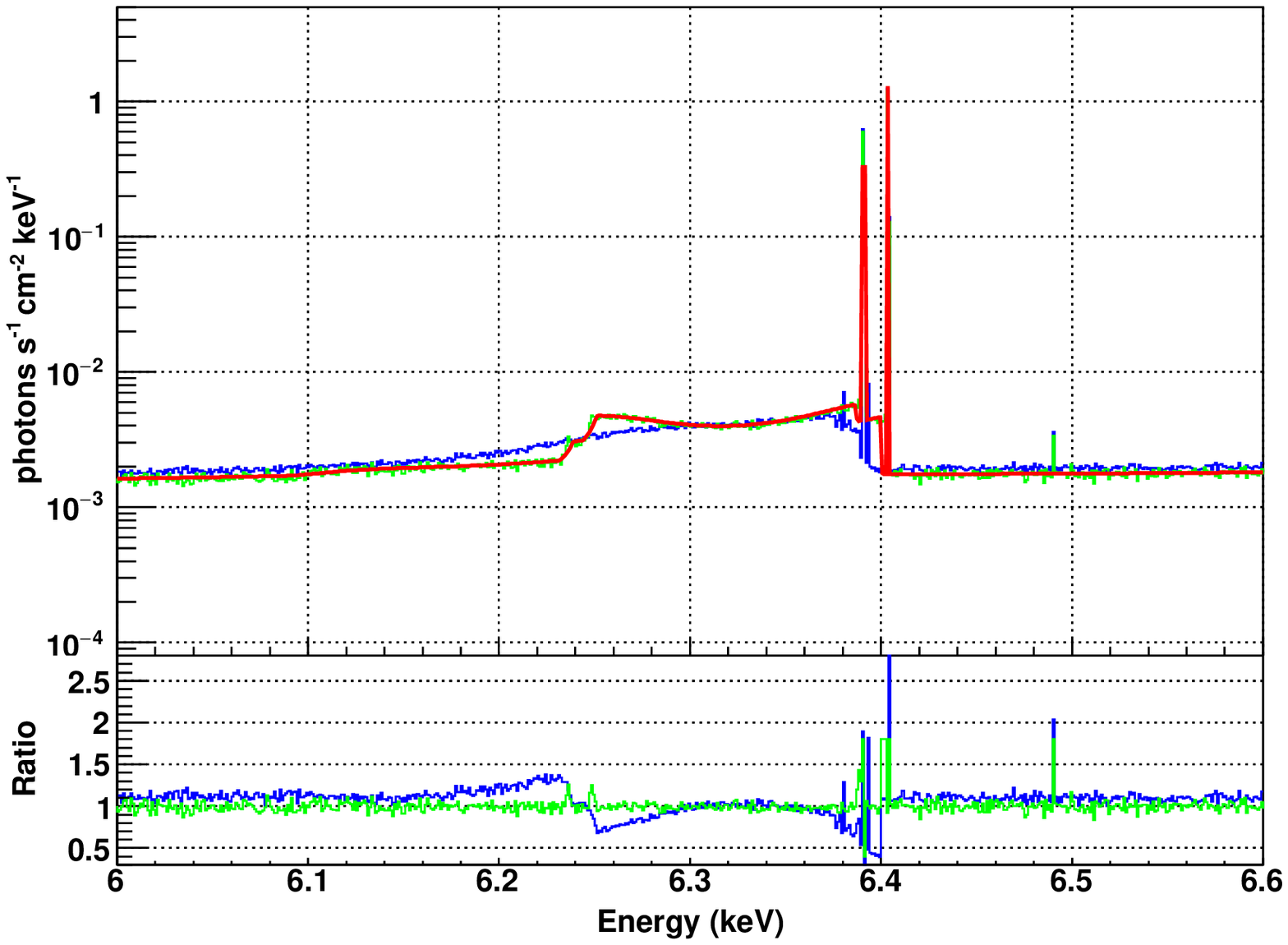}
}
\centerline{
	\includegraphics[clip, width=3.5in]{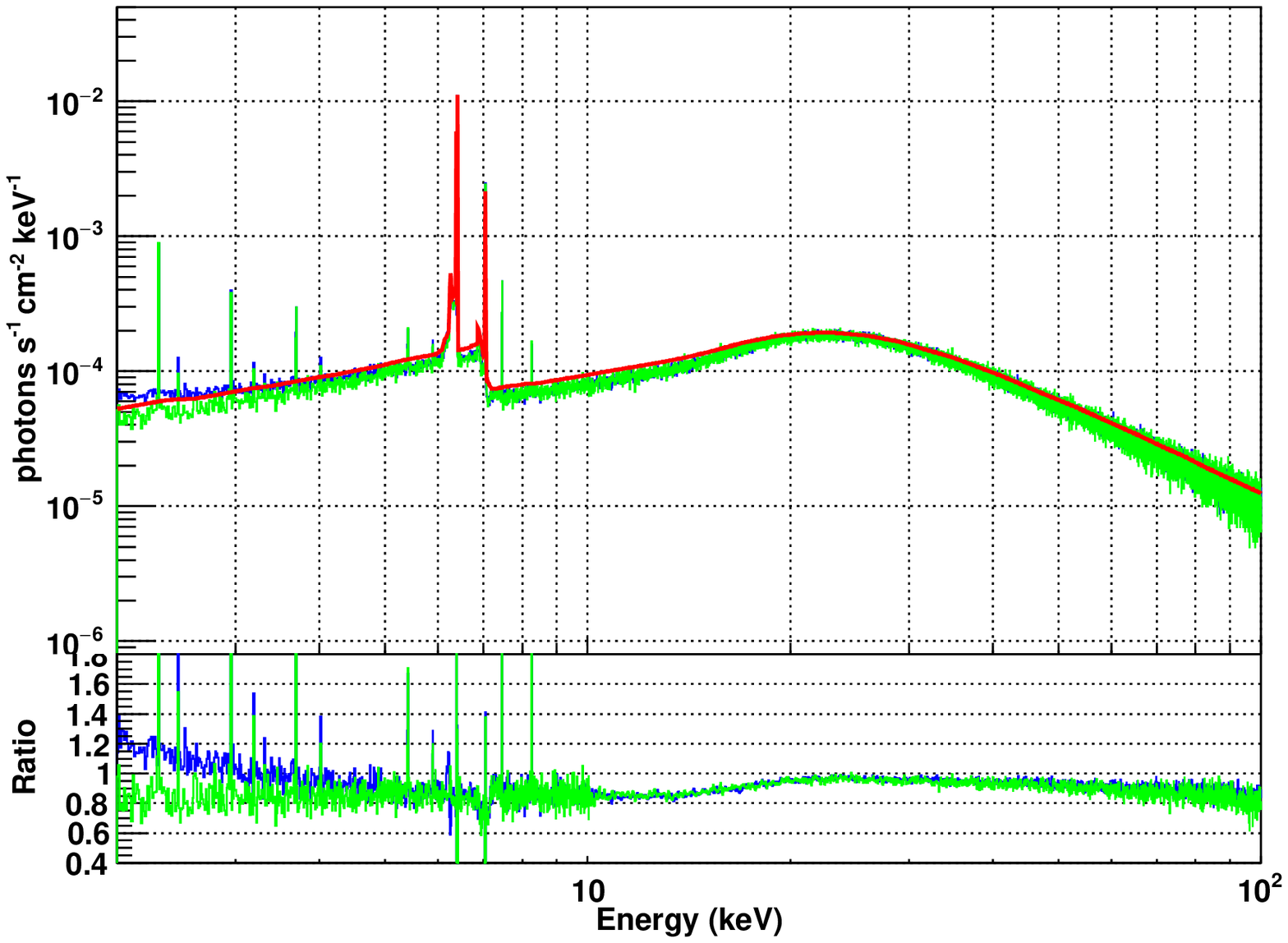}
	\hfil
	\includegraphics[clip, width=3.5in]{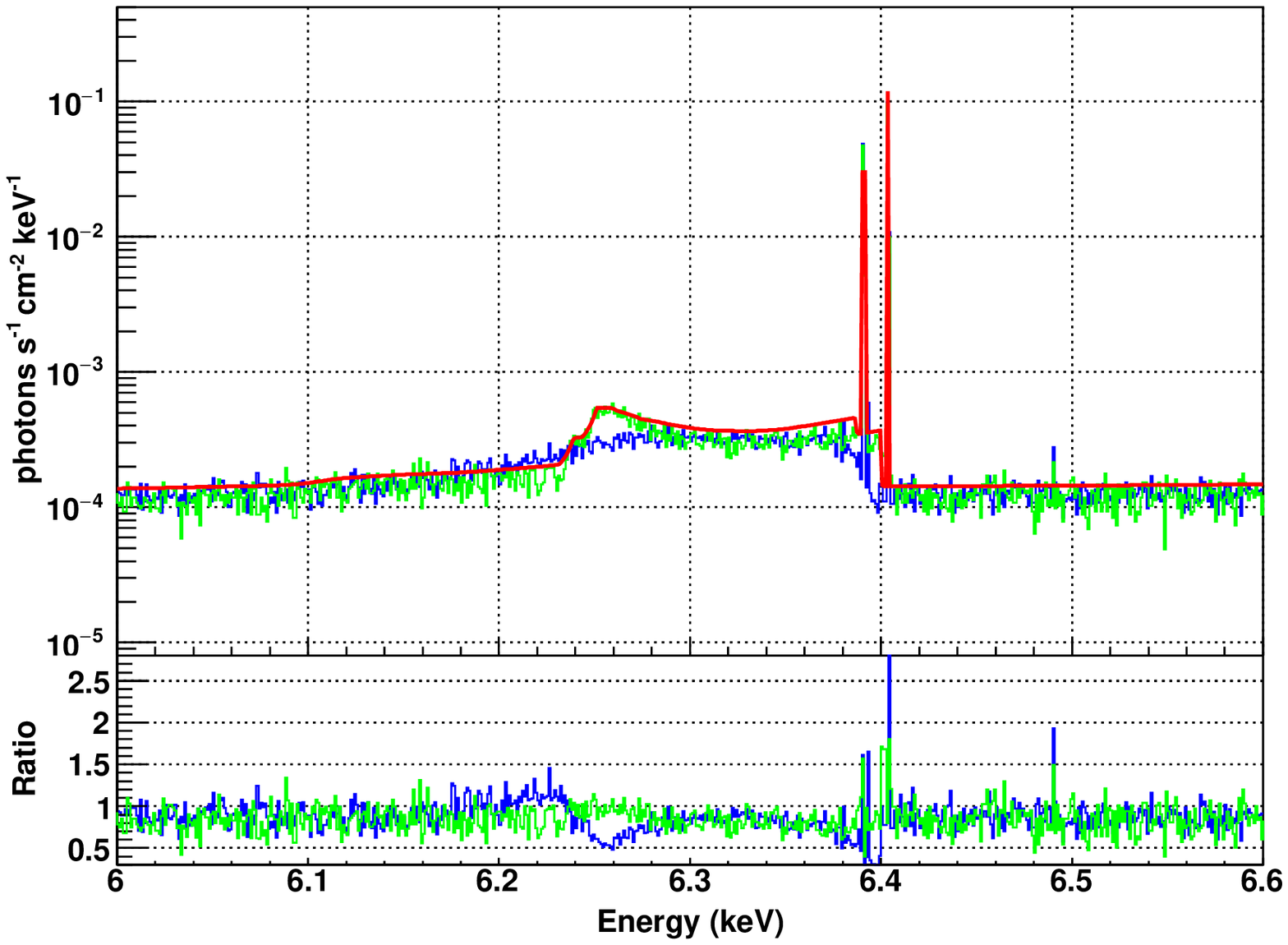}
}
        \caption{Spectra of the reprocessed component for comparison
between {\tt MONACO} and {\tt MYTorus}. The left panel is shown in 2--100 keV
 while the right panel is enlarged around the iron K$\alpha$ line at 6.4
 keV. Three models are shown: the {\tt MYTorus} model (red), the free electron
 case generated with {\tt MONACO} (green), and the bound electron case
 generated with {\tt MONACO} (blue). (See text in detail). The bottom
 panels in each figure are spectral ratios against {\tt MYTorus}. 
Top and middle are the comparison in the case of 
$0.1 \le \cos\theta_i <0.2$ (or $81.4^\circ \pm 2.9$)
and $0.6 \le \cos\theta_i <0.7$ (or $49.3^\circ \pm 3.8$) for
 $N_\mathrm{H}=1 \times10^{24}\ \mathrm{cm^{-2}}$, respectively, and bottom is 
in the case of $N_\mathrm{H}=1 \times
10^{25}\ \mathrm{cm^{-2}}$ and $0.1 \le \cos\theta_i <0.2$.
}
	\label{fig:comp_torus_ref}
\end{figure*}

Then, we compare the reprocessed component, which is best calculated by accurate Monte Carlo simulations.
Figure~\ref{fig:comp_torus_ref} shows the spectra of the reprocessed component given by the three different models.
These spectra are extracted by integrating escaping photons that have
direction within $0.1 \le \cos\theta_i <0.2$ (or $81.4^\circ \pm 2.9$)
and $0.6 \le \cos\theta_i <0.7$ (or $49.3^\circ \pm 3.8$).
Also, the comparison in the case of $N_\mathrm{H}=1 \times
10^{25}\ \mathrm{cm^{-2}}$ and $0.1 \le \cos\theta_i <0.2$ are shown.
Again, we can see excellent agreement among them in the broadband view
while the {\tt MONACO} spectra (green and blue lines) display fluorescent lines of all abundant elements in addition to iron.
The {\tt MONACO} spectrum of the bound electron scattering (blue line in the figure) shows higher than the other spectra at the lower energy band below 20 keV.
This is also because of the enhancement of Rayleigh scattering, showing consistency with the reduction seen in the direct component.

An interesting difference in the spectral shape appears in the Compton shoulder associated with the iron K$\alpha$ line, as shown in the right panels of Figure~\ref{fig:comp_torus_ref}.
The {\tt MONACO} spectrum of the free electron scattering perfectly
agrees with {\tt MYTorus}, simply because the physical conditions assumed in the two models are completely identical.
The bound electron case, however, shows a different profile of the
Compton shoulder particularly at the low energy edge of the shoulder at
6.24 keV which corresponds to the maximum energy transfer to the recoil
electron (i.e.\ at the scattering angle of 180$^\circ$ with respect to
the incident direction).
While the case of free electrons at rest shows a sharp edge at 6.24 keV, atomic or molecular binding smears the Compton shoulder profile.
This is a result of non-zero momentum of the target electron bound to an atom or molecule which broadens the energy distribution of the scattered photon.
This effect by electron binding is quite similar to what we would see if free electrons had thermal motion in a plasma (see e.g.\ Sunyaev \& Churazov, 1998).
It is essential to treat scattering by bound electrons for
high-resolution spectroscopy to evaluate the Compton shoulder seen in the AGN reflection.

%Table \ref{tb:Fe_EW} shows the equivalent width of iron line of each model.
%\begin{table}[htb]
%\begin{center}
%\begin{tabular}{ccc}
%\hline
%Model & Fe-K$\alpha_1$ & Fe-K$\alpha_2$ \\
%& (keV) & (keV) \\
%\hline
%MONACO (Blue) & 0.788 & 0.177 \\
%MONACO (Green) & 0.685 & 0.128 \\
%MYTorus (Red) & 0.695 & 0.122 \\
%\hline
%\end{tabular}
%\caption{}
%\label{tb:Fe_EW}
%\end{center}
%\end{table}

\clearpage

\section{Results}

Here, we present X-ray spectra simulated by {\tt MONACO} for an AGN
torus in smooth and clumpy cases, together with behaviors of 
Comoton shoulder of Fe-K$\alpha$ line, for various torus conditions.
We represent an intrinsic spectrum by a power-law shape with a photon index 1.9 in the range of 2--300 keV, and generated a total number of $10^9$ photons for each run.
These photons were emitted into any directions within a $4\pi$ solid
angle from the center isotropically.
The abundance are referred to Anders \& Grevesse (1989), and 
hydrogen and helium in the torus are assumed to exist as molecules and
atoms, respectively.
We sorted the photons going out of the torus into 20 bins evenly spaced in cosine of inclination angles in the range of $-1\leq \cos\theta_{i}\leq+1$, where $\theta_{i}$ is the angle between the direction of photon and the z-axis.
In this section, if we do not specify torus parameters explicitly, 
we display X-ray reflection spectra with a condition of
$0.1\leq \cos\theta_{i}\leq0.2$ ($\theta_i=78.46-84.26$), 
$N_{\rm H}=10^{24}$ cm$^{-2}$,
$\theta_\mathrm{OA}=60^{\circ}$, $A_{\rm metal}=1.0$ solar, and $V_{turb}=0 $ km s$^{-1}$.
Namely, this fiducial paramter set represents a nearly
edge-on view of a torus.
We studied the dependences on the column density for both of smooth and
clumpy cases, on the inclination and metal abundance for the smooth case,
and on the volume filling factor and clump radius for the clumpy case.

Since simulation spectra have poission noise espcially in the high
energy band, we created the simulated spectra with 1 eV bin, and then
smoothed by running average with 41 bins in $E\leq8$ keV or 
$41 + 2\sqrt{(E{\rm (keV))}-8)/0.05}$
bins in $E>8$ keV, where $E$ is a photon energy.
In this case, the smoothing precedure was applied only for the continuum, by
excluding the line region.
Then, the number of energy bins is 298000.

%smoothed it
%10 times with 353QH algorithm on the {\tt Root}\footnote{\tt
%https://root.cern.ch/} package; running medians of three, followed by
%running medians of five, followed by running medians of three..w

%%%%%%%%%%%%%%%%%%%%%%%%%%%%%%%%%%%%%%%%%%%%%%%%%%%%%%%
%%%%%%%%%%%%%%%%%%%%%%%%%%%%%%%%%%%%%%%%%%%%%%%%%%%%%%%
%%%%%%%%%%%%%%%%%%%%%%%%%%%%%%%%%%%%%%%%%%%%%%%%%%%%%%%

\subsection{Dependence of Hydrogen Column Density (Smooth, Clumpy)}

\begin{figure*}[htbp]
\centering
\subfigure{
	\includegraphics[clip, width=80mm]{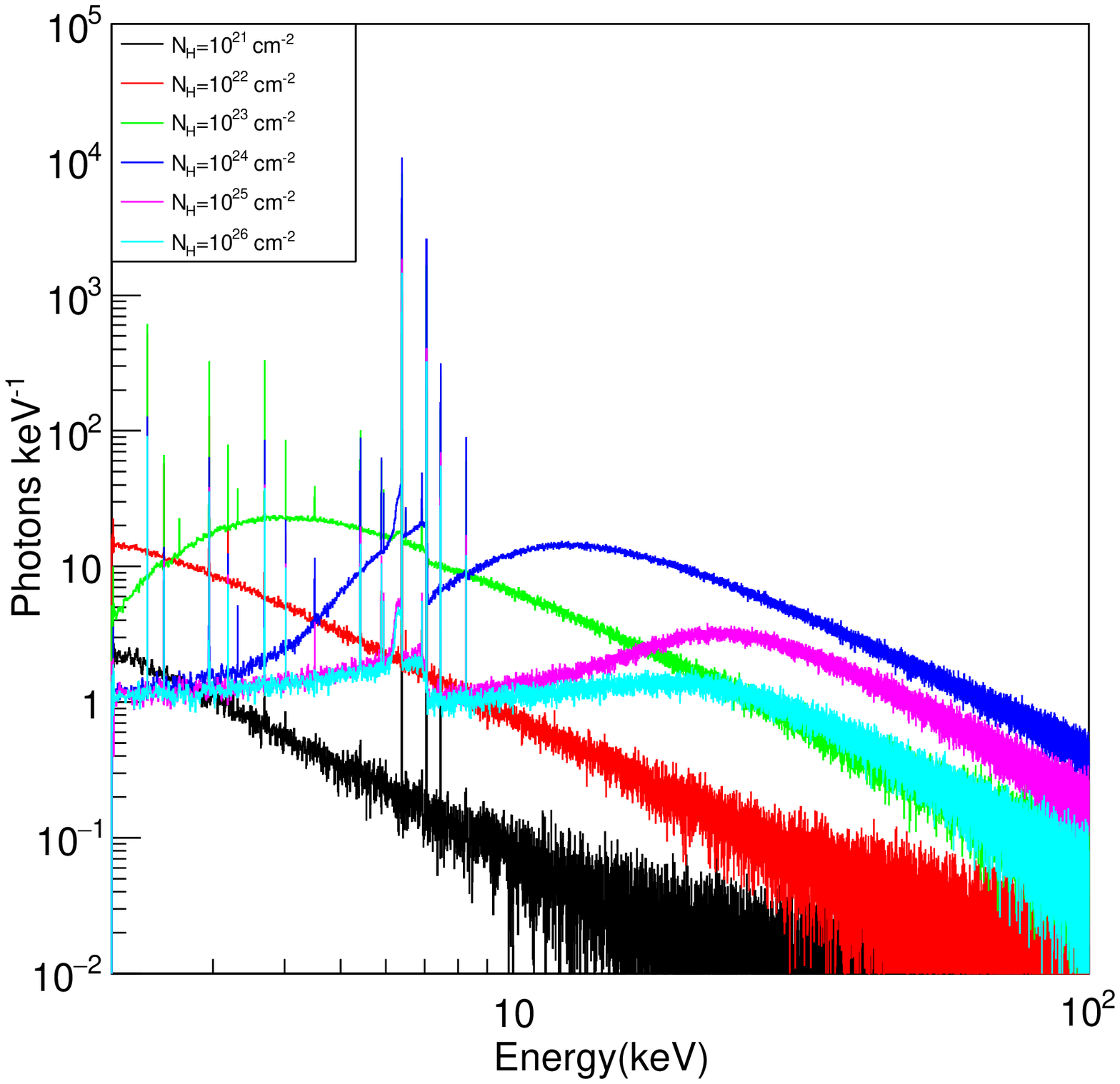}
	\includegraphics[clip, width=80mm]{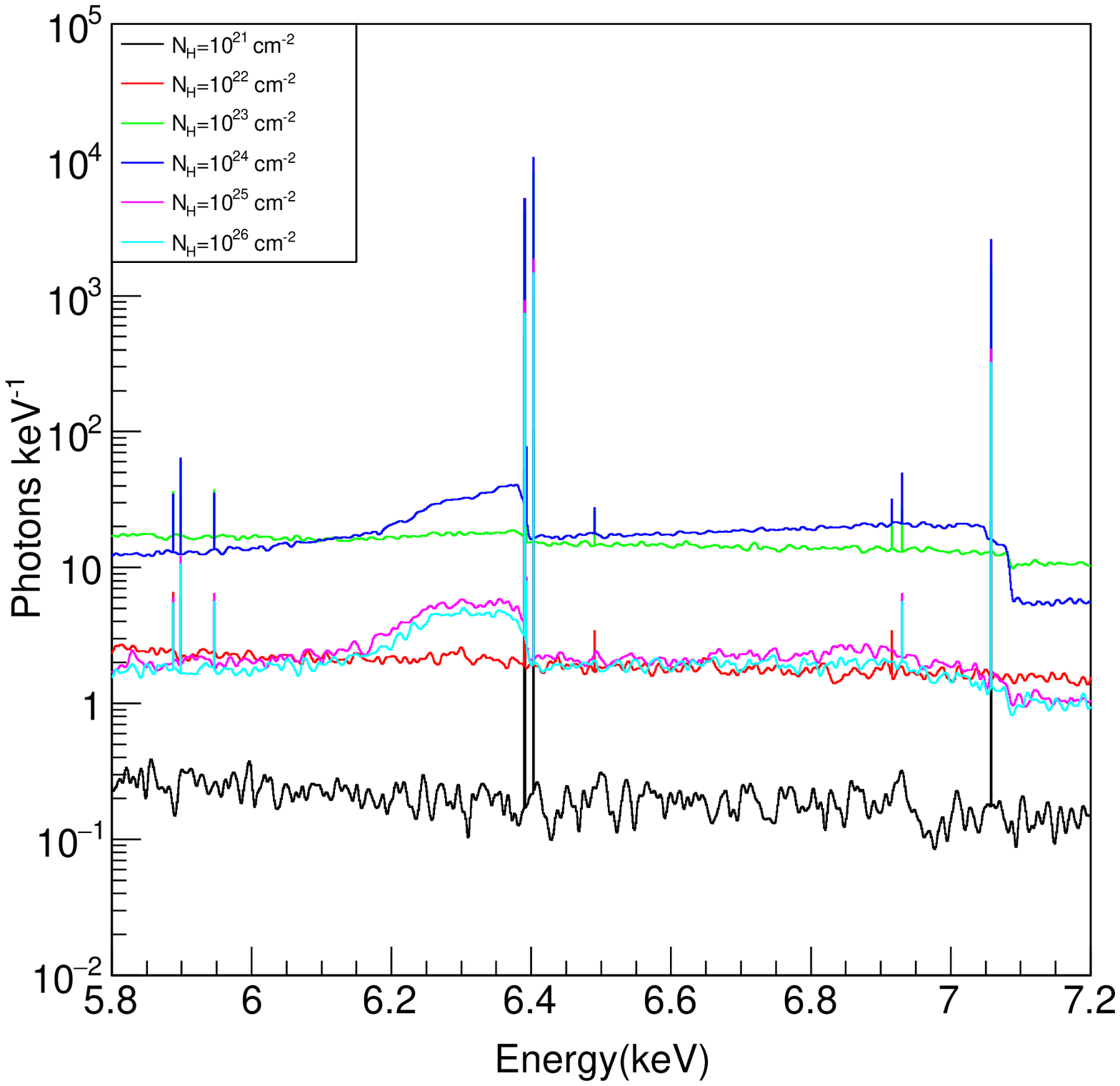}
}
	\caption{Spectra of the reprocessed component in the case of
  smooth torus with an inclination angle of $\theta_i=0.1-0.2$, 
  for various column
  densities $N_{\rm}=10^{21}, 10^{22}, 10^{23}, 10^{24}, 10^{25}$, and
  $10^{26}$ cm$^{-2}$. Right panel is an enlargement around the Fe-K line.}
	\label{fig:spectra_nh}
\end{figure*}

At first, we studied the dependence on the hydrogen column density
$N_{\rm H}$
of the torus in the range of $10^{21}-10^{26}$ cm$^{-2}$.
The simulated spectra of reprocessed component in 2--10 keV are 
shown in Figure \ref{fig:spectra_nh}.
The spectral shape is almost identical to the intrinsic one at low $N_H$
and the flux is almost proportional to $N_H$;
scattering occurs in proportional to $N_H$, and 
scattered photons mostly escape out of the torus without absorption
in such a Compton-thin regime.
Around $N_{\rm H}=10^{23-24}$ cm$^{-2}$, the low energy part becomes 
attenuated and the flux in the high energy part does not increase as
$N_H$ becomes larger.
At the Compton-thick regime of $N_{\rm H}>10^{24}$ cm$^{-2}$, the flux at the high
energy part decreases due to the effect of multi scattering.
This behavior in the Compton-thick regime is also reported in Ikeda et
al. (2009) and Murphy \& Yaqoob (2009).

Right panel of Figure \ref{fig:spectra_nh} shows the spectra around the Fe-K lines.
We can see a change of shape and strength of the Compton shoulder 
against $N_{\rm H}$.
The Compton shoulder becomes more apparent for larger $N_{\rm H}$.
The Compton shoulder monotonically decreases toward the lower energy at
$N_{\rm H}<10^{24.5}$ cm$^{-2}$, while a edge-like structure at 6.3 keV  
becomes prominent at the Compton-thick regime.
This behavior can be understood as follows.
In figure 19, we summarize the map of locations at which the last
interaction between photons and torus for various simulation conditions
and photon energies.
Looking at these figures, it is found that, at $N_{\rm H}<10^{24}$
cm$^{-2}$, 
photons in the Compton shoulder come from the whole torus region with a
wide range of scattering angle, leading to a wide range of photon energy
after Compton scattering.
At the Compton-thick regime, most of scattered photons come from the torus
behind the central engine with a large scattering angle towards the 
observer and a large
Compton loss, and thus the fraction of photons close to 6.4 keV decreases.

\begin{figure*}[htbp]
  \begin{center}
    \includegraphics[clip, width=80mm]{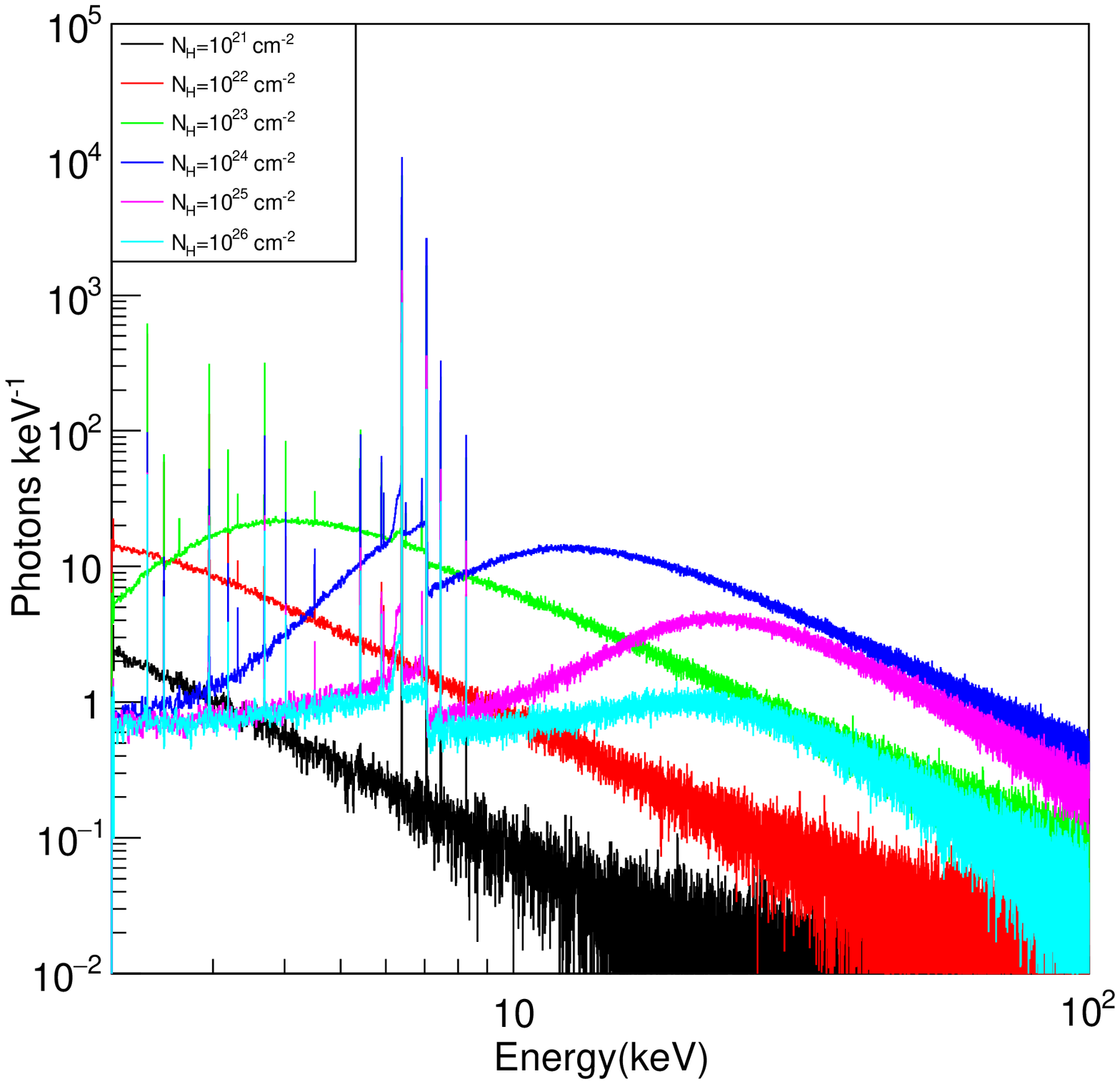}
    \includegraphics[clip, width=80mm]{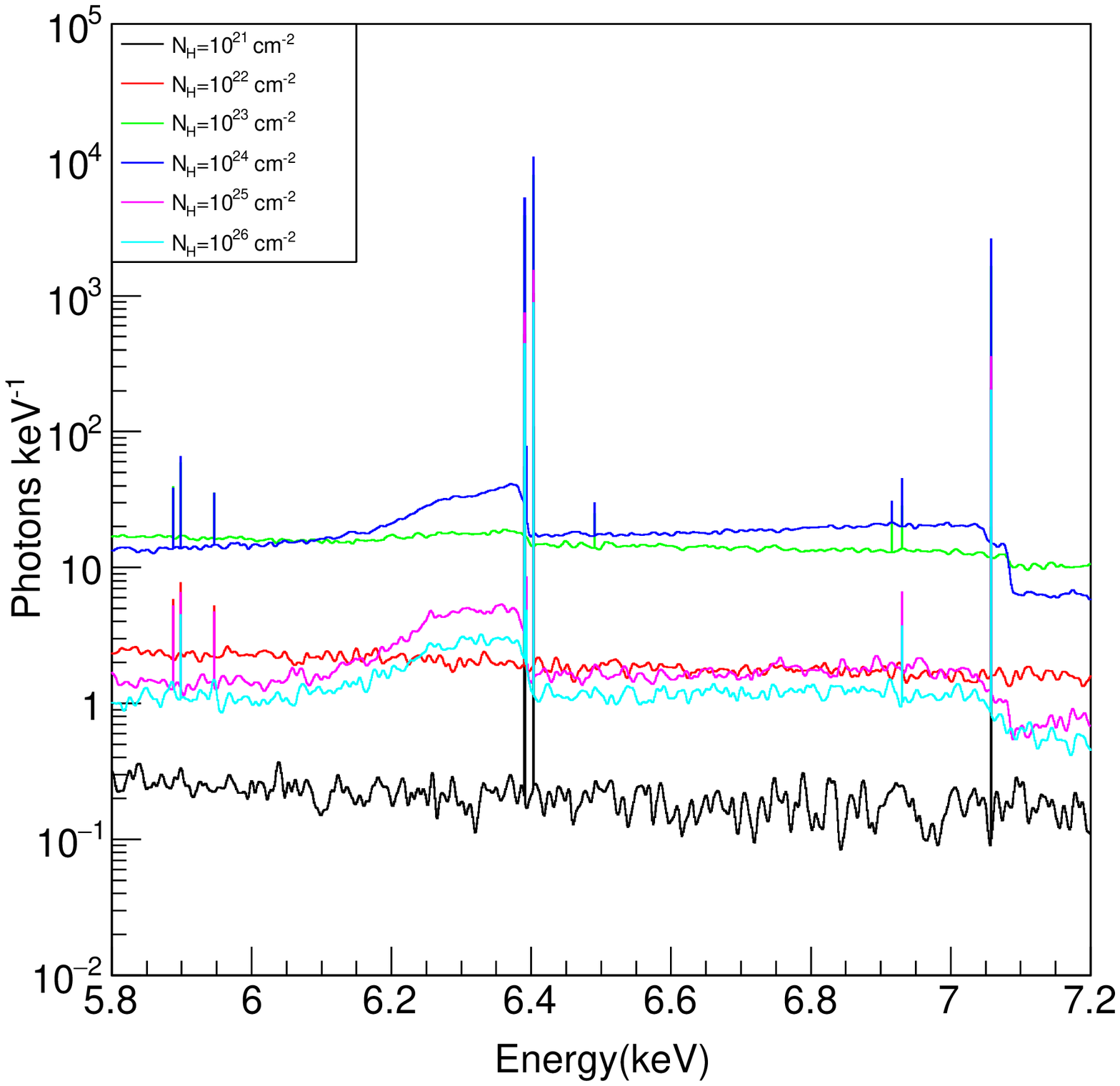}
    \includegraphics[clip, width=80mm]{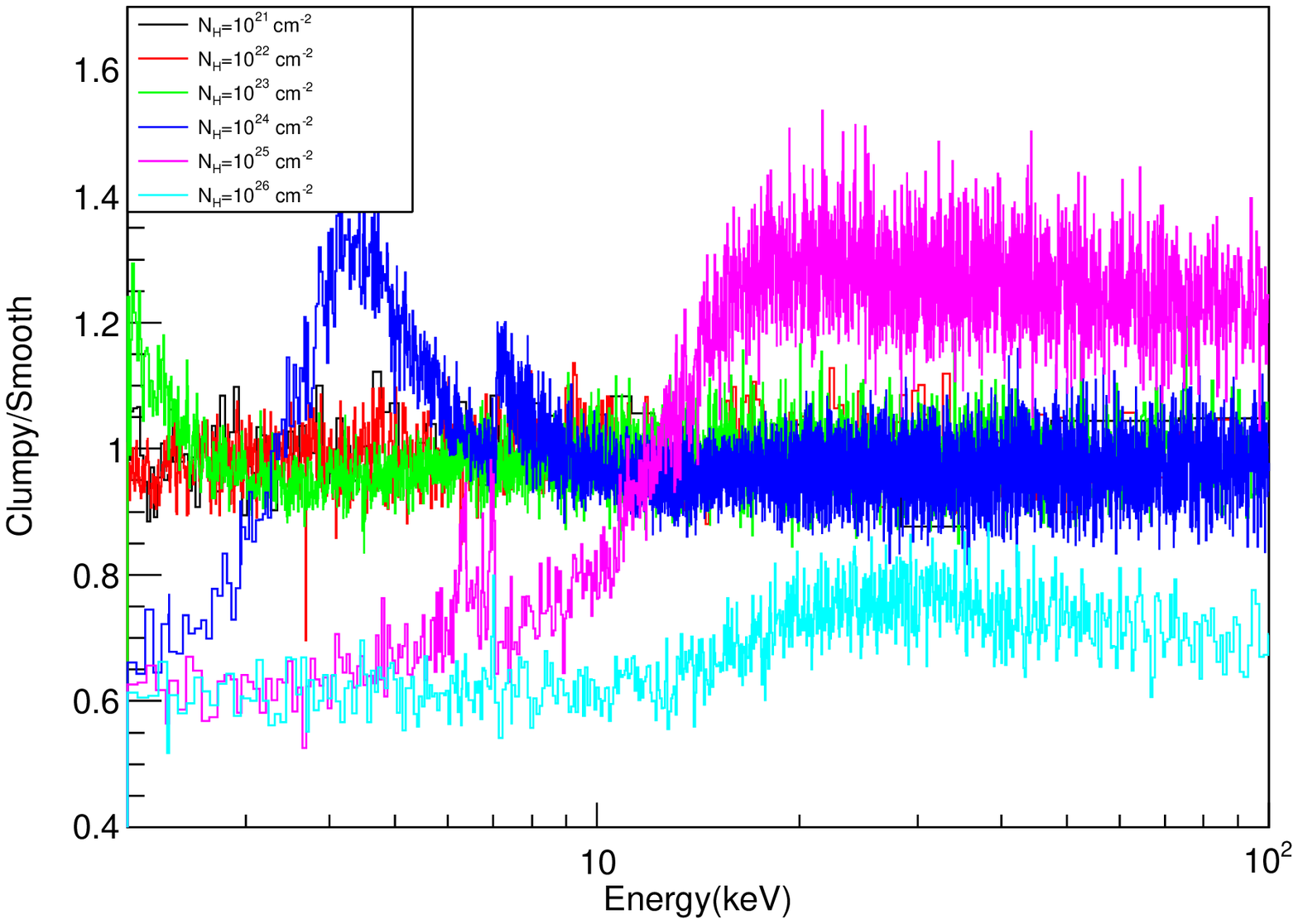}
    \includegraphics[clip, width=80mm]{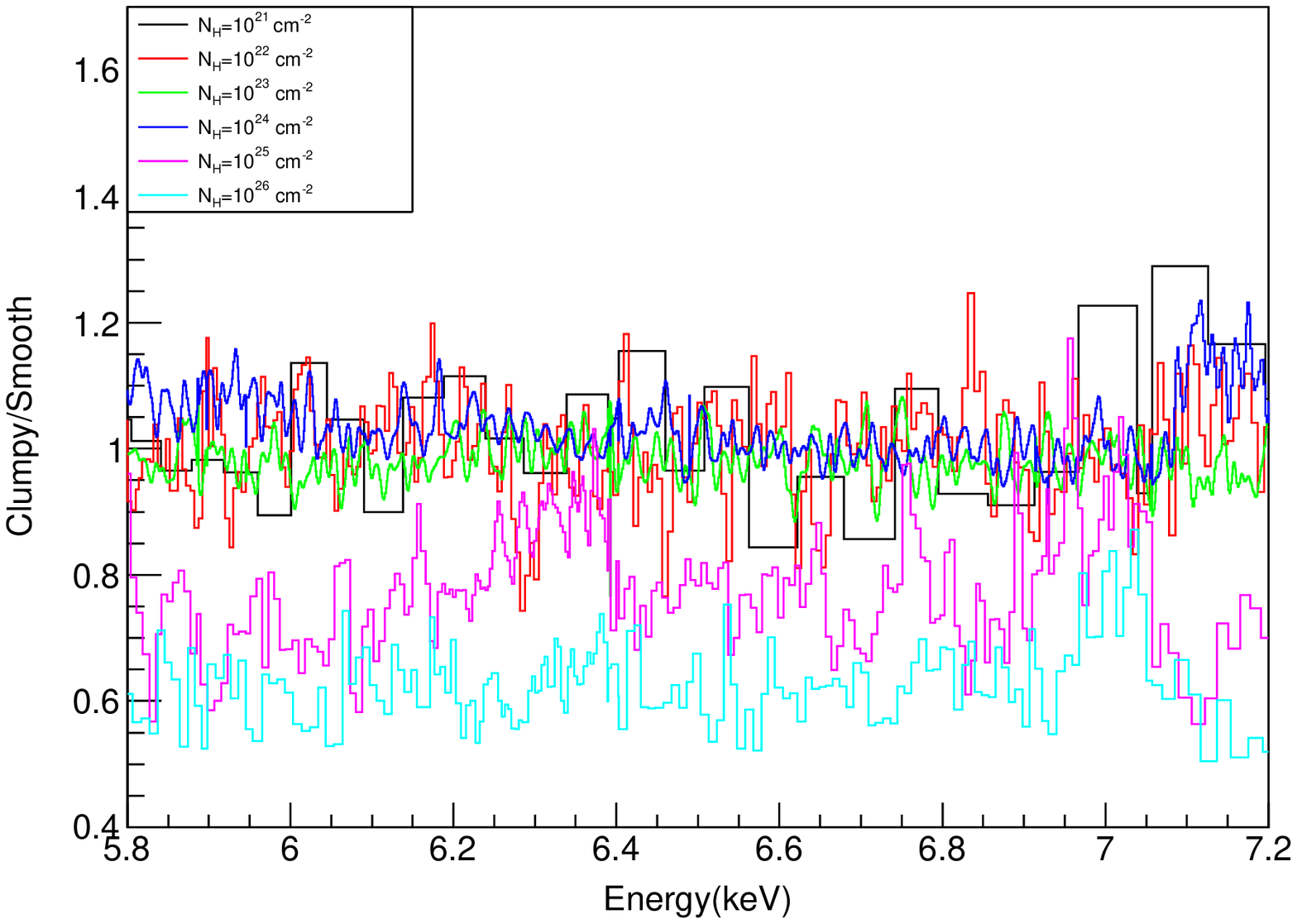}
    \caption{(Top) Spectra of the reprocessed component in the case of
  clumpy torus, for various column
  densities $N_{\rm}=10^{21}, 10^{22}, 10^{23}, 10^{24}, 10^{25}$, and
  $10^{26}$ cm$^{-2}$.
  Right panel is an enlargement around the Fe-K line. (Bottom) Spectral
   ratios of clumpy to smooth torus for the reprocessed components.}
    \label{fig:spectra_nh_clump}
  \end{center}
\end{figure*}

For the clumpy torus, we set a volume filling factor $f=0.05$ and a
clump radius $aR_{\rm torus}=0.005R_{\rm torus}$.
Figure \ref{fig:spectra_nh_clump} top shows the $N_{\rm H}$-dependence of reflection spectra in the
case of the clumpy torus, and
figure \ref{fig:spectra_nh_clump} bottom shows spectral ratios of clumpy
to smooth torus.
For the viewing angle of $\cos \theta_{i}=0.1-0.2$, the spectral shape is
somewhat different from that of the smooth torus, as seen in the
spectral ratio of clumpy to smooth torus.
As $N_{\rm H}$ increases, a part of the spectrum becomes humped in the
clumpy torus case and the hump moves to the higher energy.
From the lower energy part, the flux ratio (for the clumpy torus
compared with the smooth case) becomes smaller, and 
at last the ratio becomes almost constant at $\sim0.6$ 
at $N_{\rm H}=10^{26}$ cm$^{-2}$
for the clumpy torus than that for the smooth case.
As shown in the appendix 
(figure \ref{fig:specratio3a}--\ref{fig:specratio3c}), 
the flux of reprocessed components 
for the clumpy torus is lower than that 
for the smooth torus at most inclination angles, except the edge-on 
($\cos\theta_i=0-0.1$), in the case of $N_{\rm H}=10^{26}$ cm$^{-2}$.
This behavior is also reported by Liu \& Li (2014); note that the
average number of clumps toward the equatorial direction in our case
is 7.5 (\S2.1).
Since the density of each clump is $f^{-1}$ times as high as the average
density in the smooth torus case, an effective optical depth is larger
and thus the flux becomes lower.
The shape of Compton shoulder is also similar but photons close to the
line core are somewhat more numerous in the clumpy torus 
at $N_{\rm H}=10^{25}$ cm$^{-2}$, due to that
photons scattered at the front of the central engine with a small
scattering angle leak to the line of sight easily in the clumpy torus
than in the smooth torus around this $N_{\rm H}$.

\begin{figure}[htbp]
  \begin{center}
    \includegraphics[clip, width=80mm]{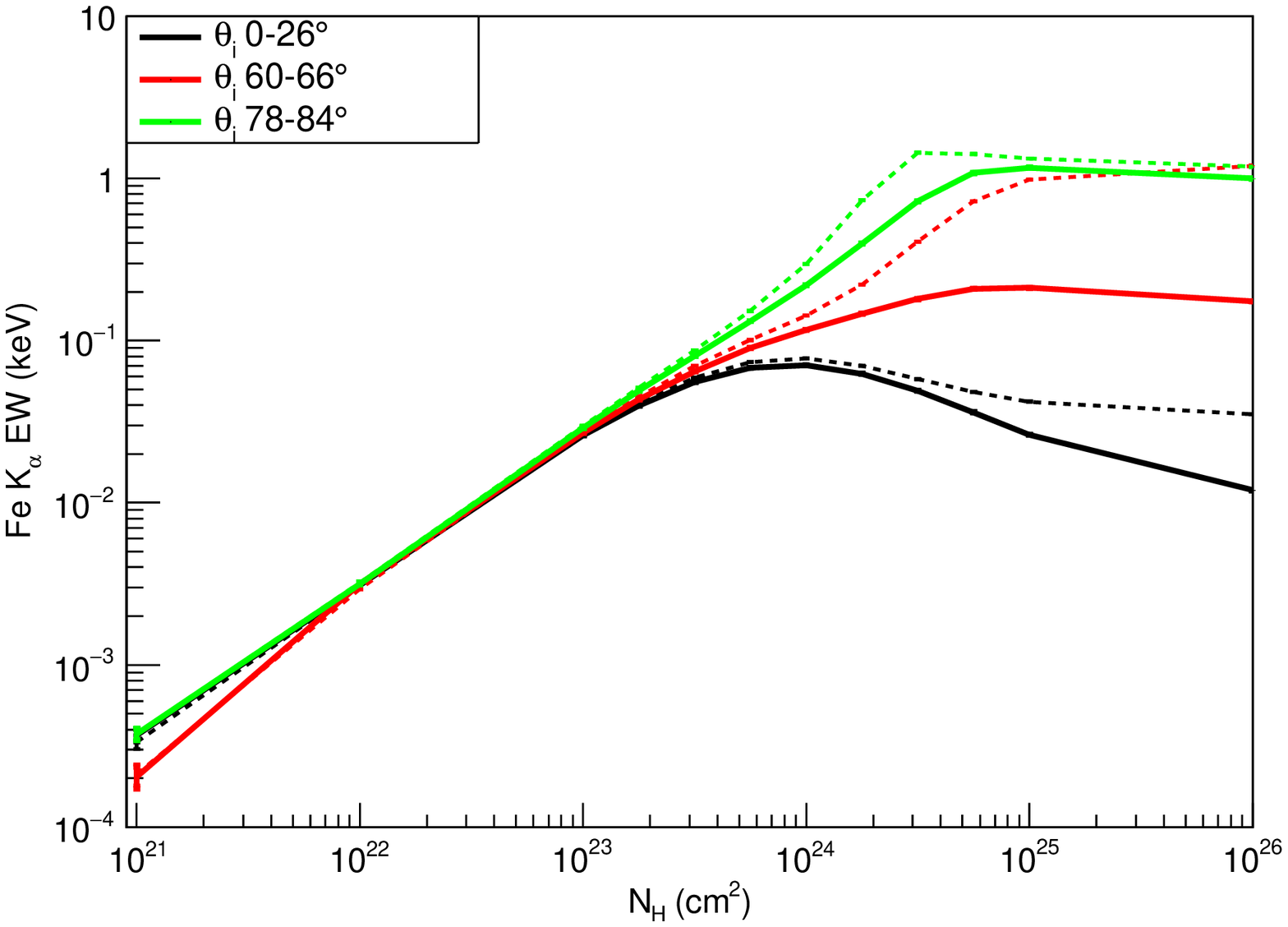}
    \includegraphics[clip, width=80mm]{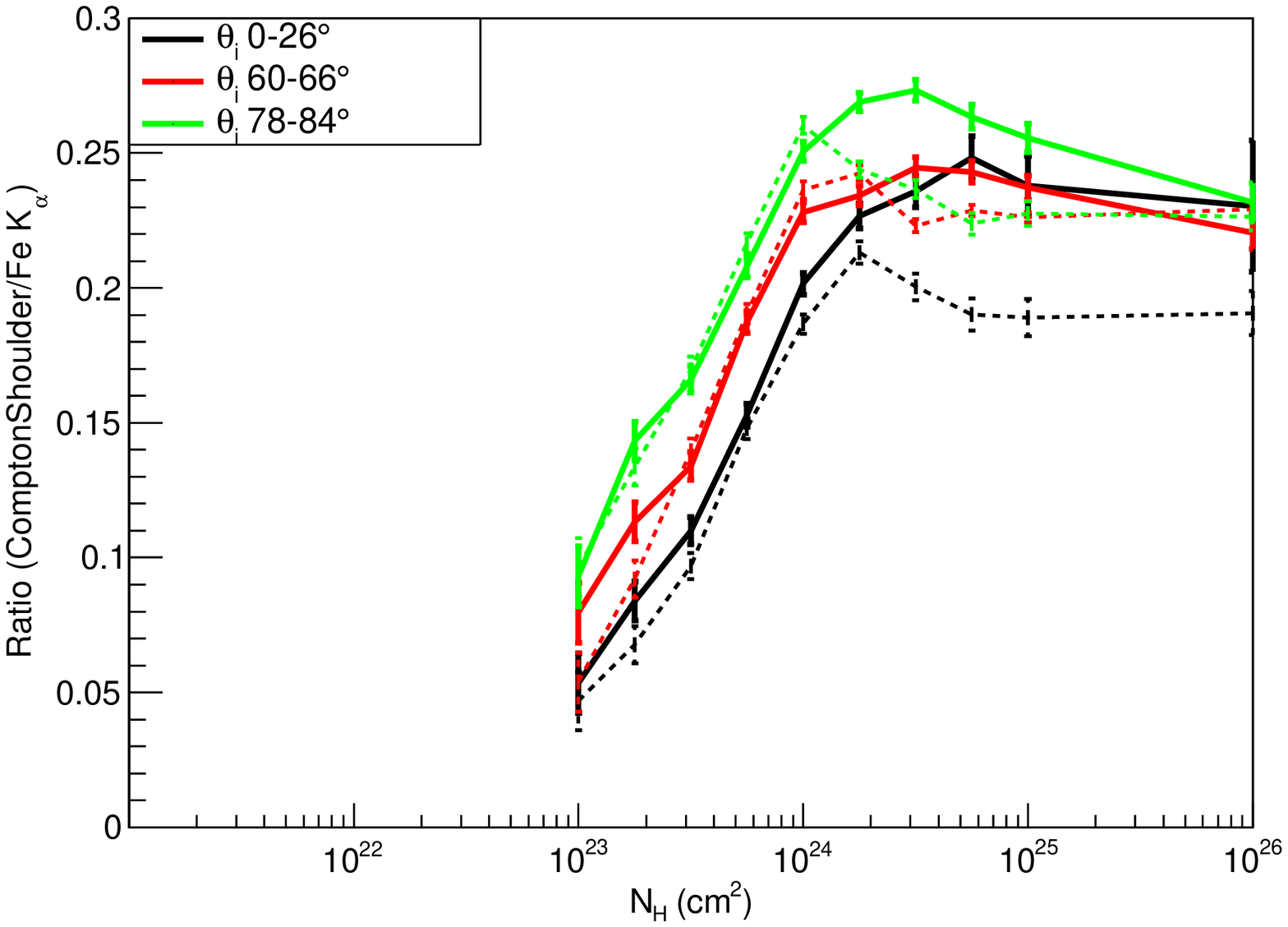}
    \caption{Left panel is a Fe-K line (core plus Compton shoulder) 
  equivalent width (EW) against the
  column density $N_{\rm H}$ for three inclination angles. 
  Right panel is a flux ratio of Fe-K
  Compton shoulder to line core 
  against $N_{\rm H}$. Solid and dashed lines represent
  clumpy and smooth torus, respectively. }
    \label{fig:dependence_nh_clump}
  \end{center}
\end{figure}

The dashed line of left panel of Figure \ref{fig:dependence_nh_clump} shows 
the equivalent width (EW) of the Fe-K$\alpha$ line (core plus Compton
shoulder); EW is the ratio relative 
to the sum of reflection and direct continuum.
This result is consistent with the previous studies (Ikeda et
al. 2009; Murphy \& Yaqoob 2009).
The dashed line of right panel of Figure \ref{fig:dependence_nh_clump} 
shows a integrated
flux ratio of the Compton shoulder to the line core.
The flux of the Compton shoulder is derived by subtracting the
continuum estimated in the wider energy bands adjacent 
to the line core and the shoulder.
The fraction of the Compton shoulder increased with the column density
$N_{\rm H}$ in the Compton-thin regime, and saturates at 0.2--0.25 
in the Compton-thick regime.
This behavior is similar to that in Matt (2002), and reasonably
understood as follows.
There is a peak at $N_{\rm H}=10^{24}$ cm$^{-2}$, the fraction decrease just
above the peaking column density is thought to be due to the
difference of effective optical depth between line core photons and
Compton shoulder photons.
The former has one interaction in the torus, while the latter has 
multiple interactions and thus a large effective optical depth than
the former.
Therefore, the Compton shoulder becomes prominent 
faster than the line core at a low $N_{\rm H}$ and its flux 
decreases faster than the core flux at a high $N_{\rm H}$. 
Then, the ratio of the Compton shoulder to 
the line core has a peak between the two regimes.

Solid lines in two panels of figure
\ref{fig:dependence_nh_clump} represent the behavior in the case of the
clumpy torus.
The dependence of the EW and the fraction of the Compton
shoulder against $N_{\rm H}$ is different from those of the smooth
torus.
The difference of the EW is, however, not so large to
distinguish by observations clearly.
On the other hand, a fraction of the Compton shoulder is larger by
several tens \% in the Compton-thick regime for the clumpy torus, 
compared with the smooth one.
This is explained in such a way that the Compton shoulder events which
experience multiple interactions could escape from the torus easier in the
clumpy torus than in the smooth torus due to intraclump spaces.

\subsection{Dependence of Inclination Angle (Smooth, Clumpy)}

\begin{figure*}[htbp]
\centering
\subfigure{
	\includegraphics[clip, width=80mm]{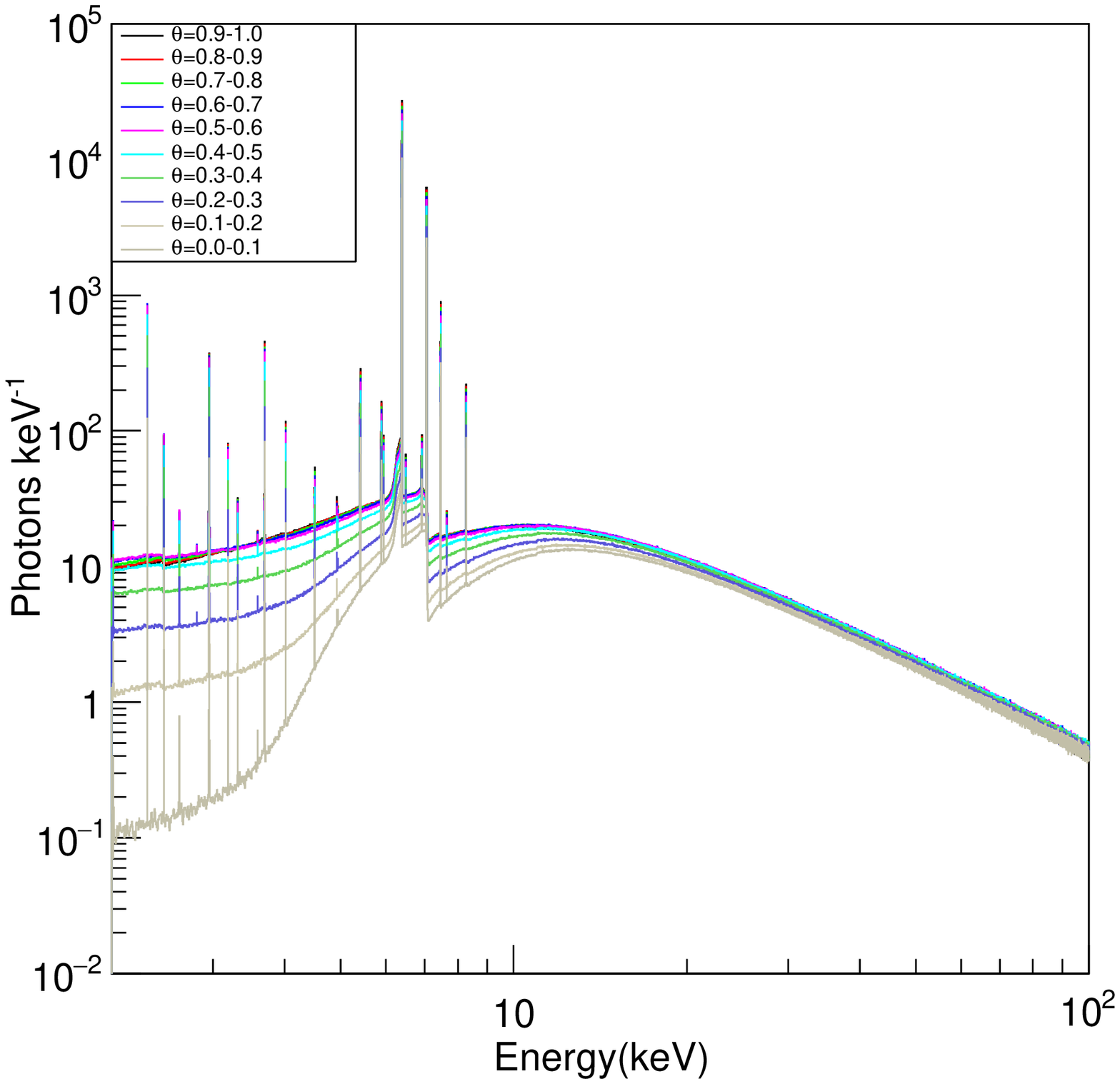}
	\includegraphics[clip, width=80mm]{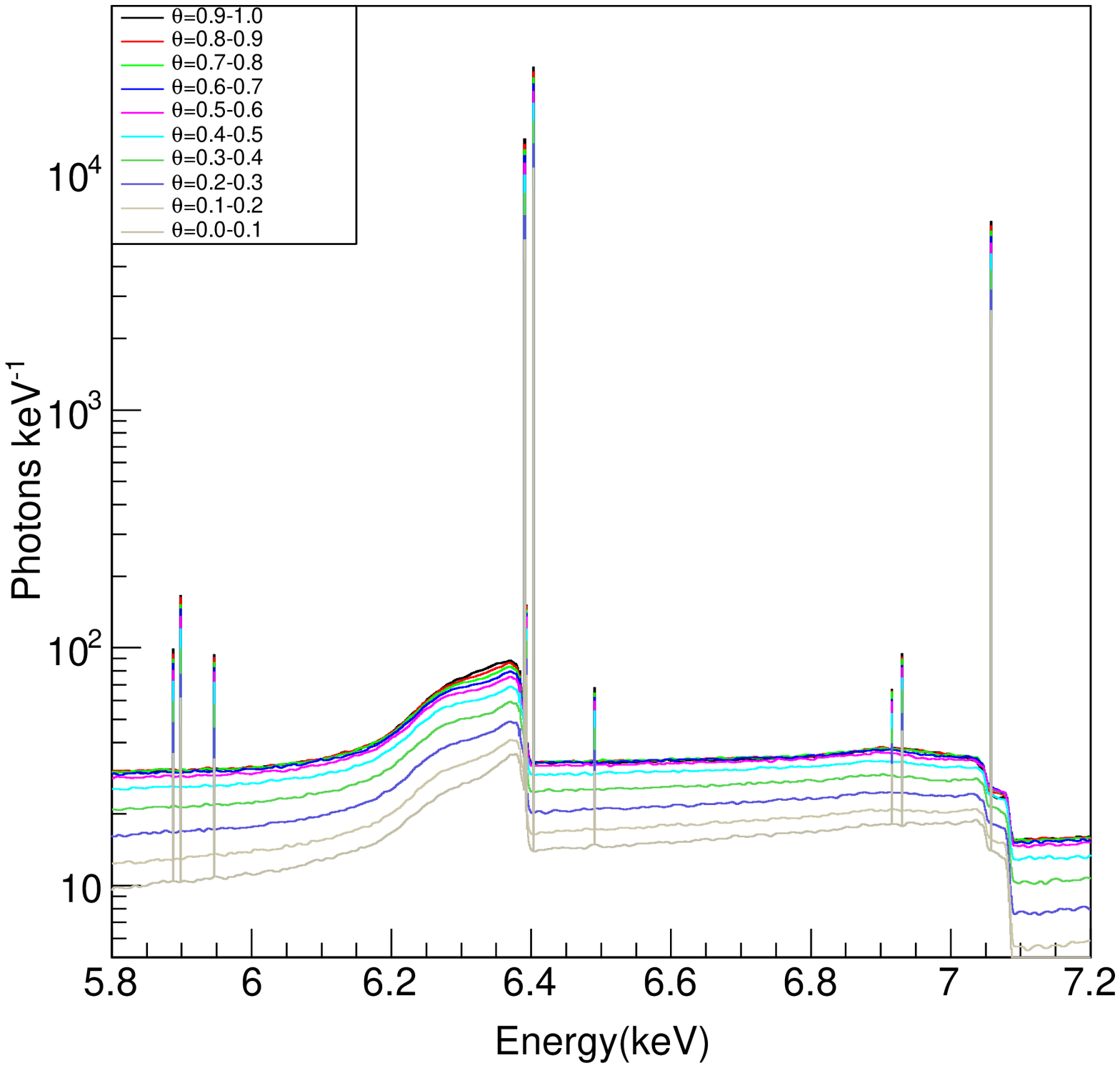}
}
	\caption{Spectra of the reprocessed component in the case of
  smooth torus with a column density of $N_{\rm H}=10^{24}$ cm$^{-2}$, 
  for various inclination angles 
  $\cos\theta_i=$0.9--1.0, 0.8--0.9, $\cdot$, and 0--0.1 from top to bottom. Right panel is an enlargement around the Fe-K line.}
	\label{fig:spectra_inc}
\end{figure*}

Figure \ref{fig:spectra_inc} shows the simulated reprocessed spectra for
$\cos \theta_{i}$= 0--1 in steps of 0.1 in case of $N_{\rm H}=10^{24}$
cm$^{-2}$ for the smooth torus.
To create this plot, we generate a total number of $10^{10}$ photons for a
run, in order to keep the statistics for arious inclination angles.
In the left panel of this figure, the spectral shape and flux are almost 
identical in high energy band for any inclination angle, 
since the absorption effect is negligible and scatterings occur with
various angles at various positions to reduce the scattering-angle dependence.
In the lower energy band, the flux decreases for the larger inclination
angle due to the absorption.
The shape of Compton shoulder becomes more upward concave at the smaller
inclination angle.
At $\cos \theta_i$ = 0.5--0.9 ($\theta_i$ is small), 
the last scattering position of photons in the Fe-K
region includes the equatorial region of the torus behind the central
engine from the observer, and the
Compton shoulder events scattered in such a region have 
a large scattering optical depth and thus a lower energy of scattered
photons due to Compton loss.
At larger inclination angle with $\cos \theta_{i}<0.5$, the equatorial region of the torus behind the central engine cannot be observed from the
observer, and thus the Compton shoulder events with lower energy
decrease.

\begin{figure*}[htbp]
\centering
    \includegraphics[clip, width=80mm]{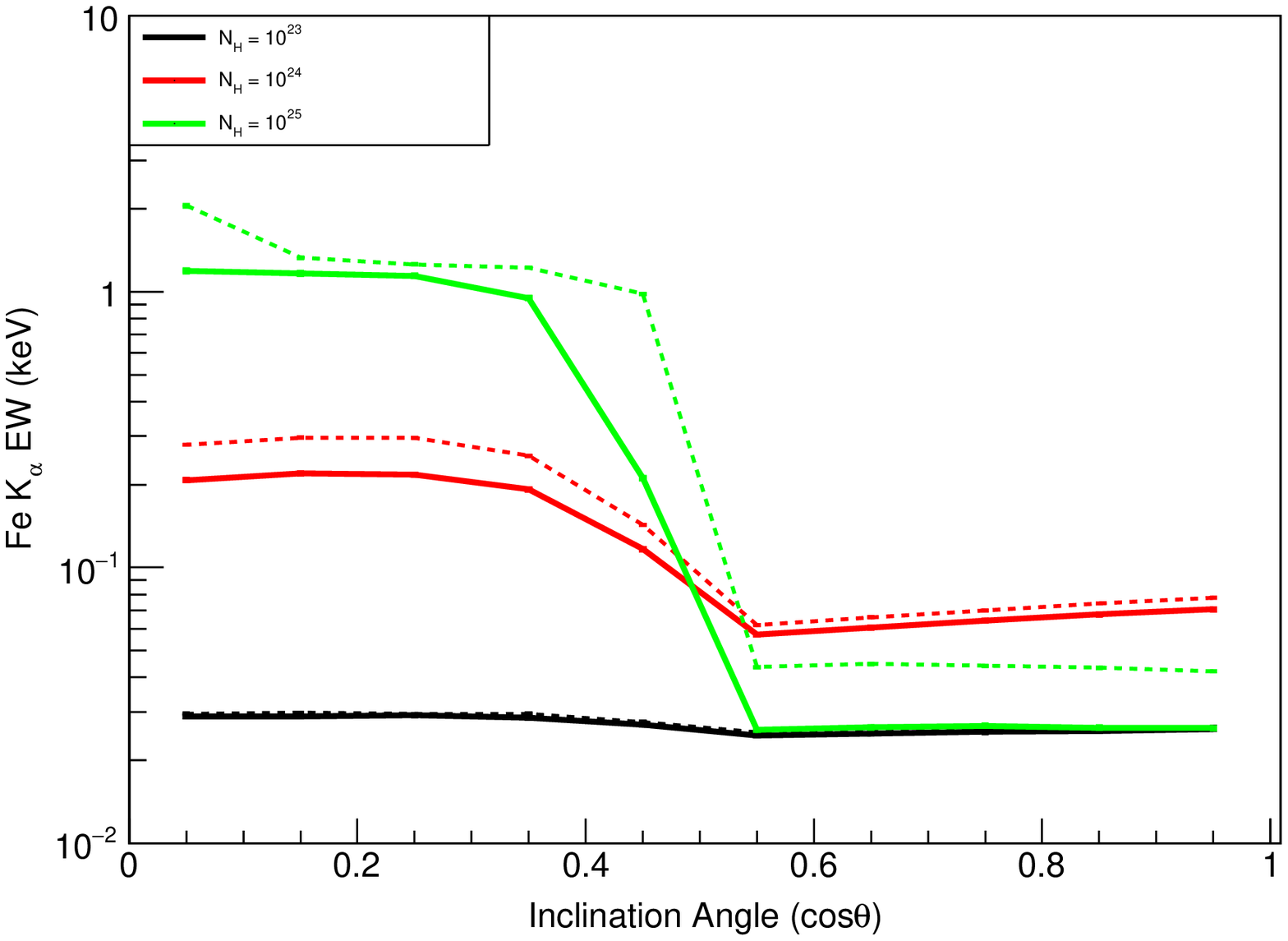}
    \includegraphics[clip, width=80mm]{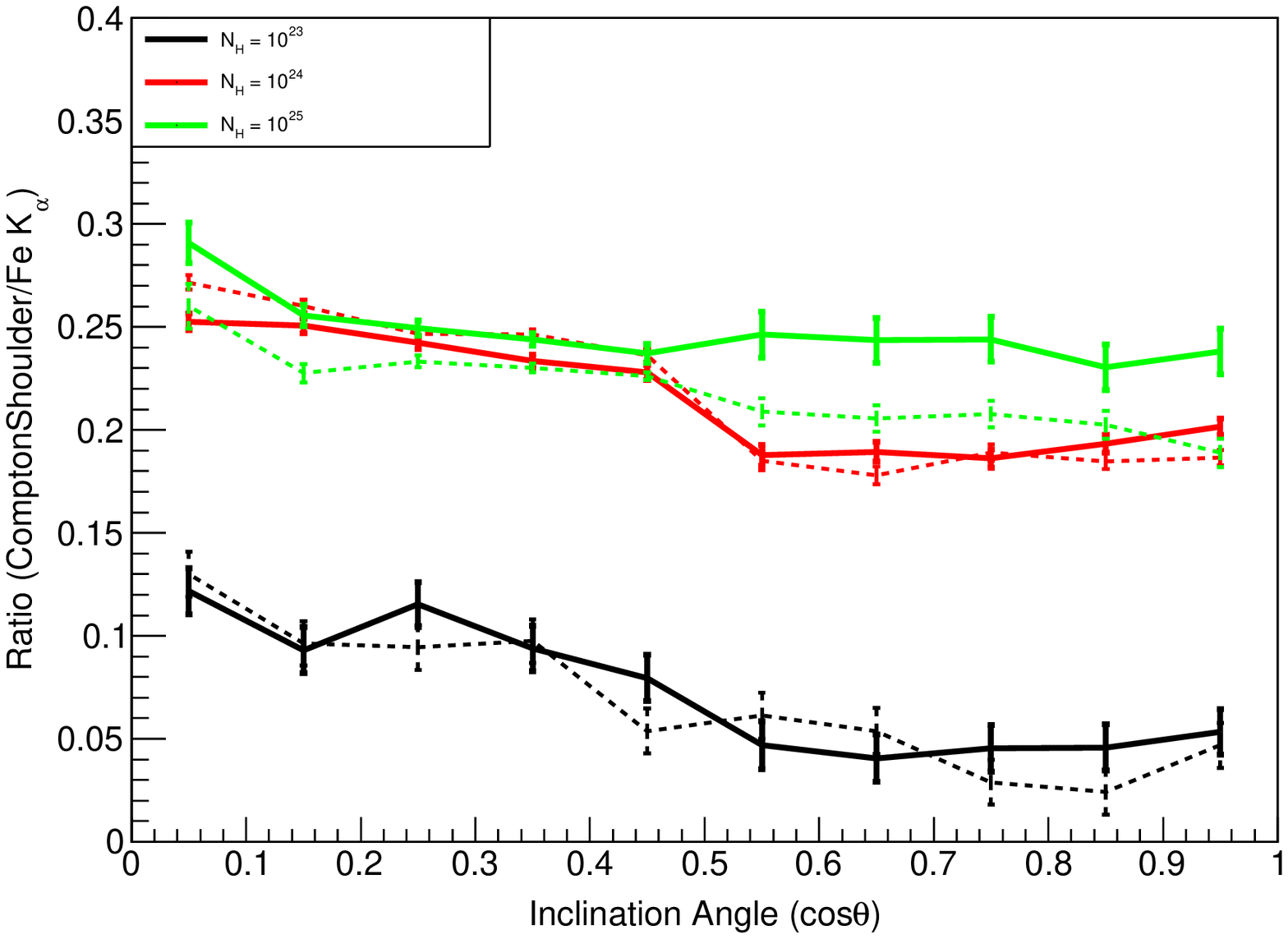}
	\caption{Left panel is a Fe-K line (core plus Compton shoulder)
 equivalent width (EW) against the cosine of 
  inclination angle $\theta_i$ for three column densities. 
  Right panel is a flux ratio of Fe-K
  Compton shoulder to line core against $\cos\theta_i$. Dashed and solid
 lines correspond to the smooth and clumpy torus, respectively.}
	\label{fig:dependence_inc}
\end{figure*}

Figure \ref{fig:dependence_inc} shows the inclination angle dependence
of the Fe-K line EW and the Compton shoulder to the line core
ratio for three cases with $N_{\rm H} = 10^{23}$, $10^{24}$, and $10^{25}$ 
cm$^{-2}$.
This figure shows results for both of smooth and clumpy cases.
The inclination angle of the EW jump corresponds to the boundary whether
the line from the center to the observer passes the torus or not.
This behavior is consistent with that of Murphy \& Yaqoob (2009).
The fraction of Compton shoulder increases with the inclination angle 
(i.e., with smaller $\cos \theta_{i}$ values).
For a larger $N_{\rm H}$, the fraction is larger for the clumpy torus
than the smooth one, as seen in the $N_{\rm H}$ dependence.


\clearpage

\subsection{Dependence of Metal Abundance (Smooth, Clumpy)}

\begin{figure*}[htbp]
\centering
\subfigure{
	\includegraphics[clip, width=80mm]{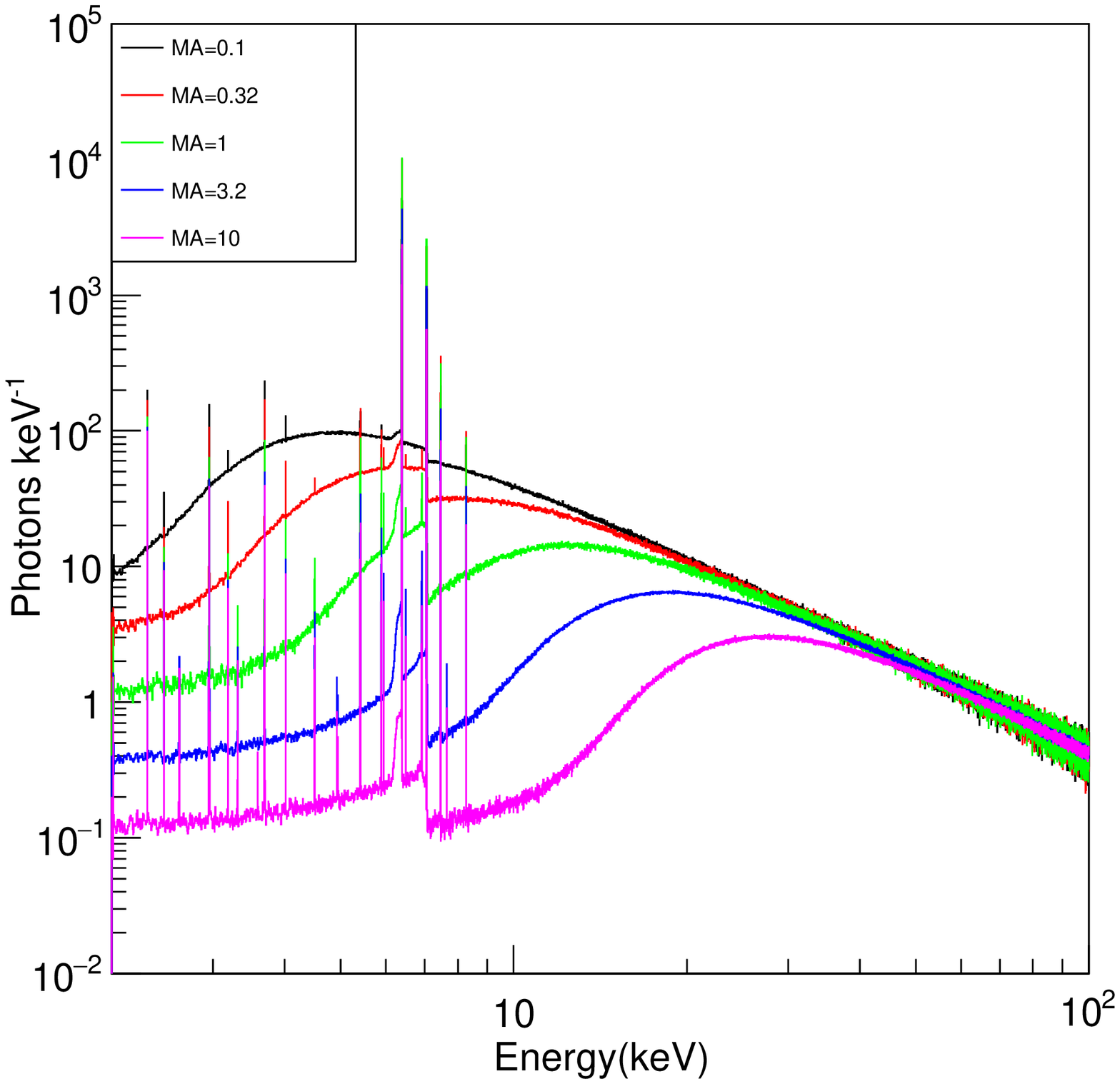}
	\includegraphics[clip, width=80mm]{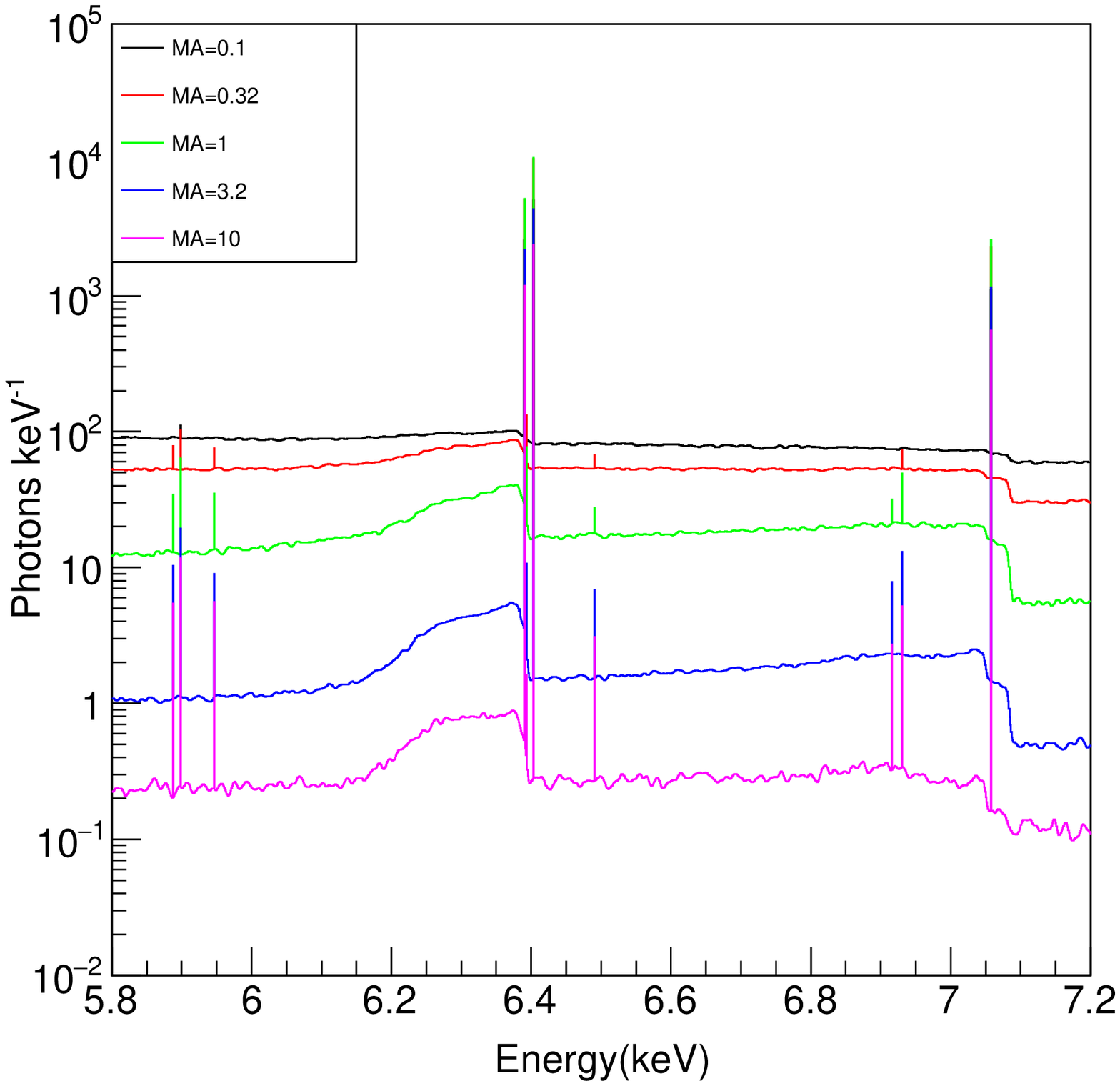}
}
	\caption{Spectra of the reprocessed component in the case of
  smooth torus with $N_{\rm H}=10^{24}$ cm$^{-2}$ and $\cos \theta_i=0.1-0.2$, 
  for various metal abundances (MA) in solar unit. 
  Right panel is an enlargement around the Fe-K line.}
	\label{fig:spectra_ma}
\end{figure*}

So far we performed simulation under the condition that metal
abundance is the same as the solar system.
In this subsection, we studied the dependence on metal abundance.
Figure \ref{fig:spectra_ma} shows the reprocessed spectra on
various metal abundances with $N_{\rm H}=10^{24}$ cm$^{-2}$ for the
smooth torus.
Metal abundances of heavy elements are kept to follow the solar
abundance ratio.
For this plot, we generate a total number of $10^{10}$ photons for
$A_{\rm metal}= 3.2$ or $10$ in order to keep the statistics, but $10^9$
photons for other $A_{\rm metal}$ values.
Metal abundance dependence on the reprocessed spectra are similar to
that on $N_{\rm H}$, but the difference is that a scattered component
in the lower energy part is weak and no flux change in the higher
energy band.
This is due to that the amount of hydrogen which are main scattering
atoms does not change in this case.

\begin{figure*}[htbp]
\centering
    \includegraphics[clip, width=80mm]{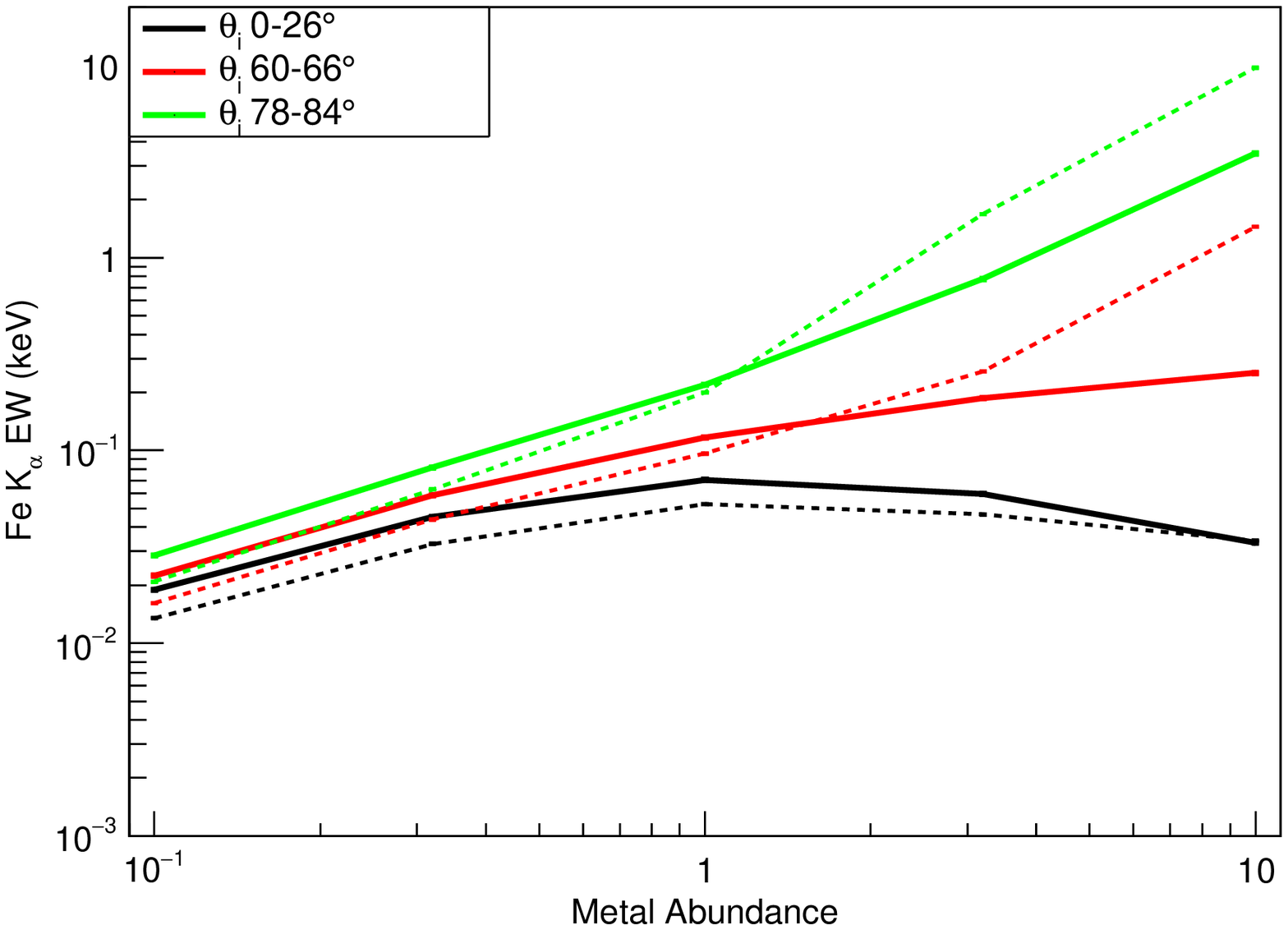}
    \includegraphics[clip, width=80mm]{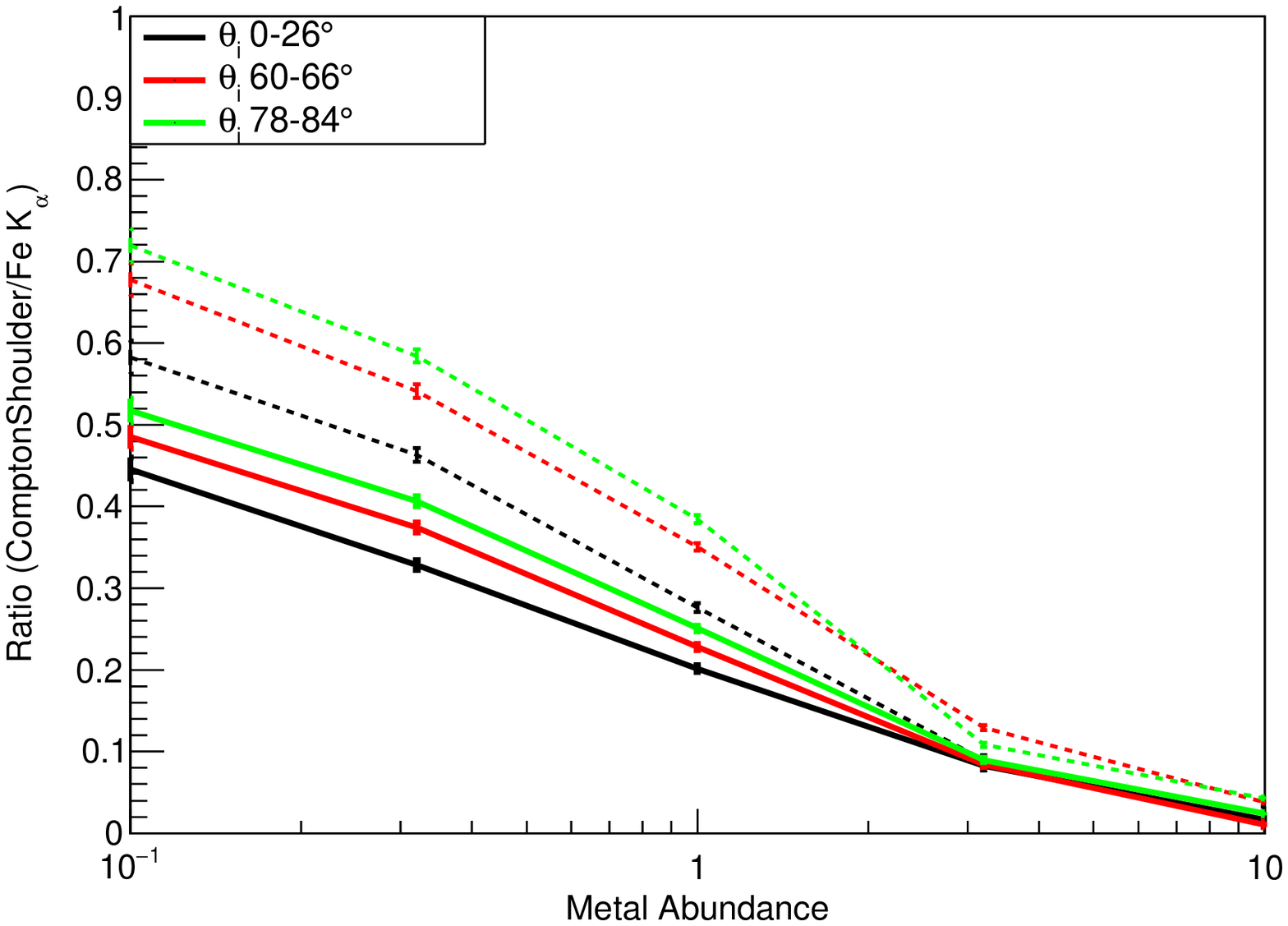}
	\caption{Left panel is a Fe-K line (core plus Compton shoulder)  equivalent width (EW) against the
  metal abundance for three inclination angles. 
  Right panel is a flux ratio of Fe-K
  Compton shoulder to line core against the metal abundance. Both panels are in
  the case of $N_{\rm H}=10^{24}$ cm$^{-2}$. Dashed and solid lines
 correspond to the smooth and clumpy torus case, respectively.}
	\label{fig:dependence_ma_NH1e24}
\end{figure*}

\begin{figure*}[htbp]
\centering
    \includegraphics[clip, width=80mm]{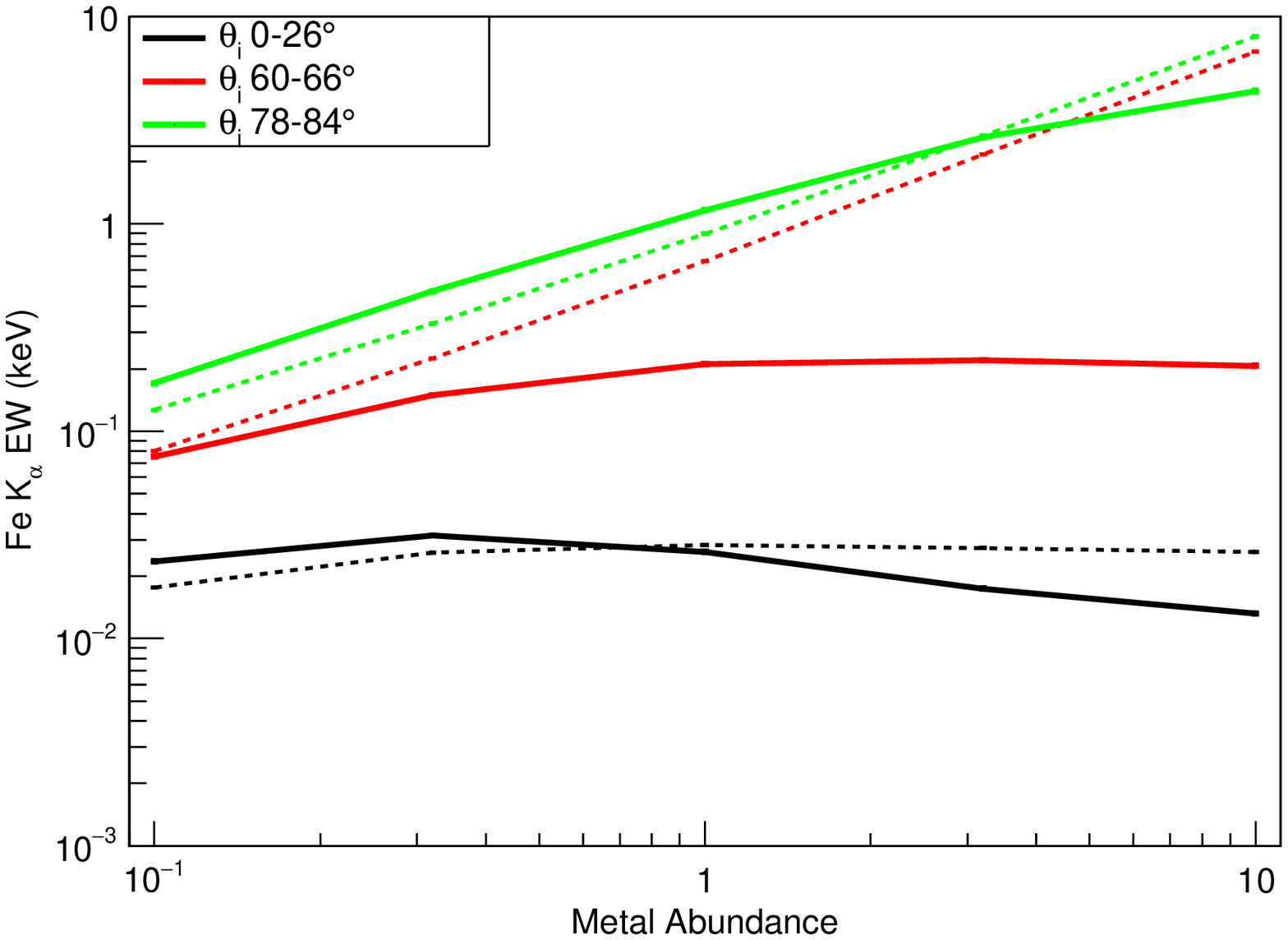}
    \includegraphics[clip, width=80mm]{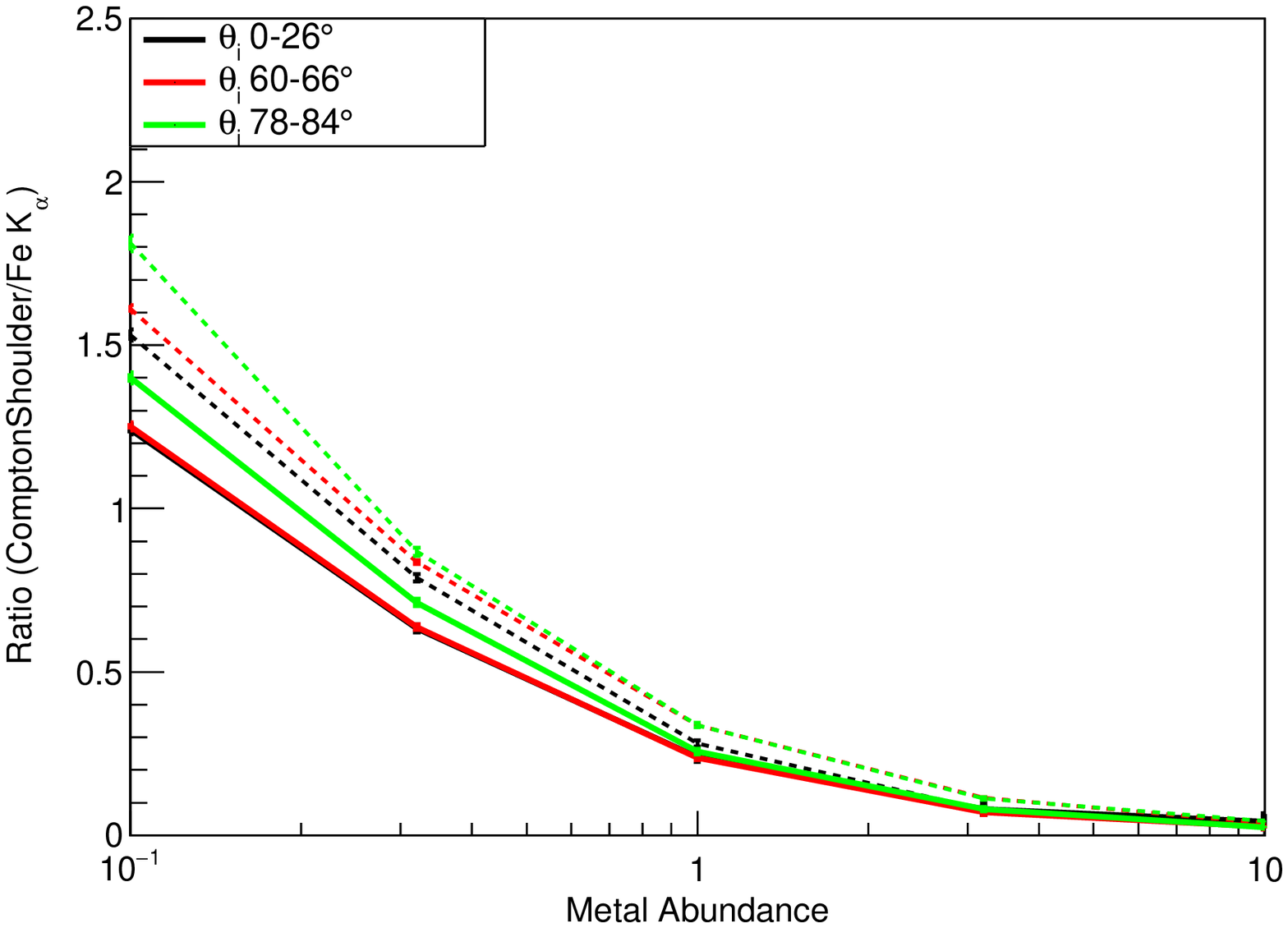}
	\caption{Same as figure \ref{fig:dependence_ma_NH1e24}, but in
  the case of $N_{\rm H}=10^{25}$ cm$^{-2}$.}
	\label{fig:dependence_ma_NH1e25}
\end{figure*}

Figure \ref{fig:dependence_ma_NH1e24} and \ref{fig:dependence_ma_NH1e25} show
the EW and the fraction of
the Compton shoulder to the line core as a function of metal
abundance for three inclination angles.
We show two cases of $N_{\rm H}=10^{24}$ cm$^{-2}$ and $10^{25}$
cm$^{-2}$ for both of smooth and clumpy cases.
At the large inclination angles, EW increases with the metal
abundance quasi-proportionally.
Looking at the Fe-K line core flux and continuum, this increase is not
due to the line flux increase but due to the continuum reduction at
high metal abundances.
At the small inclination angle, EW does not increase with the metal
abundance since the continuum reduction is weak.
Interestingly, the fraction of Compton shoulder to the line core is
smaller at higher metal abundance.
From the spectra in figure \ref{fig:spectra_ma},
the flux of Compton shoulder becomes 
more heavily reduced than the line core flux at higher metal
abundance.
At higher metal abundance, the Compton shoulder events which runs a
longer path than the line core events in the torus are more strongly 
absorbed and the flux reduction is large.

For lower metal abundances, the clumpy case gives a higher EW and a lower
Compton shoulder to line core ratio than the smooth case.
On the other hand, for higher metal abundances, it gives a lower EW.
The density in the clumpy case is higher than that in the smooth case
with the same $N_{\rm H}$ in our definition.
Consider a torus size $R_{\rm torus}$, a clump size $aR_{\rm
torus}=0.005R_{\rm torus}$,
and a mean free path $l_{\rm Fe}=0.035A_{\rm metal}^{-1}R_{\rm torus}$ 
at the Fe-K$\alpha$ line energy 6.4 keV under the metal abundance 
$A_{\rm metal}$ and the volumn filling factor $f=0.05$.
The last one is derived by using the relation between $N_{\rm H}$ and the
hydrogen number density $n_{\rm H}$ in section 2.1.
In the lower metal abundances, a mean free path $l_{\rm Fe}$ 
is larger than a clump size $aR_{\rm torus}$ but smaller than a torus
size $R_{\rm torus}$.
For the clumpy torus,
X-rays can efficiently pass through the intraclump space from the
central source to the torus, generate the Fe-K lines almost everywhere 
in each torus, and then efficiently escape out of
the torus.
In other words, an effective volume to generate K$\alpha$ line photons
is larger for the clumpy torus, and thus EW is larger and Compton
shoulder to line core ratio is smaller.
For higher metal abundances, $l_{\rm Fe}$ becomes comparable to or smaller than
the clump size $aR_{\rm torus}$.
In this case, $l_{\rm Fe}$ of the clumpy trus is smaller than that of
the smooth torus by a factor of $f$ due to higher density.
Since observed K$\alpha$ line photons come from the region within 
$l_{\rm Fe}$ of the surface of smooth torus or each clump.
As a result, the effective volume to generate K$\alpha$ line photons
is smaller for the clumpy torus, and thus EW is smaller.
In addition, the scattering efficiency of Fe-K$\alpha$ photons also
become higher for the clumpy case, the Compton shoulder to line core
ratio is similar to that of the smooth torus.


\clearpage

\subsection{Dependence of Turbulent Velocity (Smooth, Clumpy)}

\begin{figure*}[htbp]
\centering
\subfigure{
	\includegraphics[clip, width=80mm]{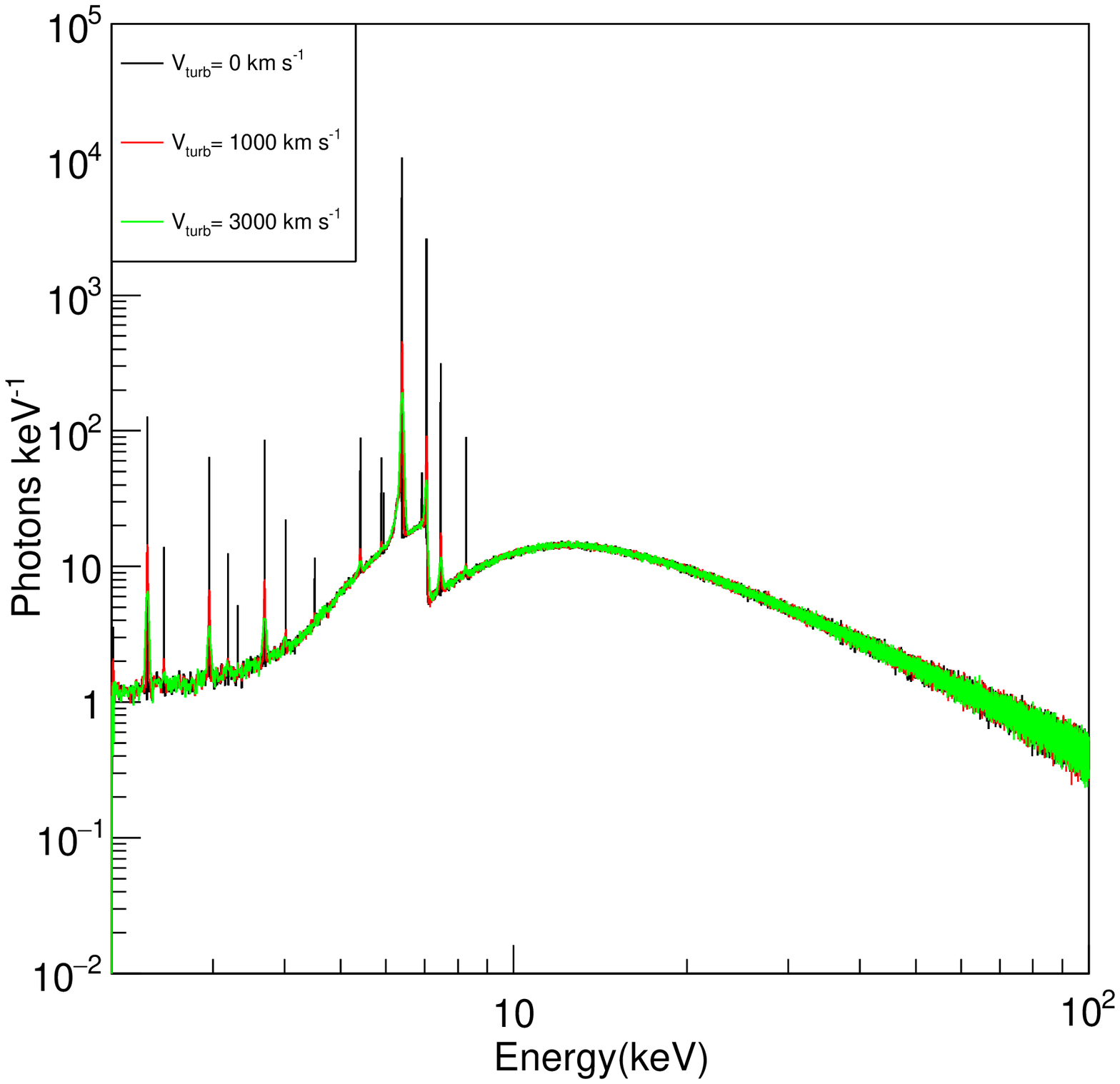}
	\includegraphics[clip, width=80mm]{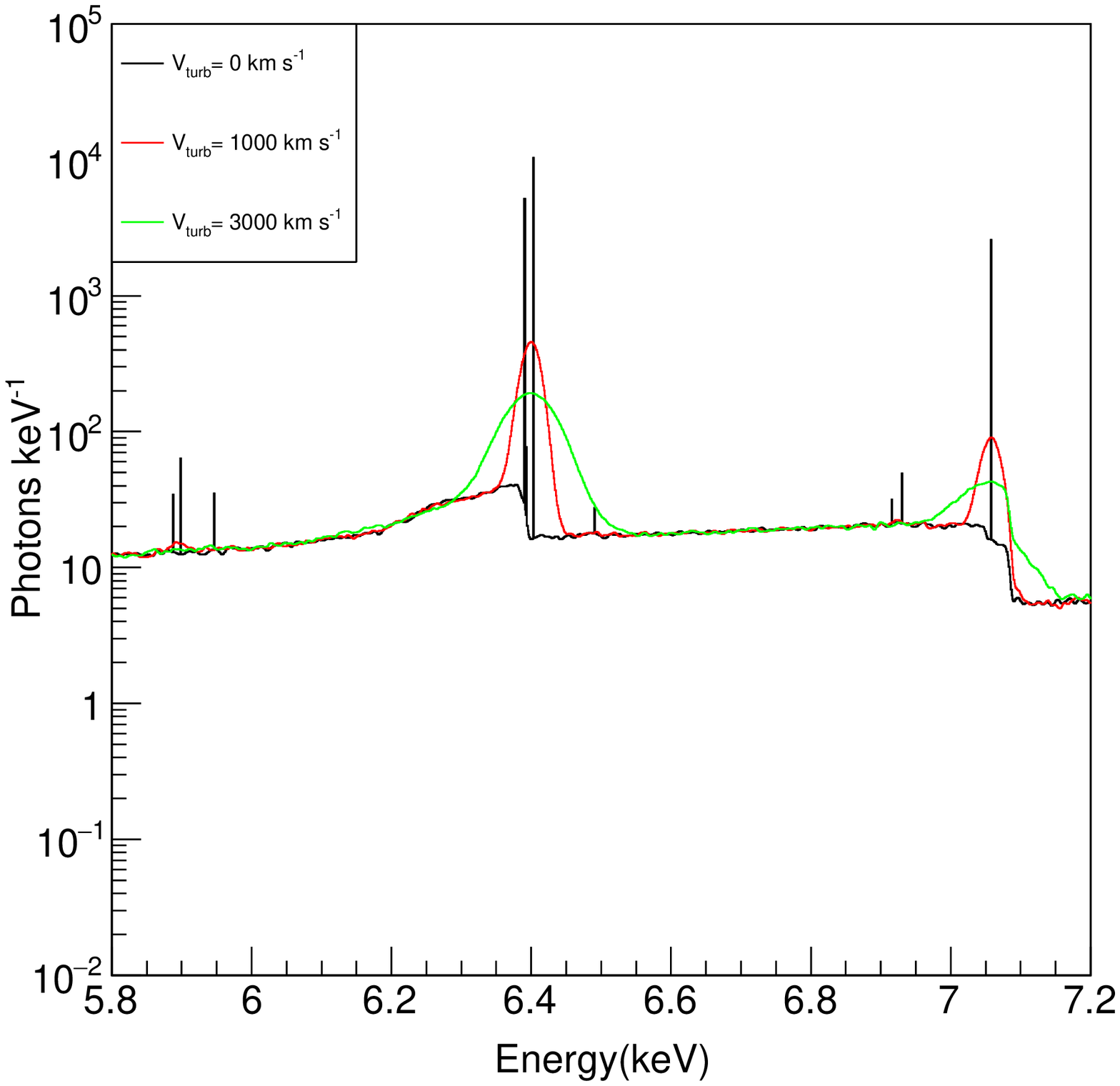}
}
	\caption{Spectra of the reprocessed component in the case of
  smooth torus with $\cos \theta_i=0.1-0.2$, 
  for various turbulence velocities. 
  Right panel is an enlargement around the Fe-K line.}
	\label{fig:spectra_v}
\end{figure*}

\begin{figure*}[htbp]
\centering
    \includegraphics[clip, width=80mm]{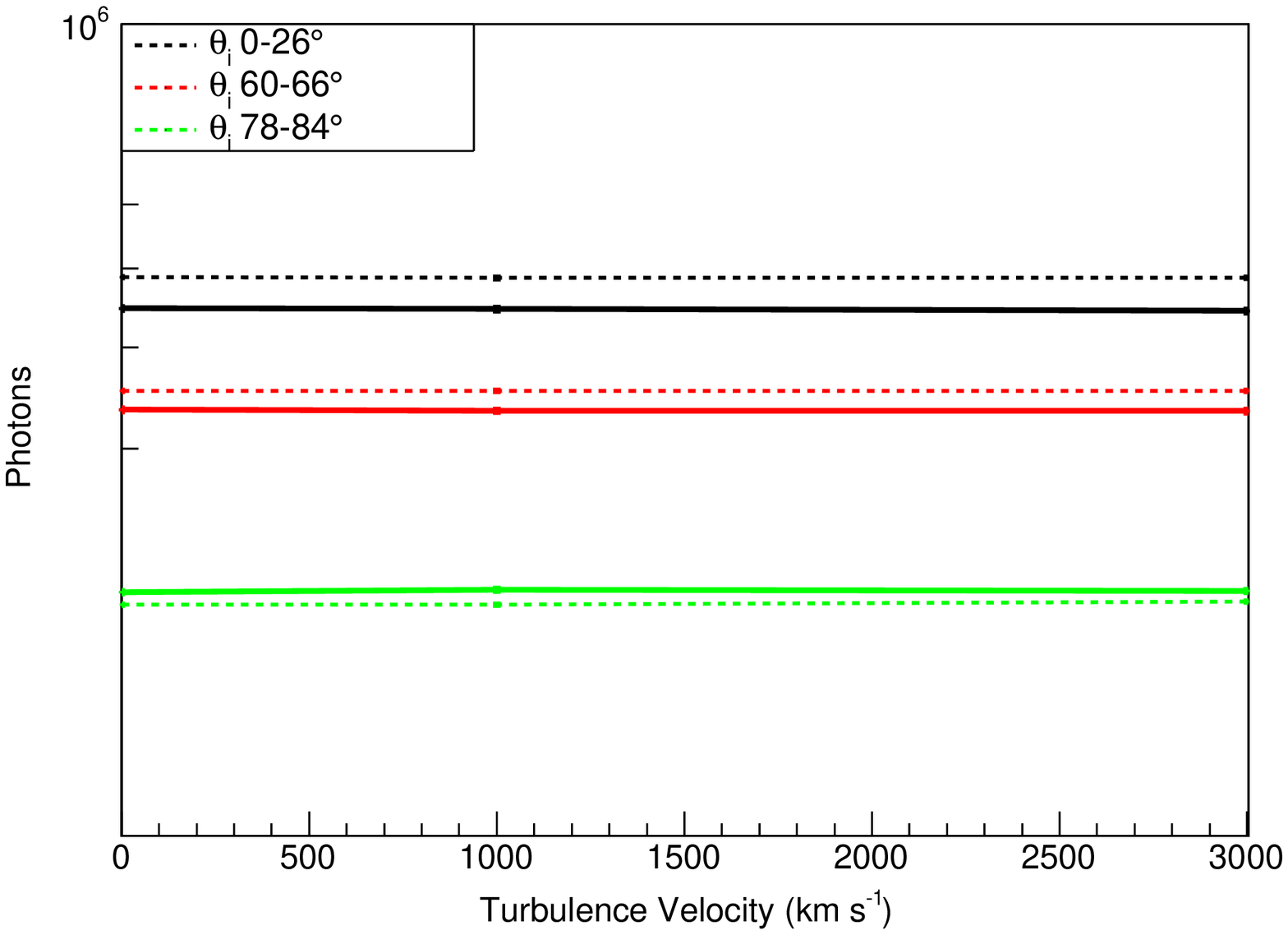}
    \includegraphics[clip, width=80mm]{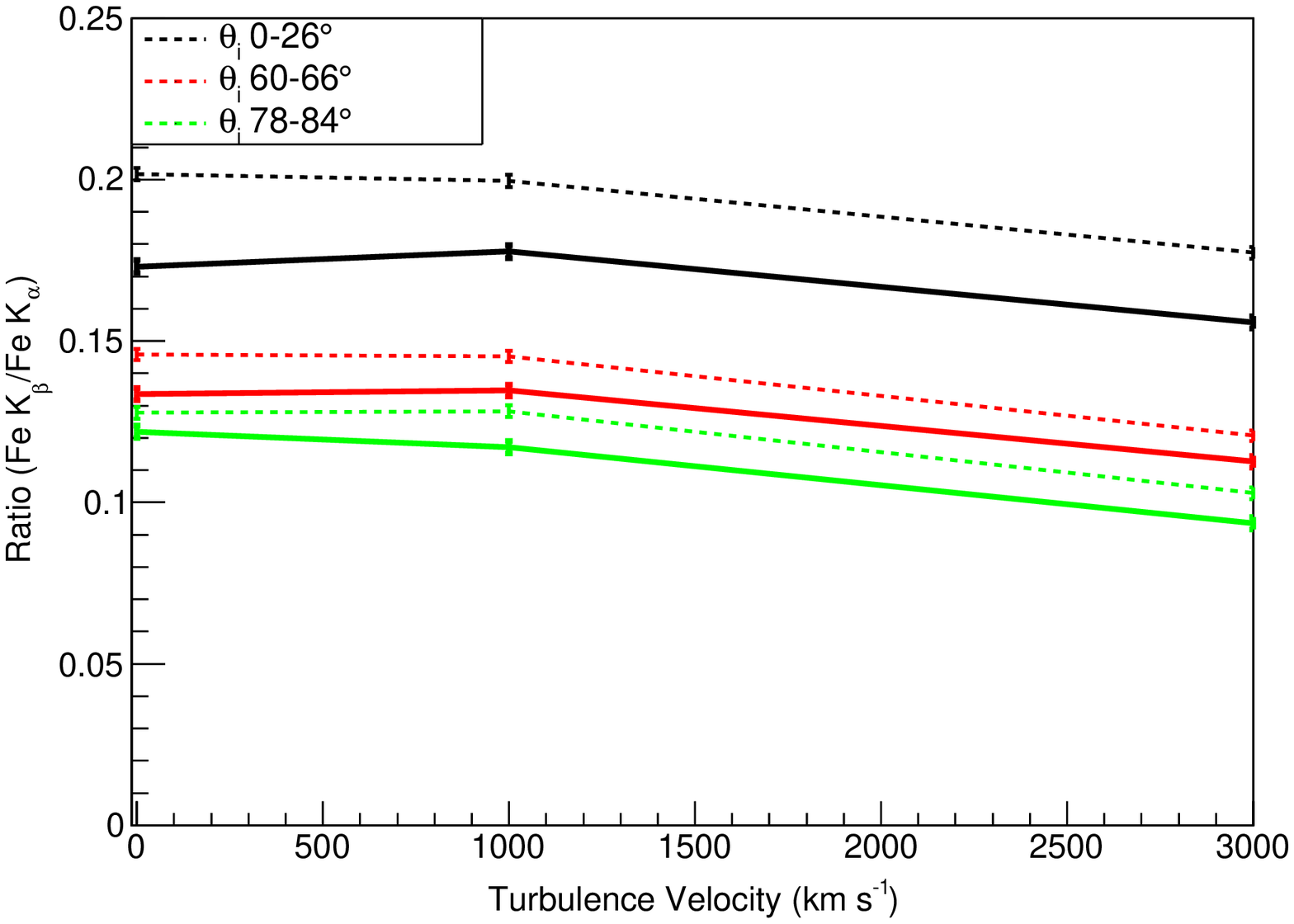}
	\caption{Left panel is a Fe-K line (core plus Compton shoulder) equivalent width (EW) against the
  turbulence velocity for three inclination agn;es. 
  Right panel is a flux ratio of Fe-K$\beta$
  to Fe-K$\alpha$ against the turbulence velocity. 
  Dashed and solid lines
  correspond to the smooth and torus case, respectively.}
	\label{fig:dependence_v}
\end{figure*}

The matter in the torus might have a Keplarian and random velocity,
but here we artificially give a random motion.
Figure \ref{fig:spectra_v} shows the simulated reprocessed spectra 
for three $v_{\rm turb}$ for the smooth torus.
Fe-K line width is certainly broadened, and the interesting feature is
that a part of broadening K$\beta$ line exceeds the absorption edge
energy at 7.112 keV (X-ray Transition Energy Database
\footnote{\tt http://www.nist.gov/physlab/data/xraytrans/index.cfm}).
We can see this fact quantitatively in the K$\beta$ to K$\alpha$ line
ratio in Figure \ref{fig:dependence_v}.
This ratio decreases toward a high random velocity.
Of course, this behavior could be dependent on the velocity field.
In other words, we can extract the information of the velocity field in
the torus from the K$\beta$ to K$\alpha$ line ratio.

Figure \ref{fig:dependence_v} shows that the K$\beta$ to K$\alpha$ line
ratio of the clumpy torus is systematically smaller than that of the 
smooth one for any velocity dispersion; 
this difference is due to the difference of torus structure, smooth or
clumpy.
The difference is more prominent for a small inclination angle.
It is found that 
line intensity ratio of clumpy to smooth torus is smaller for K$\beta$
line than for K$\alpha$ line.
This could be explained as follows.
Contrary to the smooth case where line photons escaped from torus can
be directly observed, line photons escaped from one clump are absorbed
or scattered by front clumps before observed.
As a result, the line intensity of the clumpy torus becomes smaller 
than that of the smooth one.
For a smooth torus, observed line photons are generaed mainly at the
surface region with a thickness $l_{\rm Fe}$.
In this case, K$\beta$ photons can escape more efficiently than
K$\alpha$ photons.
On the other hand, for a clumpy torus, generated K$\alpha$ line 
photons can escape as efficienctly as K$\beta$ photons, 
since a clump size $aR_{\rm torus}$ is smaller than a mean free path 
$l_{\rm Fe}$.
Then, the K$\beta$ to K$\alpha$ line ratio becomes relatively smaller in
the clumpy torus.


\clearpage

\subsection{Dependence of volume filling factor and radius of clumps (Clumpy)}

\begin{figure*}[htbp]
  \begin{center}
    \includegraphics[clip, width=80mm]{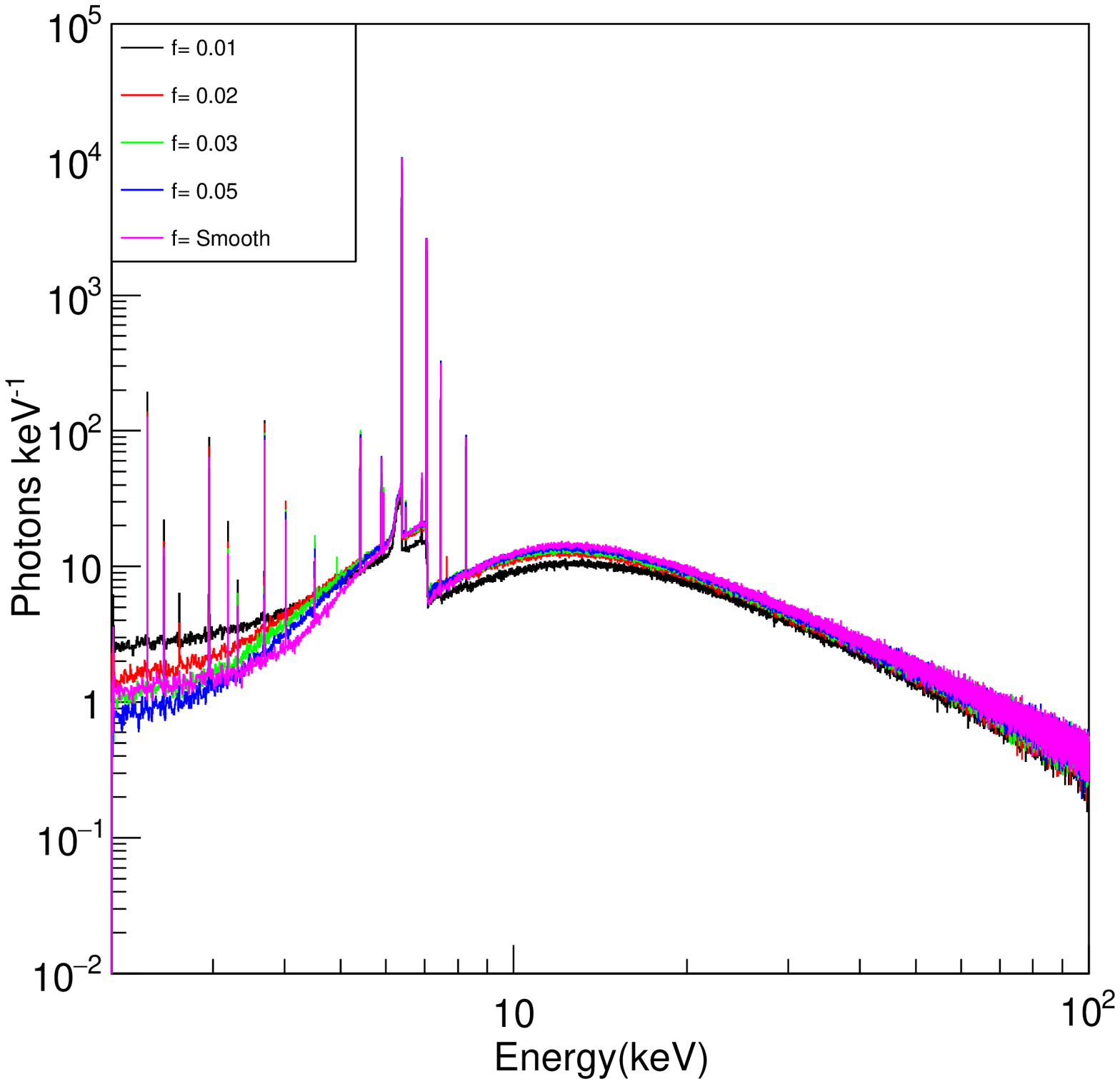}
    \includegraphics[clip, width=80mm]{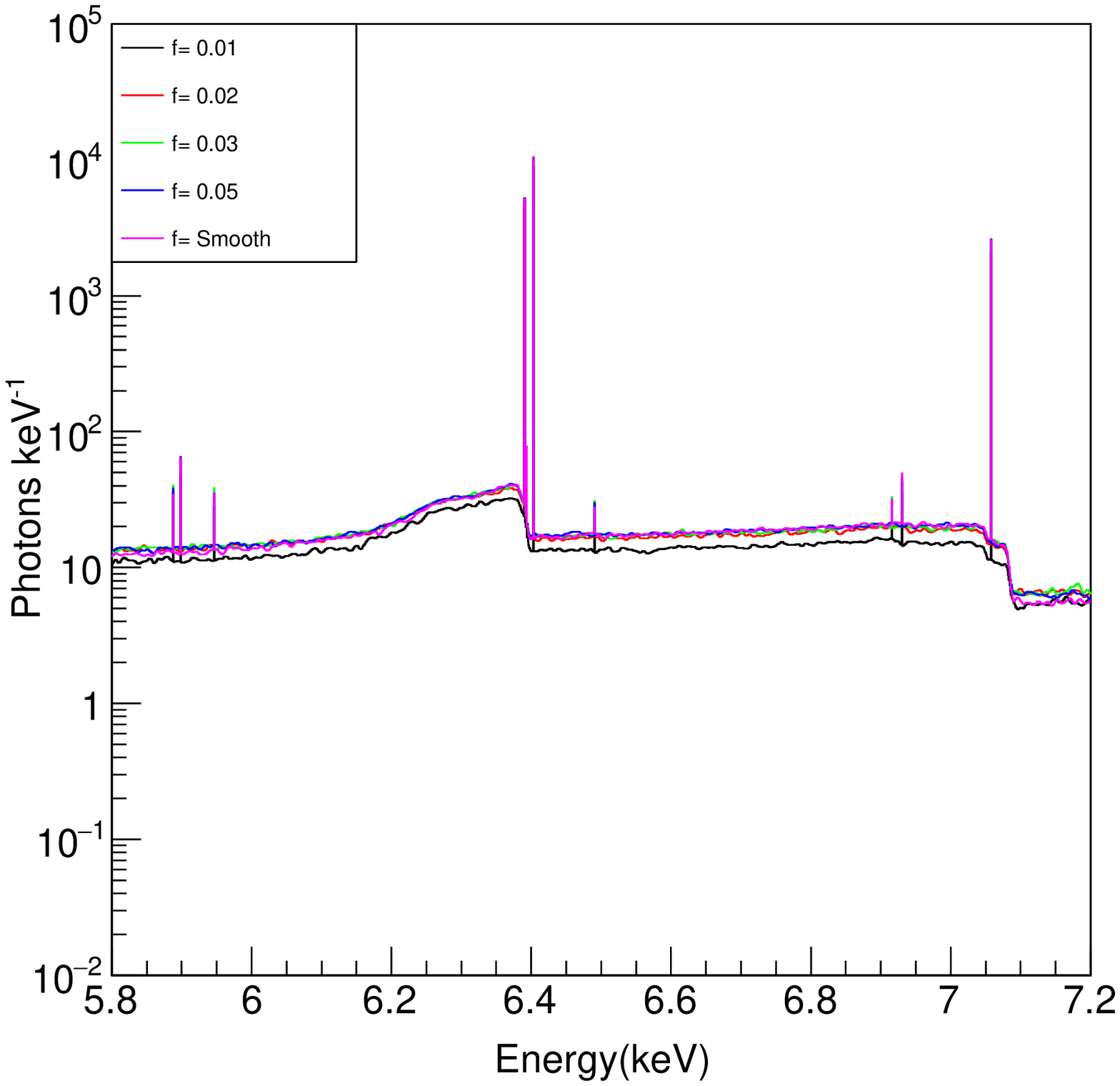}
    \caption{Spectra of the reprocessed component in the case of
  clumpy torus for various volume filling factor $f$. 
  Right panel is an enlargement around the Fe-K line.}
    \label{fig:spectra_vff_clump_0.1_0.2}
  \end{center}
\end{figure*}

\begin{figure}[htbp]
  \begin{center}
    \includegraphics[clip, width=80mm]{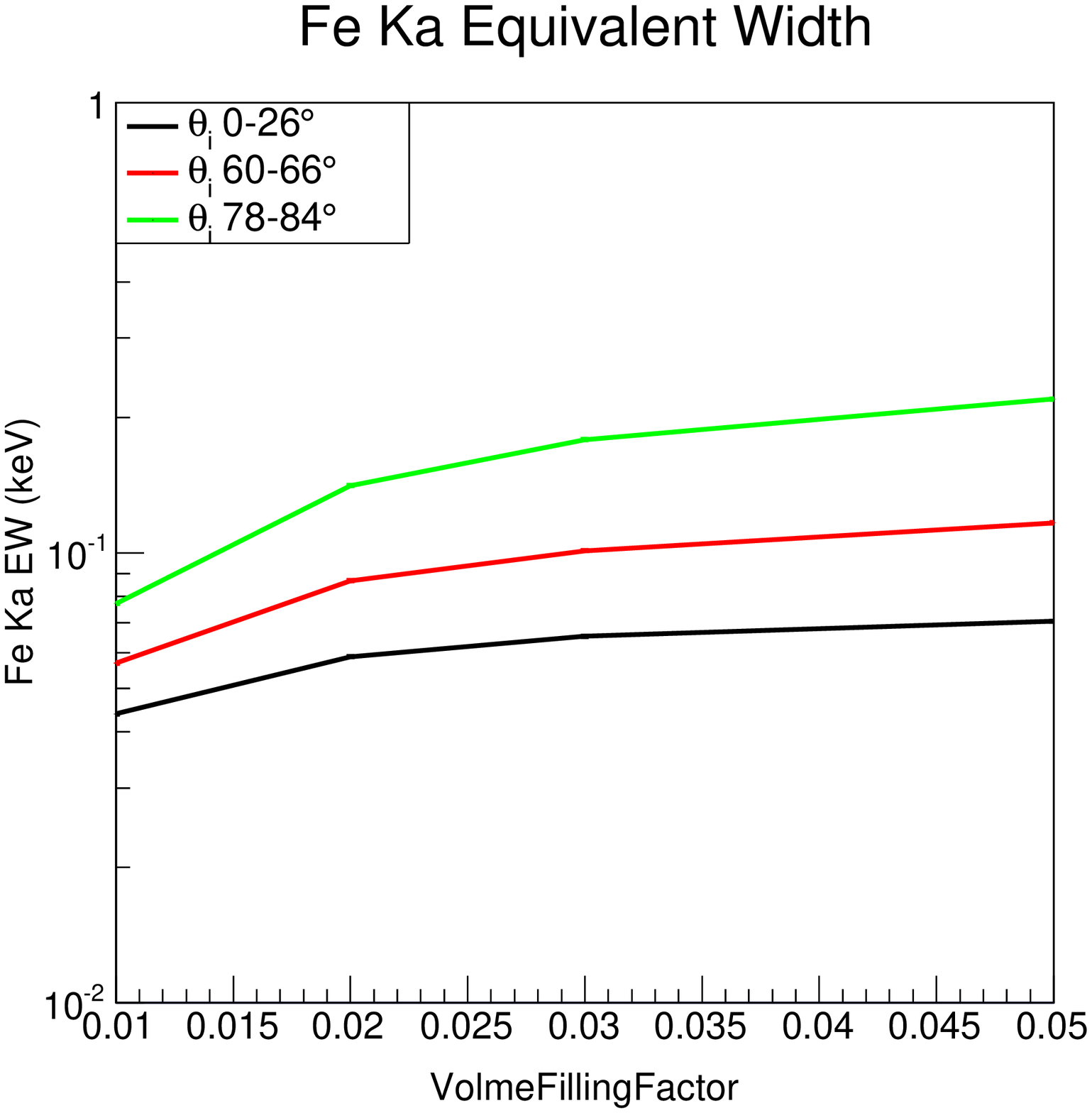}
    \includegraphics[clip, width=80mm]{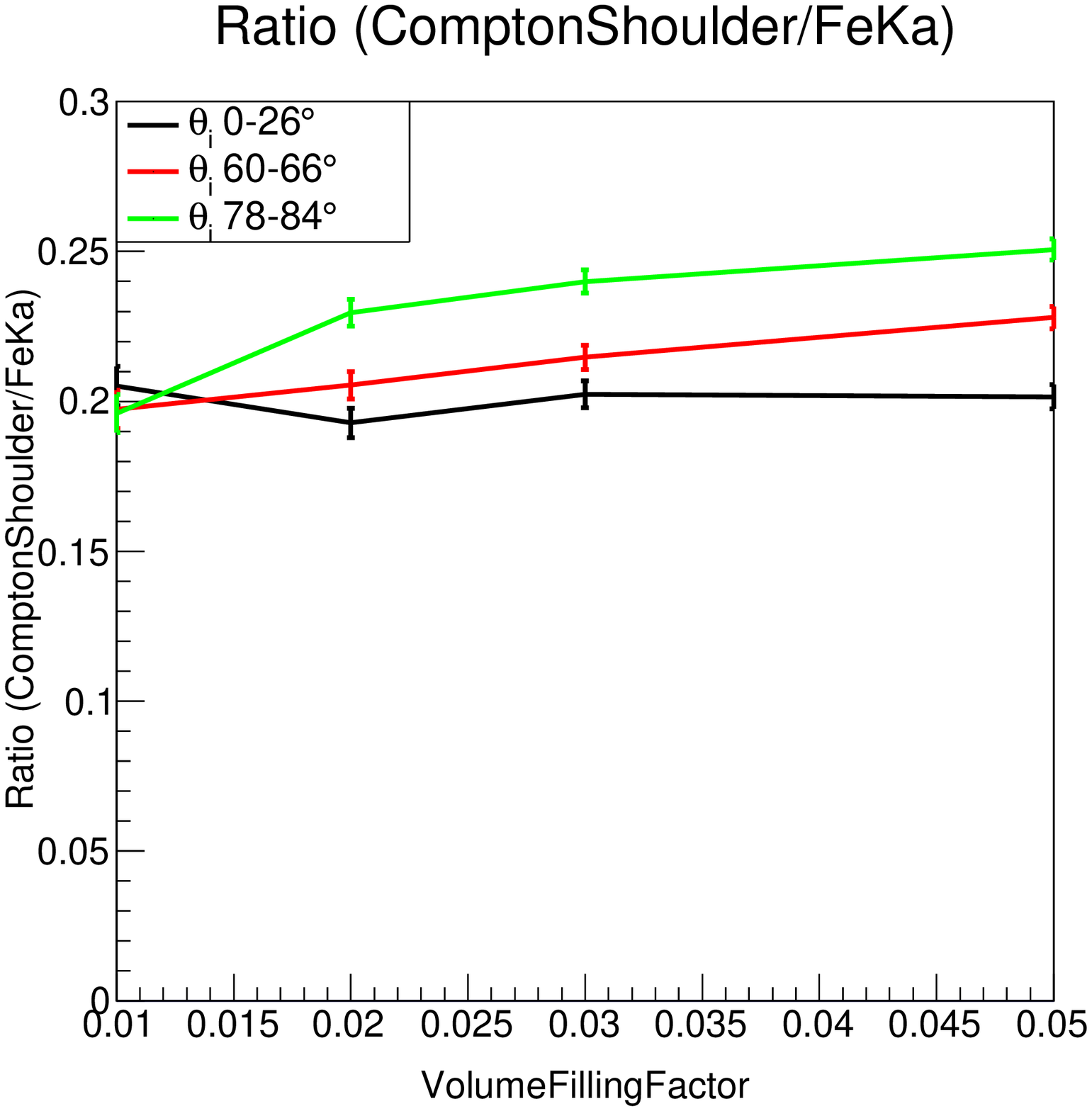}
    \caption{Left panel is a Fe-K line (core plus Compton shoulder) equivalent width (EW) against the
  volume filling factor with $\theta_i=0.1-0.2$. 
  Right panel is a flux ratio of Fe-K
  Compton shoulder to line core against the volume filling factor for
   three inclination angles. Both panels are in
  the case of clumpy torus.}
    \label{fig:dependence_vff_clump}
  \end{center}
\end{figure}

We studied the dependence of volume filling factor under the condition
that the total $N_{\rm H}$ at the torus mid-plane is fixed to be 
$10^{24}$ cm$^{-2}$ and 
the clump radius is fixed to $aR_{\rm torus}=0.005R_{\rm torus}$.
Accordingly, the number of clump is proportional to the volume filling
factor $f$ while the column density of each clump is inversely
proportional.
Figure \ref{fig:spectra_vff_clump_0.1_0.2} shows the reprocessed spectra.
We see a difference among spectra in the low and high energy.
In the high energy, the flux is smaller for a smaller $f$, due to a
lower scattering efficiency in the torus.
On the other hand, in the low energy; the flux is higher for a smaller $f$ 
since the scattered low-energy
photons can easily escape from the torus for a small $f$.
On the other hand, the shape of Compton shoulder is almost identical
among different $f$.
We also show the case of the smooth torus for reference, but the shape
is almost the same.
Quantitatively as shown in figure \ref{fig:dependence_vff_clump},
there is only a weak dependence on $f$ for the EW and the
Compton shoulder fraction.

\begin{figure*}[htbp]
  \begin{center}
    \includegraphics[clip, width=80mm]{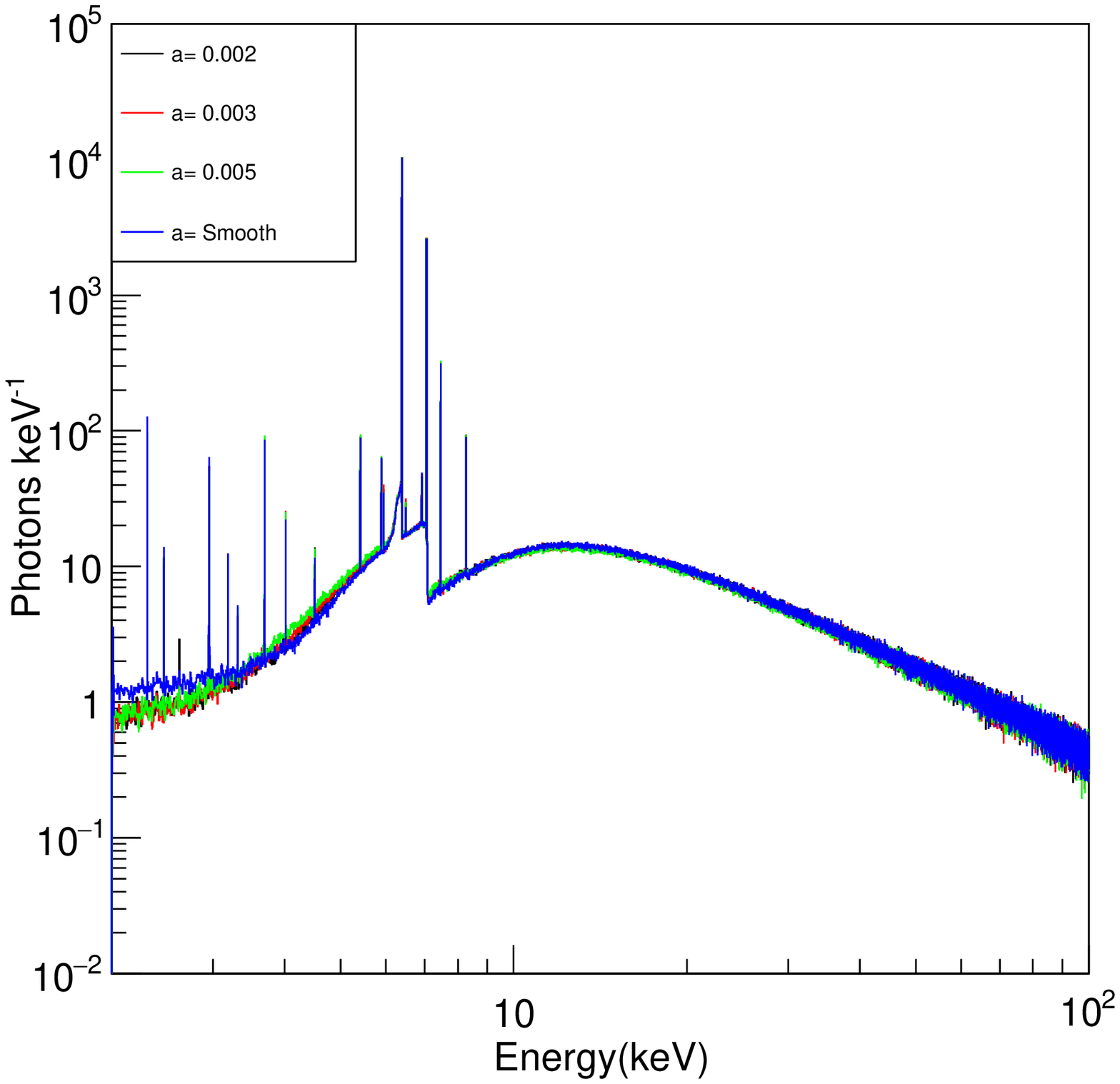}
    \includegraphics[clip, width=80mm]{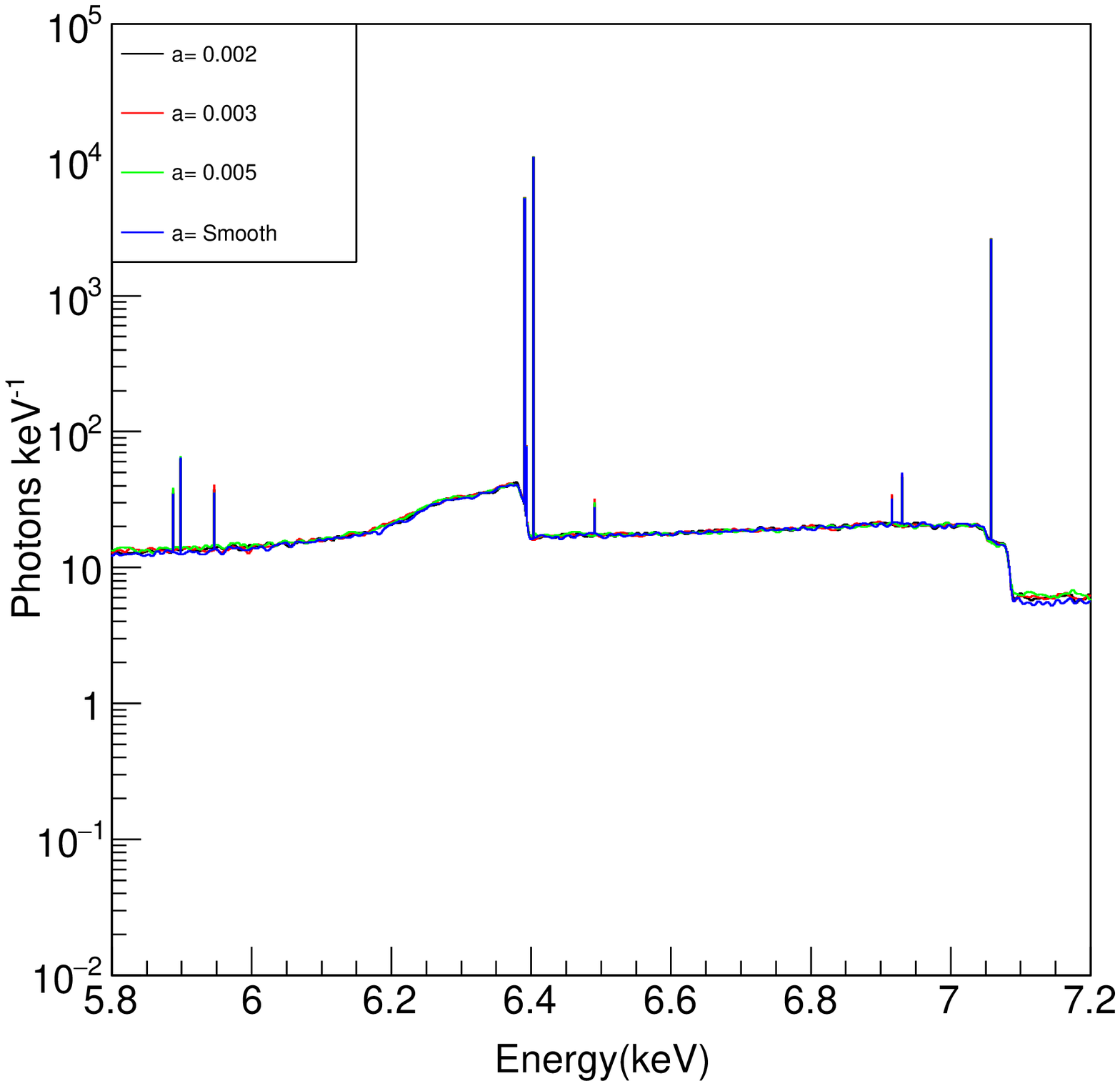}
    \caption{Spectra of the reprocessed component in the case of
  clumpy torus for various clump radii $a$ with $\cos \theta_i=0.1-0.2$. 
  Right panel is an enlargement around the Fe-K line.}
    \label{fig:spectra_clumpscale_clump}
  \end{center}
\end{figure*}

We next varied a clump radius $aR_{\rm torus}$ as $a=$ 0.005, 0.003,
0.002 under the condition that the total $N_{\rm H}$ at the torus mid-plane
and the volume filling factor $f$ are constant at $10^{24}$
cm$^{-2}$ and 0.05, respectively.
Then, the number of clumps in the torus varies as $\propto a^{-3}$.
As shown in figure \ref{fig:spectra_clumpscale_clump}, the reprocessed
spectra at higher energy bands are almost identical except for the lowest energy part.
As a result, we do not see any dependence on the clump radius for the
EW and the fraction of Compton shoulder.

\clearpage

\section{Discussion}

{\it ASTRO-H} SXS (Takahashi et al. 2014) will for the first time enable us to perform 
unprecedentedly fine spectroscopy around the Fe-K line.
Compton shoulder and absorption edge structures of the reflection
component are interesting and important targets for SXS.
Chandra HETG has resolved Compton shoulder clearly for X-ray binaries
(Watanabe et al. 2003; Torrej\'on et al. 2010).
On the other hand, 
Compton shoulder was marginally resolved with HETG for a Seyfert 2
NGC 3783 (Kaspi et al. 2002; Yaqoob et al. 2005) with a very long exposure of 800--900 ks,
and for a Seyfert 2
galaxy NGC 4507 with an exposure of 140 ks but with a low significance
(Matt et al. 2004).
Therefore, SXS will open a new window of Fe-K line spectroscopy with
Compton shoulder for AGNs with a normal exposure.

Here we demonstrate simulated spectra of SXS with our constructed
X-ray reflection simulator.
Models and parameters of simulated spectra are based on the Suzaku
observation of a Compton-thick Seyfert 2 galaxy Mrk 3 (Awaki et
al. 2008).
We model the AGN spectrum with the model {\tt phabs*(powerlaw +
zvphabs*powerlaw + reflection)} in the {\tt XSPEC} model.
Here, {\tt reflection} is a table model for a clumpy torus, 
generated by our AGN X-ray reflection simulator.
Table \ref{table:param_nh_inc} summarizes model parameters.
In this case, the X-ray flux in 2--10 keV is $8\times10^{-12}$ erg
cm$^{-2}$ s$^{-1}$.
We simulated the SXS spectrum with 500 ks exposure, by using SXS response
matrices.

Since the advantage of SXS spectroscopic power is to resolve the Compton
shoulder with a good signal-to-noise ratio for the first time, 
we try to constrain torus parameters by
spectral fitting only around the Fe-K line.
The spectral model is the same as the input model; an absorbed power-law 
and a smooth torus reflection or a clumpy torus reflection.
The parameters of absorption and powerlaw are fixed to the input
value, except for a powerlaw normalization (the 6th parameter in table \ref{table:param_nh_inc}).
Metal abundance, turbulence velocity, and volume filling factor are also
fixed to the input value.
That is, free parameters are a powerlaw normalization, torus column
density, torus inclination angle, and reflection normalization.
We limited the energy band to 6.0--6.7 keV, and obtained the
confidence contour.

\begin{figure*}[htbp]
  \begin{center}
    \includegraphics[width=80mm]{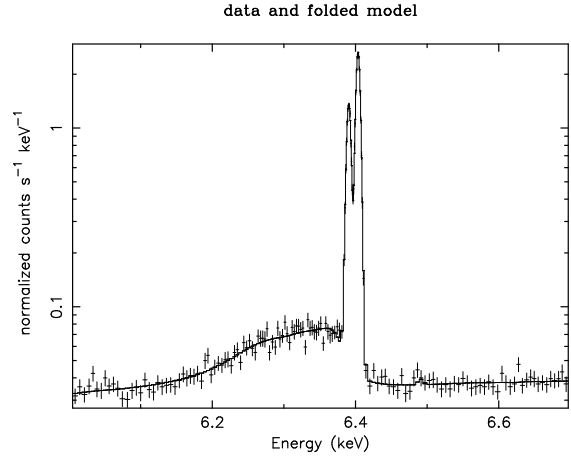}
   \vspace*{1cm}
    \caption{Simulated {\it ASTRO-H} SXS spectra for Mrk 3 around the Fe-K
   line (6.0--6.7 keV), based on our clumpy torus model. Solid line represents the
   best-fit model. See the detail in the text.}
    \label{fig:astroh_spectra}
  \end{center}
\end{figure*}

\begin{figure*}[htbp]
  \begin{center}
    \includegraphics[width=80mm]{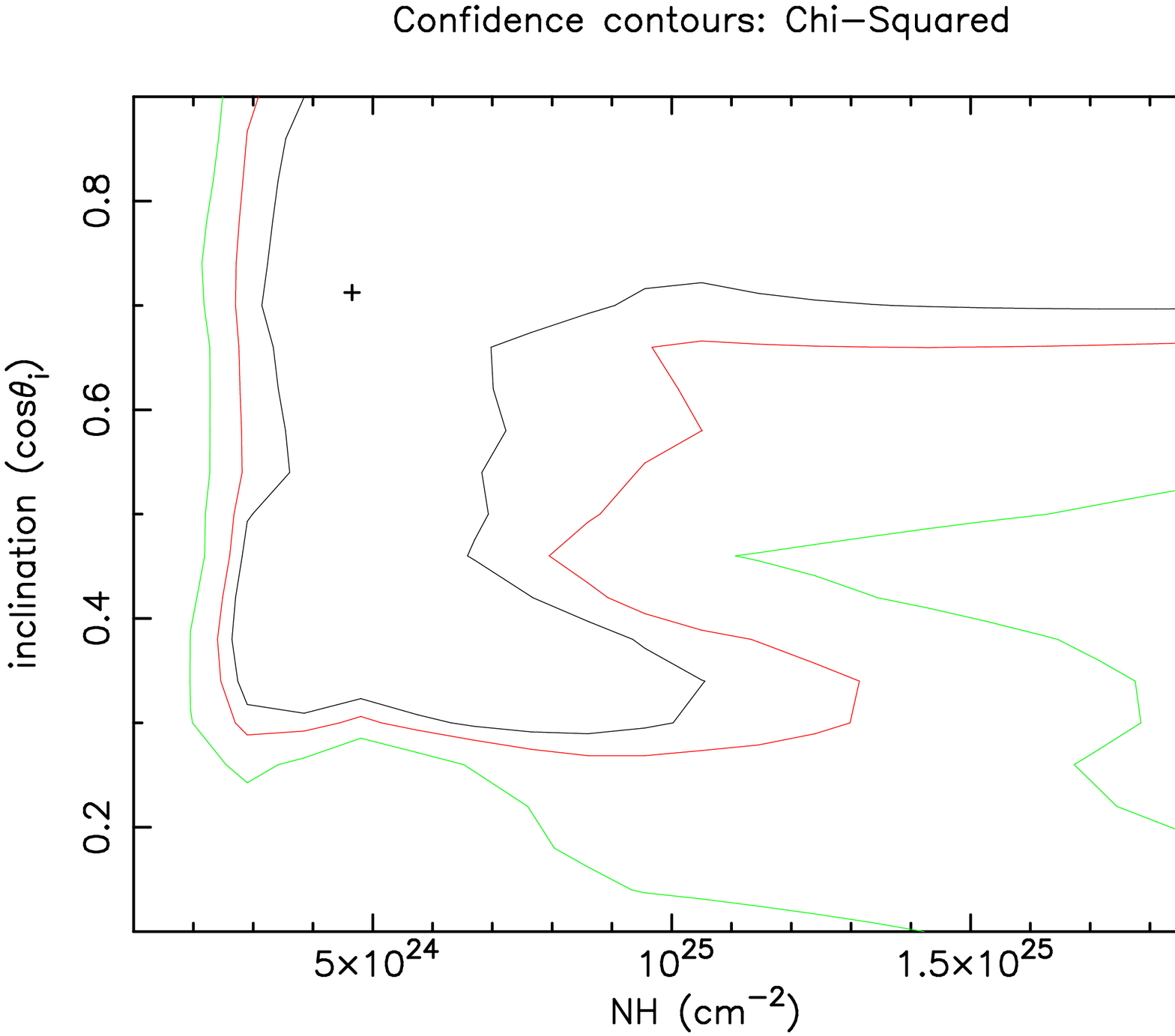}
    \includegraphics[width=80mm]{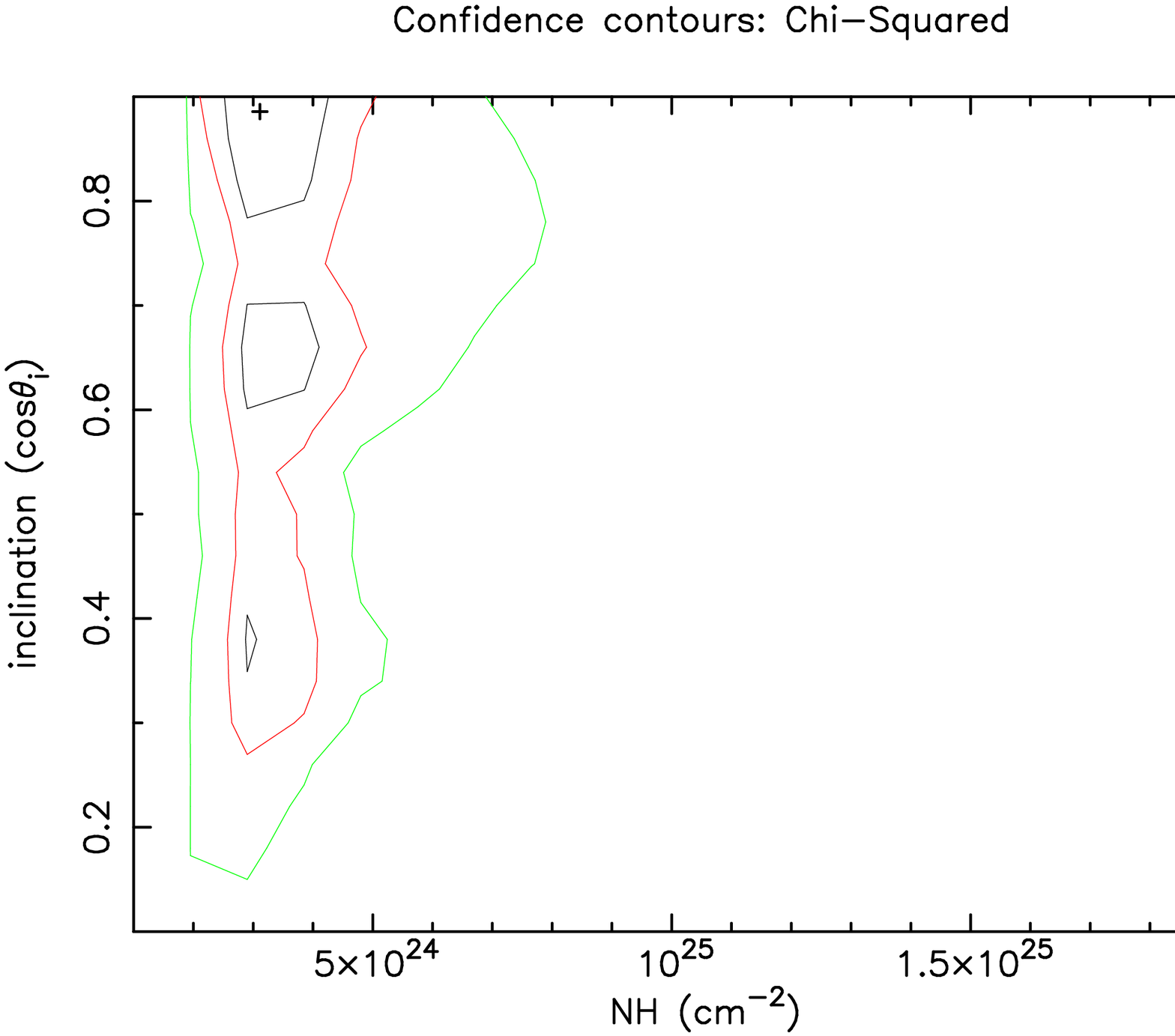}
\vspace*{1cm}
    \caption{Confidence contours between column density and inclination
   angle 
   for spectral fitting of the simulated
   {\it ASTRO-H} SXS spectrum only around the Fe-K line. 
   Black, red, and green lines represent 68\%, 90\% and 99\% confidence levels, respectively.
   See the detail in the
   text. Left and right are for the smooth and clumpy cases,
   respectively. ``+'' reoresents a location of the best-fit values.}
    \label{fig:astroh_contour}
  \end{center}
\end{figure*}

Figure \ref{fig:astroh_spectra} shows a simulated SXS spectrum around the Fe-K
line. Compton shoulder is clearly resolved, and also K$\alpha_1$ and
K$\alpha_2$ are separated, demonstrating the SXS power.
Figure \ref{fig:astroh_contour} shows confidence contours 
between column density and
inclination angle for fitting with smooth or clumpy torus.
These figures demonstrate that spectral fitting only around Fe-K$\alpha$ 
line can constrain reflection parameters by using the Compton shoulder,
and thus gives us a new tool to study the torus structure.
Thus, together with the broad-band spectral fitting of continuum, better
constraint could be available.
The behavior of contours are different between smooth and clumpy models.
This is due to the different dependence on the column density for the
Compton shoulder to core intensity ratio between two models as shown in
figure \ref{fig:dependence_nh_clump} right.

Therefore, this demonstrates that the correct modeling of the Compton
shoulder is very important to probe the torus condition by using the
Compton shoulder.
In terms of this view point, our model considers the different shape
of Compton shoulder between bound and free electrons 
and also that between atoms and molecules.

%$\theta_\text{oa}=60^{\circ}$、$R_\text{torus}=2\times 10^6$ cm、$MA=1$、$V_\text{turb}=0$ km s$^{-1}$、$a=0.005$。

\begin{table}[htb]
\begin{center}
\caption{Model parameters for {\it ASTRO-H} SXS simulation for the {\tt XSPEC}
 model:  {\tt phabs*(powerlaw + zvphabs*powerlaw + reflection)}}
\begin{tabular}{cccc} \hline\hline
component & parameter & unit & value\\ \hline
zvphabs & $N_{\rm H}$ & $10^{22}$ cm$^{-2}$ & 8.7$\times10^{-2}$ \\
powerlaw & Index &  & 1.8 \\
powerlaw & norm$^a$ & & $1.27\times10^{-4}$ \\
phabs & $N_{\rm H}$ & $10^{22}$ cm$^{-2}$ & 110 \\
powerlaw & Index &  & 1.8 \\
powerlaw & norm$^a$ &  & $5\times10^{-3}$ \\
Reflection & $N_{\rm H}$ & $10^{22}$ cm$^{-2}$  & $3\times10^{2}$ \\
Reflection & Inclination $\cos\theta_i$ &  & 0.5 \\
Reflection & Metal Abundance & solar & 1.0 \\
Reflection & Turbulence Velocity & km s$^{-1}$ & 0 \\
Reflection & Volume Filling Factor &  & 0.005 \\
\hline
\multicolumn{4}{l}{$a$: Normalization in unit of photons cm$^{-2}$ s$^{-1}$ keV$^{-2}$ @ 1 keV. }
\end{tabular}
\label{table:param_nh_inc}
\end{center}
\end{table}

The {\tt MONACO} frame work on which our model is constructed has more
advantages to explore the torus model furthermore.
For example, X-ray reflection spectral
modeling by ionized material and velocity structure of torus material 
can be implemented, as constructed for simulation of AGN outflow 
(Hagino et al. 2015).
Time history of reflection spectra after flares of the central engine 
can be also tracked, as applied for the Galactic center (Odaka et al. 2011).
In contrast with a smooth torus model, 
a clumpy torus likely exhibits
a fluctuation of spectral shape for different azimuthal angles
even with the same inclination angle.
In a forthcoming paper, we will investigate to what
extent such variations will appear.
In this paper, we construct a simple torus geometry, but more complex
structure of torus that has been recently suggested can be modeled.
If the UV from the central accretion disk controls the 
innermost edge of a dusty tori, the
shape of the inner part of the torus must be significantly
different from that of the donut-like tori.
Namely, a strong anisotropy of the disk (Netzer 1987)
makes the innermost edge of the torus concave
(Kawaguchi \& Mori 2010; 2011).
At low latitudes of the torus (close to the midplane
of the disk), the dusty torus exists very close to the
outermost edge of the accretion disk.
This region, where the rotational and turbulent velocity
are likely large (some thousands km s$^{-1}$; Kawaguchi 2013),
may affect the line profile of emission lines (e.g., core to
wing ratio), time response of the line flux and profile,
and the viewing angle dependence of the Fe-K lines.
In a future paper, we will examine how the
innerpart structure of the torus changes the
overall results.

\section{Conclusions}

We construct an X-ray spectral model of a torus in AGN 
with a Monte Carlo simulation framework {\tt MONACO}.
Two torus geometries of smooth and clumpy cases are considered.
In order to reproduce a Compton shoulder accurately, {\tt MONACO} 
includes not only free electron scattering but also bound electron scattering.
Raman and Reyleigh scattering are also treated, and scattering cross
 sections dependent on chemical states of hydrogen and helium are
 included.
Doppler broadening by turbulence velocity can be implemented.
We compared our simulation spectra with a widely used X-ray spectral
model {\tt MYTorus}, and found almost consistent results.
We studied the dependence of reprocessed X-ray spectra, especially for 
Fe-K line Compton shoulder on various torus parameters, such as Hydrogen column
density along a line on the equitorial plane, inclination angle from the
line of sight, metal abundance, turbulence velocity for both smooth and
clumpy cases, volumn filling factor and clump scale radius for the clumpy case.
The fraction and shape of Compton shoulder depends on the column
density, and the dependence is different between smooth and clumpy
cases.
Also, a weak dependence of Compton shoulder shape on the inclination
angle is seen.
An equivalent width of Fe-K core and Compton shoulder and 
a fraction of Compton shoulder largely depend on the metal abundance.
Only a weak dependence of Compton shoulder fraction is seen for
turbulence velocity, filling factor, and clump scale radius.
We found that a intensity ratio of Compton shoulder to line core mainly
 depends on the column density, inclination angle, and metal abundance.
For instance, an increase of metal abundance makes
the Compton shoulder relatively weak.
Also, shape of Compton shoulder depends on the column density.
Then, we present the {\it ASTRO-H} SXS simulated spectra of Mrk 3 
and found to a clear Compton shoulder in the spectrum.
Even with a narrow-band spectral fitting only around the Compton shoulder,
we will be able to put some constraints on the torus geometry.

%\bibliography{clumpy_torus}




\clearpage

\appendix

\section*{Position maps of last scattering of Compton shoulder photons}

Please see the original journal for figure \ref{table:posmap_6-6.39keV},
since the file size exceeds the astro-ph limit.

\begin{figure*}[htbp]
    \caption{Position maps of last scattering of Compton shoulder
 photons in 6--6.39 keV for various column densities $N_{\rm H}$. 
 The observer locates toward the inclination angle of $\cos\theta_i=0.1-0.2$
 in the X-Z plane. Left
 panle is a map in the X-Y plane, where the
observer is located towards the right direction. 
Middle-left, middle-right, are 
 right are maps within the torus integrated over the 1st quadrant ($y>-x$ and $y<x$; i.e., close to the observer),
 2nd quadrant ($y>-x$ and $y>x$), and 3rd quadrant ($y<-x$ and $y>x$; i.e., behind the central
engine with respect to the observer), respectively.}
\label{table:posmap_6-6.39keV} 
\end{figure*}

\clearpage

\section*{Comparison of Reprocessed Spectra between Clumpy and Smooth Torus}

\begin{figure*}[htbp]
	\includegraphics[width=5in]{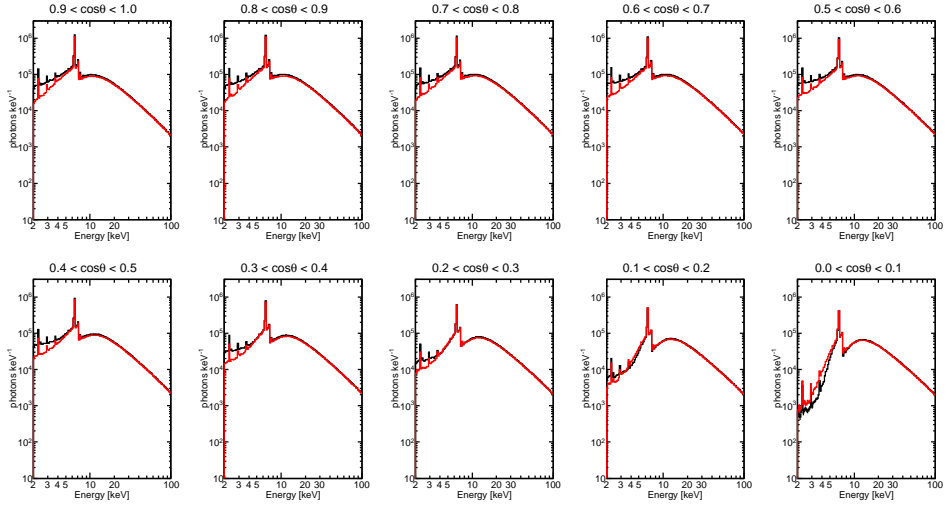}
	\caption{Comparison of Reprocessed Spectra between Clumpy and
 Smooth Torus for various inclination angles of $\cos\theta_i=0-1$ with
 a step of 0.1 in the case of $N_{\rm H}=10^{24}$ cm$^{-2}$. Black and red spectra correspond to smooth and clumpy
 torus, respectively.
}
	\label{fig:specratio3a}
\end{figure*}

\begin{figure*}[htbp]
	\includegraphics[width=5in]{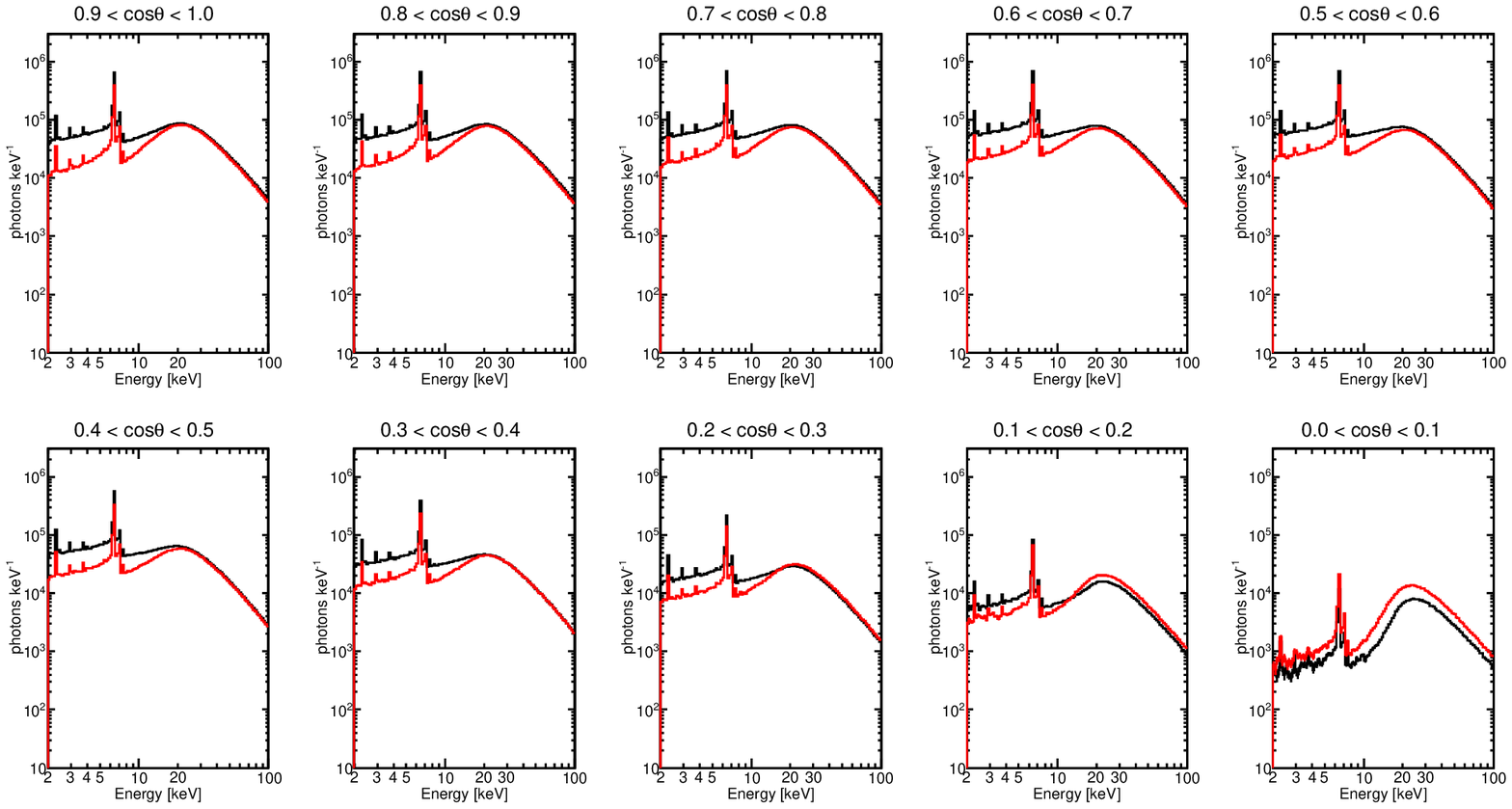}
	\caption{Same as figure \ref{fig:specratio3a} but in the case of $N_{\rm H}=10^{25}$ cm$^{-2}$.
}
	\label{fig:specratio3b}
\end{figure*}

\begin{figure*}[htbp]
	\includegraphics[width=5in]{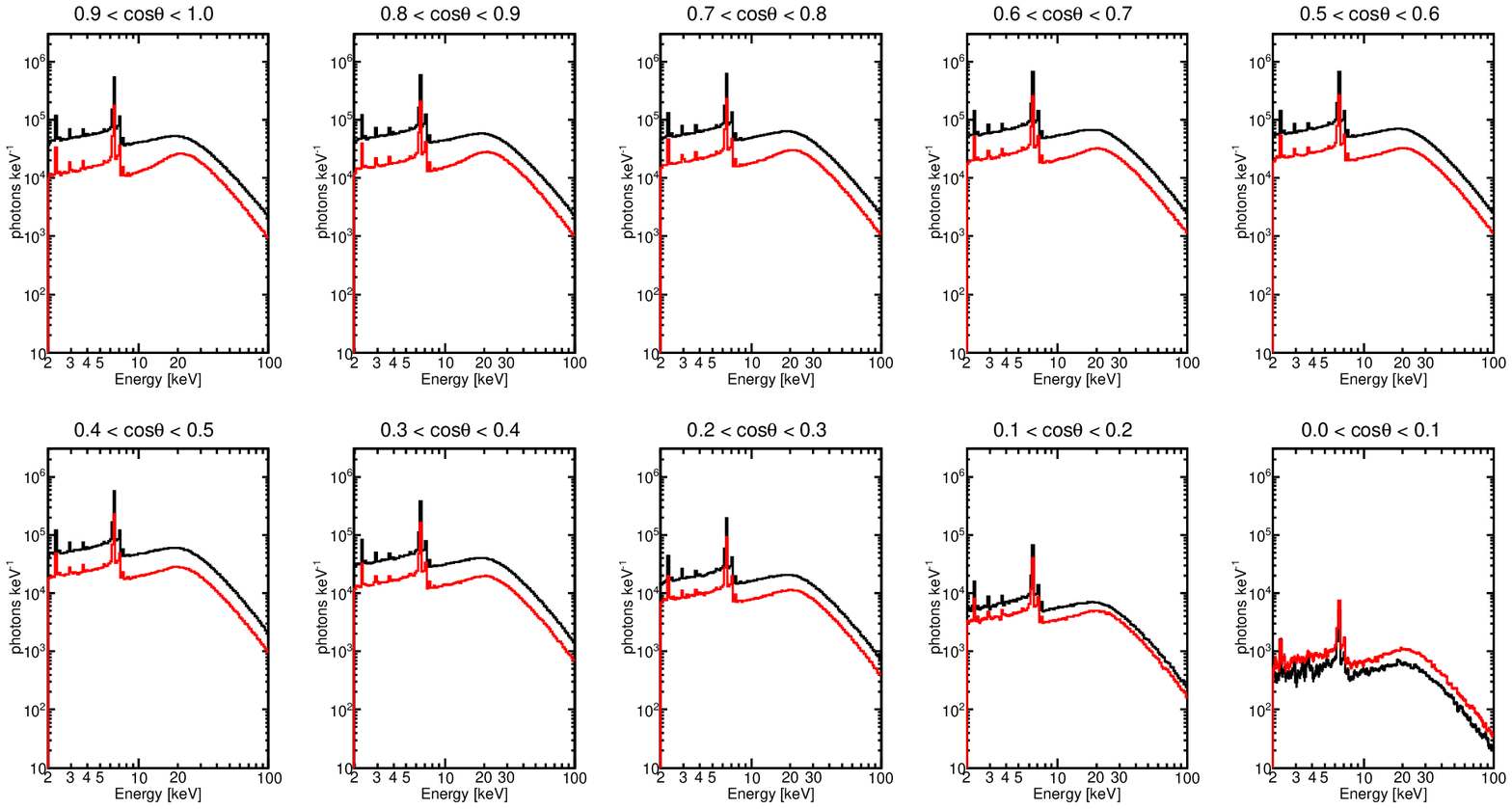}
	\caption{Same as figure \ref{fig:specratio3a} but in the case of $N_{\rm H}=10^{26}$ cm$^{-2}$.
}
	\label{fig:specratio3c}
\end{figure*}

\end{document}